\documentclass[aps,pra,amssymb,superscriptaddress,twocolumn]{revtex4-1}
\usepackage[pdftex]{graphicx}
\usepackage[dvipsnames]{xcolor}
\usepackage{bm}
\usepackage{amsmath}
\usepackage{ascmac}
\usepackage{color}
\usepackage{setspace}
\usepackage{amssymb}
\usepackage{latexsym}
\usepackage{mathtools}
\usepackage{lipsum}
\usepackage{braket}
\usepackage[colorlinks,linkcolor=RoyalBlue,citecolor=RoyalBlue,urlcolor=RoyalBlue]{hyperref}
\usepackage{MnSymbol}

\usepackage{amsthm}
\theoremstyle{plain}
\newtheorem*{conj}{Conjecture}
\allowdisplaybreaks

\begin{document}

\title{Quantum Transport in Interacting Spin Chains: Exact Derivation of the GUE Tracy-Widom Distribution} 

\author{Kazuya Fujimoto}
\affiliation{Department of Physics, Institute of Science Tokyo, 2-12-1 Ookayama, Meguro-ku, Tokyo 152-8551, Japan}

\author{Tomohiro Sasamoto}
\affiliation{Department of Physics, Institute of Science Tokyo, 2-12-1 Ookayama, Meguro-ku, Tokyo 152-8551, Japan}

\date{\today}

\begin{abstract}
We theoretically study quantum spin transport in a one-dimensional folded XXZ model with an alternating domain-wall initial state via the Bethe ansatz technique, exactly demonstrating that a probability distribution of finding a left-most up-spin with an appropriate scaling variable converges to the Tracy-Widom distribution for the Gaussian unitary ensemble (GUE), which is a universal distribution for the largest eigenvalue of GUE under a soft-edge scaling limit. Our finding presented here offers a first exact derivation of the GUE Tracy-Widom distribution in the dynamics of the interacting quantum model not being mapped to a noninteracting fermion Hamiltonian via the Jordan-Wigner transformation. On the basis of the exact solution of the folded XXZ model and our numerical analysis of the XXZ model, we discuss a universal behavior for the probability of finding the left-most up-spin in the XXZ model.
\end{abstract}

\maketitle

{\it Introduction.--}
Transport of a physical quantity is ubiquitous both for classical and quantum systems, having played pivotal roles in deepening our understanding of many-body dynamics over decades~\cite{Dhar2008,Nagaosa2010,Marchetti2013,Bertini1}. One of the notable achievements in classical transport is the establishment of the celebrated Kardar-Parisi-Zhang (KPZ) universality~\cite{Sasamoto2007,Krug2010,CORWIN2012,Quastel2015,takeuchi2018}, which was originally developed in classical statistical mechanics for growing surface physics~\cite{barabasi1995} and transport of stochastic processes~\cite{Schmittmann1995,liggett2013}. When a stochastic system belongs to the KPZ universality, the integrated particle current are universally characterized by the Tracy-Widom distribution of random matrix theory, which is a universal distribution for the largest eigenvalue of random matrices~\cite{Tracy1993,Tracy1996,Forrester1993,forrester2010}. Recently, such universal transport featuring random matrix theory and the KPZ universality is intensively explored in quantum regimes from theoretical~\cite{Nahum2017, Ljubotina2019, Dupont2020, Nardis2020, Fujimoto2020, Bingtian2022, Moca2023, Nardis2023, Sen2024, Cecile2024, Gopalakrishnan2024_r, Aditya2024} and experimental~\cite{Scheie2021, Wei2022, Rosenberg2024} perspectives, having been recognized as an important research subject in quantum many-body systems. 

One of intriguing exact results for quantum transport featuring random matrix theory is emergence of the Tracy-Widom distribution in a one-dimensional XX model being equivalent to noninteracting fermions~\cite{Eisler2013,Saenz2022}. The previous works of Refs.~\cite{Eisler2013,Saenz2022} consider quantum spin transport starting from a domain-wall state, uncovering that a probability $P(x,t)$ for the farthest up-spin at site $x$ and time $t$ obeys the GUE Tracy-Widom distribution~\cite{Tracy1993,Forrester1993,forrester2010}, which is the universal distribution for the largest eigenvalue in the Guassian unitary ensemble (GUE) of random matrix theory. After this finding, several numerical works~\cite{Collura2018,Stephan2019} studied impact of interactions on the quantum dynamics using a one-dimensional XXZ model, which is mapped into an interacting fermions, and then reported a signature for absence of the  GUE Tracy-Widom behavior. On the other hand, Bulchandani and Karrasch reported tendencies for presence of the GUE Tracy-Widom behavior~\cite{Bulchandani2019}. On the mathematical side, Saenz, Tracy, and Widom conducted pioneering and laborious analysis for the probability $P(x,t)$ in the XXZ model via the Bethe ansatz~\cite{Takahashi1999,Franchini2017}, proposing an important conjecture concerning a scaling limit for $P(x,t)$~\cite{Saenz2022}. However, exact derivation of the GUE Tracy-Widom distribution in the XXZ model has yet to be completed. Therefore, it has been elusive whether the GUE Tracy-Widom behavior can survive in interacting quantum many-body systems. 

\begin{figure}[b]
\begin{center}
\includegraphics[keepaspectratio, width=8.7cm]{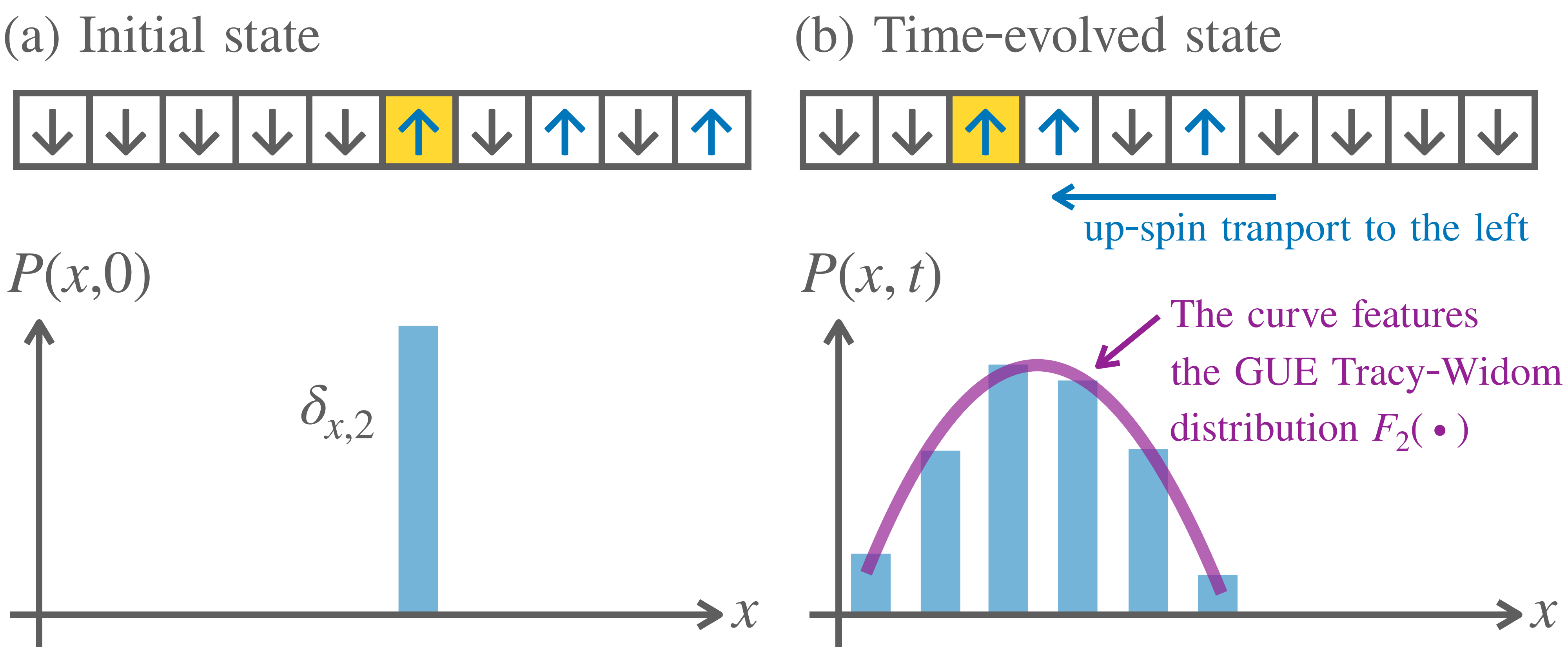}
\caption{
Schematic illustration for the main result of this Letter. 
The prime quantity of our interest is a probability $P(x,t)$ of finding a left-most up spin at site $x$ and time $t$, which is emphasized by the yellow-color cell. (a) Initial spin configuration and probability $P(x,0)$. The initial state is an alternating domain-wall state, where the up spins occupy every other sites in half of the system ($ x \geq 2$). By definition, we have $P(x,0) = \delta_{x,2}$ (see Eqs.~\eqref{IS} and \eqref{Plmup}). (b) Time-evolved spin configuration and probability $P(x,t)$ at time $t$. After the unitary time-evolution with the folded XXZ model, the up-spins are transported to the left region ($x<2$). In this Letter, we demonstrate that the rescaled probability $P(x,t)$ converges to a probability density function for the GUE Tracy-Widom distribution function $F_2(\bullet)$~\cite{Tracy1993,Forrester1993,forrester2010} in the long-time limit. 
} 
\label{fig1} 
\end{center}
\end{figure}

In this Letter, we present a first example for exactly deriving the GUE Tracy-Widom distribution in an interacting quantum spin model on a one-dimensional lattice, namely a folded XXZ model~\cite{Yang2020,Zadnik2021_1,Zadnik2021_2,Pozsgay2021,Pozsgay2021_2,Borsi2023}, which cannot be mapped into a noninteracting fermion Hamiltonian via the Jordan-Wigner transformation~\cite{Sachdev1999,Lewenstein2012}. This theoretical model is originally derived as an effective Hamiltonian for the XXZ model with the large anisotropic interaction. Using the folded XXZ model, we theoretically study quantum spin transport starting from an alternating domain-wall state where the up-spin occupies half of the system every other site as depicted in Fig.~\ref{fig1}(a). We employ exact analysis based on the Bethe ansatz~\cite{Takahashi1999,Franchini2017}, analytically showing that the rescaled probability $P(x,t)$ of finding the left-most up-spin at site $x$ and time $t$ converges to a probability density function for the GUE Tracy-Widom distribution function $F_2(\bullet)$ in the long-time limit. Figure~\ref{fig1}(b) schematically illustrates this result. Beyond the folded XXZ model, we numerically investigate the XXZ model via the time-evolving decimation method (TEBD)~\cite{TEBD1,TEBD2,TEBD3,TEBD4}, discussing signatures for a universal behavior of $P(x,t)$ in the XXZ model.

{\it Setup.--}
We consider an infinite lattice, sites of which are labeled by $\mathbb{Z}$, and denote spin-1/2 operators at site $w \in \mathbb{Z} $ by $\hat{X}_w, \hat{Y}_w$ and $\hat{Z}_w$ in the $x, y,$ and $z$ directions, respectively. These operators satisfy SU(2) commutation relations, e.g., $[\hat{X}_x, \hat{Y}_{y}] = {\rm i} \hat{Z}_x \delta_{x,y}$, where we set the Dirac constant $\hbar$ to be unity. Under this setup, we consider the Hamiltonian of the folded XXZ model~\cite{Yang2020,Zadnik2021_1,Zadnik2021_2,Pozsgay2021,Pozsgay2021_2,Borsi2023}, which is defined by
\begin{align} 
\hat{H}_{\rm fXXZ} \coloneqq \sum_{x \in \mathbb{Z}} \left( \hat{X}_{x} \hat{X}_{x+1} + \hat{Y}_{x} \hat{Y}_{x+1}  \right) \dfrac{ 1 + 4\hat{Z}_{x-1} \hat{Z}_{x+2} }{2}. 
\label{fXXZ}
\end{align}
This is the effective Hamiltonian for the XXZ model with the large anisotropic interaction $\Delta \gg 1$. 
Here, the Hamiltonian for the XXZ model is given by $\hat{H}_{\rm XXZ} \coloneqq \sum_{x \in \mathbb{Z} } \left( \hat{X}_{x} \hat{X}_{x+1} + \hat{Y}_{x} \hat{Y}_{x+1} + \Delta \hat{Z}_{x} \hat{Z}_{x+1}  \right)$ with the parameter $\Delta$ being responsible for the spin interaction in the $z$-direction.
We denote the quantum state by $\ket{\phi(t)}$ and assume that it obeys the Schrödinger equation, ${\rm i} d \ket{\phi(t)}/dt = \hat{H}_{\rm fXXZ} \ket{\phi(t)}$. 
The initial state considered in this work is the alternating domain-wall state defined by 
\begin{align} 
\ket{\phi(0)} = \prod_{x=1}^{N} \hat{R}_{2x} \ket{0}
\label{IS}
\end{align}
with the vacuum $\ket{0}$ representing all the down-spin state, the raising operator $\hat{R}_x \coloneqq \hat{X}_x + {\rm i} \hat{Y}_x$, and the total number $N$ of the up-spins. Figure~\ref{fig1}(a) displays the schematic illustration for this initial state. Since $\hat{H}_{\rm fXXZ}$ conserves the total up-spin, we can expand the quantum state as $\ket{\phi(t)} = \sum_{ \bm{x}: x_j < x_{j+1} } \Phi(x_1,...,x_N,t) \ket{x_1,...,x_N}$, where $x_j~ (j \in \{1,..., N\})$ is a lattice site occupied by an up-spin and $\Phi(x_1,...,x_N,t)$ is the many-body wavefunction in the basis $ \ket{x_1,...,x_N} \coloneqq \prod_{j=1}^N \hat{R}_{x_j} \ket{0}$.

The quantity of our interest is the probability $P(x,t)$ of finding the left-most up-spin at site $x$ and time $t$, which is defined by
\begin{align} 
P(x,t) \coloneqq  \sum_{y_2=1}^{\infty} \cdots \sum_{y_N=1}^{\infty}  \left| \Phi \left(x, x+y_2,...,x+ \sum_{j=2}^N y_j,t \right) \right|^2. 
\label{Plmup}
\end{align}
The corresponding complementary cumulative distribution function $F(x,t)$ is defined by 
\begin{align} 
F(x,t) \coloneqq \sum_{y=x}^{\infty} P(y,t).
\end{align} 
In what follows, we shall prove that $F(x,t)$ converges to the GUE Tracy-Widom distribution function in the long-time limit. 

{\it Determinantal formula of $F(x,t)$ via the Bethe ansatz.--}
We shall derive a determinantal expression for $F(x,t)$ using the Bethe ansatz~\cite{Takahashi1999,Franchini2017} because the folded XXZ model is Bethe solvable~\cite{Zadnik2021_1,Zadnik2021_2,Pozsgay2021}.

We first derive an integral formula for $\Phi(x_1,...,x_N,t)$ by using the Bethe-ansatz method developed by Schütz~\cite{Schutz1997}, Tracy and Widom~\cite{Tracy2008}. 
As described in Sec. I of the Supplemental Material (SM)~\cite{SM}, we obtain
\begin{align} 
\Phi(x_1,...,x_N,t) &= \int_{C_r} d \bm{\xi} \sum_{\sigma \in \mathbb{S}_N }  A_{\sigma}( {\bm \xi} )  \prod_{j=1}^N\xi_{\sigma_j}^{x_{j} - 2 \sigma_j  -1} e^{- {\rm i}E_{\xi_j} t}
\label{Esol1}
\end{align}
with the set $\mathbb{S}_N$ for $N$th permutations and the multiple complex integral $\int_{C_r} d \bm{\xi} \coloneqq (2\pi {\rm i})^{-N} \int_{C_r} d \xi_1 \cdots \int_{C_r} d \xi_N  $, and $E_{\xi} \coloneqq(\xi + \xi^{-1})/2$. The contour $C_r$ is a circle encircling the origin in the complex plane and its radius $r$ is strictly smaller than unity. The coefficient $A_{ \sigma}({\bm \xi})$ is defined by $A_{ \sigma}({\bm \xi}) \coloneqq \prod_{ (j,k) \in \mathcal{A}_{\sigma} } S (\xi_{j},\xi_{k})$, where $\mathcal{A}_{\sigma}$ is a set for $(j,k)$ such that $(j,k)$ are an inversion in a given element $\sigma$ of $\mathbb{S}_N$.
The scattering amplitude is given by $S(\xi_{j},\xi_{k}) \coloneqq - \xi_{j} /\xi_{k}$~\cite{Zadnik2021_1,Zadnik2021_2,Pozsgay2021}.

We next calculate $F(x,t)$ by using Eq.~\eqref{Esol1}. By definition, we get the expression of $P(x,t)$ as
\begin{widetext}
\begin{align} 
P(x,t) = \sum_{\sigma \in \mathbb{S}_N} \sum_{\mu \in \mathbb{S}_N}  \int_{C_r} d\bm{\xi} \int_{C_r} d\bm{\eta}~  \dfrac{ A_{\sigma}({\bm \xi}) A_{\mu}({\bm \eta})  \prod_{j=1}^{N-1} \left( \xi_{\sigma_{j+1}} \eta_{\mu_{j+1} } \right)^{j}   }
{ \prod_{j=1}^{N-1} \left(1 -  \prod_{k=j+1}^N \xi_{\sigma_k }  \eta_{\mu_k} \right) } \prod_{j=1}^N \left[ \left( \xi_j \eta_j \right)^{x - 2j - 1} e^{-{\rm i}t(E_{\xi_j} - E_{\eta_j})} \right]. 
\label{Esol2}
\end{align}
\end{widetext}
To derive this expression, we use the fact that $\Phi(x_1,...,x_N,t)$ vanishes if there exists a site label $j$ such that $x_j = x_{j+1} + 1$ is satisfied (see the derivation of this property in Sec. II of SM~\cite{SM}). To compute the summations over $\sigma$ and $\mu$ in Eq.~\eqref{Esol2}, we note the following identity~\cite{Cantini2020,Petrov2021,Saenz2022} being related to the Izergin-Korepin determinant of the six-vertex model~\cite{Korepin1982,Izergin1987,Izergin1992,Korepin_Bogoliubov_Izergin_1993}, 
\begin{align} 
&\sum_{\sigma \in \mathbb{S}_N} \sum_{\mu \in \mathbb{S}_N}  ~ \dfrac{ B_{\sigma}({\bm \xi}) B_{\mu}({\bm \eta})  \prod_{j=1}^{N-1} \left( \xi_{\sigma_{j+1}} \eta_{\mu_{j+1} } \right)^{j}   }
{ \prod_{j=1}^{N-1} \left(1 -  \prod_{k=j+1}^N \xi_{\sigma_k }  \eta_{\mu_k} \right) } \nonumber \\
&= \dfrac{ \left( 1 - \prod_{j=1}^N \xi_j \eta_j \right)  \prod_{j,k=1}^N \left(  \xi_j + \eta_k -2 \Delta \xi_j \eta_k \right)  }
{ \prod_{j<k} \left( 1 + \xi_j \xi_k - 2 \Delta \xi_j \right) \left( 1 + \eta_j \eta_k - 2 \Delta \eta_j \right)  } D_{N}(\bm{\xi}, \bm{\eta}),
\label{IKI1}
\end{align}
where we define $D_{N}(\bm{\xi}, \bm{\eta}) \coloneqq \det( d(\xi_j, \eta_k) )_{j,k \in \{1,...,N\}}$ with $d(\xi_j, \eta_k)\coloneqq (1-\xi_j \eta_k)^{-1} (\xi_j + \eta_k - 2 \Delta \xi_j \eta_k)^{-1}$. The function $B_{\sigma}(\bm{\xi})$ is defined by $B_{\sigma}(\bm{\xi}) \coloneqq \prod_{ (j,k) \in \mathcal{A}_{\sigma} } S_{\rm XXZ} (\xi_{j},\xi_{k}) $ with the scattering amplitude $S_{\rm XXZ} (\xi_{j},\xi_{k}) \coloneqq - ( 1 + \xi_j \xi_k - \Delta \xi_{j} )/( 1 + \xi_j \xi_k - \Delta  \xi_{k}  )$ for the XXZ model~\cite{Takahashi1999,Franchini2017}. We can easily show $\displaystyle \lim_{\Delta \rightarrow \infty} B_{\sigma}(\bm{\xi}) = A_{\sigma}(\bm{\xi})$. Thus, taking the limit $\Delta \rightarrow \infty$ in Eq.~\eqref{IKI1}, we derive
\begin{align} 
&\sum_{\sigma \in \mathbb{S}_N} \sum_{\mu \in \mathbb{S}_N}  ~  \dfrac{ A_{\sigma}({\bm \xi}) A_{\mu}({\bm \eta})  \prod_{j=1}^{N-1} \left( \xi_{\sigma_{j+1}} \eta_{\mu_{j+1} } \right)^{j}   }
{ \prod_{j=1}^{N-1} \left(1 -  \prod_{k=j+1}^N \xi_{\sigma_k }  \eta_{\mu_k} \right) } \nonumber \\
&= \left( 1 - \prod_{j=1}^N \xi_j \eta_j \right) \det \left( \dfrac{ (\xi_j)^{j-1} (\eta_k)^{k-1} }{ 1 - \xi_j \eta_k } \right)_{j,k \in \{1,...,N\}}.
\label{IKI2}
\end{align}
Plugging Eq.~\eqref{IKI2} into Eq.~\eqref{Esol2} and taking the summation $\sum_{y=x}^{\infty} P(y,t)$, we get the following determinantal formula of $F(x,t)$:
\begin{align} 
F(x,t)      &= \det \left( K(x,t,j,k) \right)_{j,k \in \{1,...,N\}}, \label{Esol3}
\end{align}
where $K(x,t,j,k)$ is defined by $ K(x,t,j,k) \coloneqq - \int_{C_r} d\eta \int_{C_r} d\xi~  \xi^{x-j-2} \eta^{x-k-2} e^{ {\rm i}t (E_{\eta} - E_{\xi}) } / (4 \pi (1 - \xi \eta) )$. 
This determinantal form of Eq.~\eqref{Esol3} is compatible with random matrix theory for GUE because many formulae for GUE are given by determinants~\cite{mehta1,forrester2010}. 

\begin{figure*}[t]
\begin{center}
\includegraphics[keepaspectratio, width=18.0cm]{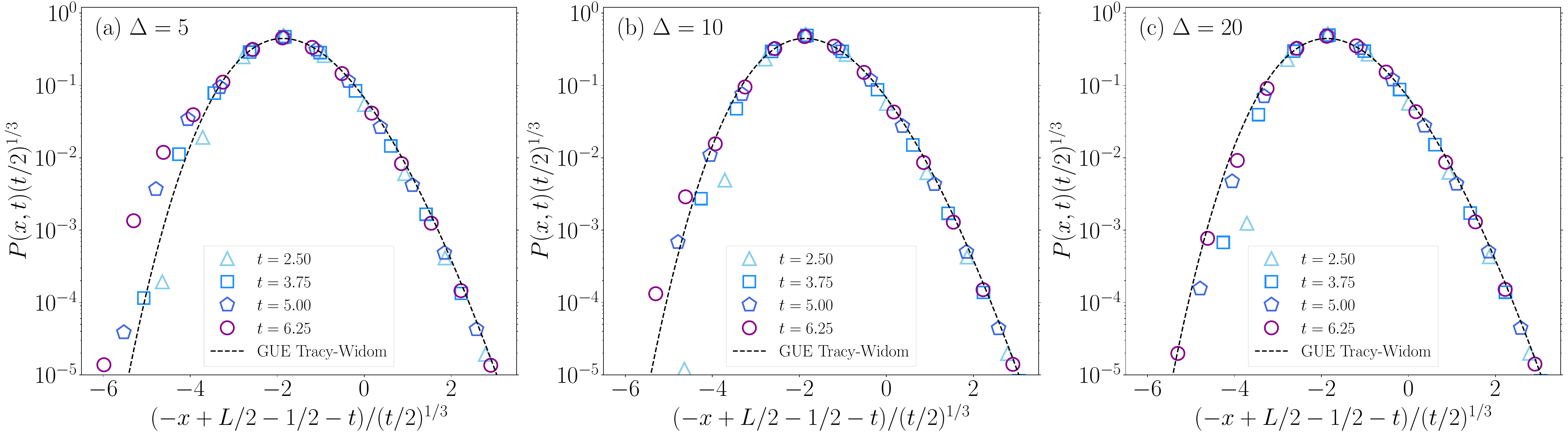}
\caption{Numerical results for the probability $P(x,t)$ of finding the left-most up-spin at site $x$ and time $t$ in the XXZ model with the alternating domain-wall initial state. The anisotropic interaction parameter is $\Delta=$ (a) $5$, (b) $10,$ and (c) $20$ and the system is $L=100$. The time evolution of $P(x,t)$ is numerically computed by the TEBD method~\cite{TEBD1,TEBD2,TEBD3,TEBD4}. In order to compare the numerical data with the GUE Tracy-Widom distribution function $F_2(\bullet)$, the ordinate is divided by $(t/2)^{1/3}$ and the abscissas is rescaled by  using a result for fast convergence~\cite{Ferrari2011}, which is described in Sec. IV of SM~\cite{SM}. The dashed lines in the panels indicates the probability density function $dF_2(s)/ds$ for the GUE Tracy-Widom distribution. The numerical convergences associated with truncation of a matrix product state are discussed in Sec. V of SM~\cite{SM}.  
} 
\label{fig2} 
\end{center}
\end{figure*}

{\it Derivation of the GUE Tracy-Widom distribution.--}
Using Eq.~\eqref{Esol3}, we shall show that $F(x,t)$ converges to the GUE Tracy-Widom distribution function $F_2(\bullet)$ in the long-time limit. A determinant with a function being similar to $K(x,t,j,k)$ was investigated with techniques of Toeplitz operators~\cite{bottcher2012} in Ref.~\cite{Saenz2022}. Following the same techniques with $N \rightarrow \infty$, we obtain
\begin{align} 
F(x,t) = {\rm det} \left( 1 - K_{\rm B} (t,m,n) \right)_{l^2 (\{ 2-x, 3-x, ...\}) } \label{Esol6}
\end{align}
with $K_{\rm B} (t,m,n) \coloneqq   t \left( J_{m}(t) J_{n+1}(t) - J_{m+1}(t) J_{n}(t) \right)/(2m-2n)$.
We apply the asymptotic analysis with a scaling variable $s$ defined through $x = 2 + \lfloor -t - s(t/2)^{1/3} \rfloor $, deriving
\begin{align} 
\lim_{t \rightarrow \infty} F\left( 2 + \lfloor -t - s (t/2)^{1/3} \rfloor, t \right) = F_2(s). \label{Esol8}
\end{align}
Here the GUE Tracy-Widom distribution function is given by $F_2(s) = {\rm Det} \left[ 1 - K_{\rm Ai}(x,y) \right]_{\mathbb{L}^2(s,\infty)}$ with the Airy Kernel $K_{\rm Ai}(x,y) \coloneqq \left( {\rm Ai}(x) {\rm Ai}'(y) - {\rm Ai}'(x) {\rm Ai}(y)\right)/(x-y)$. From Eq.~\eqref{Esol8} and the relation $P(x,t) = F(x,t) - F(x+1,t)$, we obtain for $t \gg 1$
\begin{align} 
P\left( 2 + \lfloor -t - s (t/2)^{1/3} \rfloor, t \right) \simeq \left( \dfrac{2}{t} \right )^{1/3} \dfrac{F_2(s)}{d s}. \label{Esol9}
\end{align}
Therefore, we analytically prove the emergence of the GUE Tracy-Widom distribution in the quantum spin model that cannot be mapped to a noninteracting fermion Hamiltonian via the Jordan-Wigner transformation~\cite{Sachdev1999,Lewenstein2012}. 

{\it Relation between exact many-body wavefunctions of the folded XXZ model and the XX model.--}
We explain that the GUE Tracy-Widom distribution in the folded XXZ model is related to a many-body wavefunction of the XX model. 

Let us consider the XX model, Hamiltonian of which is defined by $\hat{H}_{\rm XX} \coloneqq \sum_{x \in \mathbb{Z}} \left( \hat{X}_{x} \hat{X}_{x+1} + \hat{Y}_{x} \hat{Y}_{x+1}  \right)$~\cite{Takahashi1999,Franchini2017}. We denote the many-body wavefunction for this model at time $t$ by $\Phi_{\rm XX}(x_1,...,x_N,t)$ with a constraint $(x_{j} < x_{j+1})$ and assume that the initial state is the domain-wall state, $\Phi_{\rm XX}(x_1,...,x_N,0) = \prod_{j=1}^N \delta{x_j, j}$. As explained in Sec. III of SM~\cite{SM}, the exact form of the many-body wavefunction reads
\begin{align} 
\Phi_{\rm XX} (x_1,...,x_N,t) &= \int_{C_r} d \bm{\xi} \det \left(  \xi_{k}^{x_{j} - k -1} e^{- {\rm i}E_{\xi_j} t} \right)_{j,k \in \{1,...,N\} }.
\label{MBW4}
\end{align}
On the other hand, the exact many-body wavefunction for the folded XXZ mode with the alternating domain-wall initial state is expressed by
\begin{align} 
\Phi(x_1, ... ,x_N,t) = \int_{C_r} d\bm{\xi} ~\det \left(  \xi_{k}^{ x_j - j  - k - 1} e^{- {\rm i} E_{\xi_j} t}  \right)_{j,k \in \{1,...,N\} }, 
\label{MBW1}
\end{align}
which is proved in Sec. II of SM~\cite{SM}. 

We find that the many-body wavefunction of Eq.~\eqref{MBW1} for the folded XXZ model has the determinantal structure similar to that of Eq.~\eqref{MBW4} for the XX model. As discussed in Ref.~\cite{Saenz2022}, the left-most up-spin for the XX model with the domain-wall initial state obeys the GUE Tracy-Widom distribution. Hence, a mathematical origin for the GUE Tracy-Widom distribution in the folded XXZ is close to that for the XX model. 

{\it Numerical study for $P(x,t)$ of the XXZ model.--}
So far, we have analytically studied the folded XXZ model, which is the effective description for the XXZ model with the large anisotropic interaction ($\Delta \gg 1$).
Then, it is natural and intriguing to explore the GUE Tracy-Widom distribution in the XXZ model itself from the theoretical viewpoint, and such exploration is important for discussing experimental possibilities of observing our theoretical prediction since the XXZ model has been experimentally studied~\cite{Wei2022,Jepsen2020,Jepsen2021,Jepsen2022,Rosenberg2024}. 
We here show our numerical study concerning this issue. 

The model used in the numerical simulation is the XXZ model with an open boundary condition. The Hamiltonian is defined by $\hat{H}_{\rm XXZ}^{\rm (open)} \coloneqq \sum_{j =1 }^{L-1} \left( \hat{X}_{x} \hat{X}_{x+1} + \hat{Y}_{x} \hat{Y}_{x+1} + \Delta \hat{Z}_{x} \hat{Z}_{x+1}  \right)$ with the total number $L$ of the one-dimensional lattice. Here, we assume $L$ to be multiples of four. We denote the quantum state at time $t$ by $\ket{\psi(t)}$ and the initial state is assumed to be the alternating domain-wall state $\ket{\psi(0)} = \prod_{x=L/4}^{L/2} \hat{R}_{2x} \ket{0}$. We numerically solve the Schrödinger equation using the TEBD method~\cite{TEBD1,TEBD2,TEBD3,TEBD4}, computing the probability $P(x,t)$ of finding the left-most up-spin at site $x$ and time $t$.

Figure~\ref{fig2} displays numerical results of $P(x,t)$ with $\Delta=5, 10$, and $20$, where we rescale the abscissas and the ordinates by following Eq.~\eqref{Esol9} and a result for fast convergence described in Sec. IV of SM~\cite{SM}. We find that deviations between the numerical data and the probability density function $dF_2(s)/ds$ become large when $\Delta$ is small. Thus, we speculate the absence of the GUE Tracy-Widom distribution for the XXZ model in the {\it long-time limit}. 

We, however, find the signature that the probability $P(\lfloor -t - s (t/2)^{1/3} \rfloor,t)$ in the right region ($s > 0$) of Fig.~\ref{fig2} exhibits the universal curve being independent of $\Delta$. This curve is characterized by the diagonal Airy kernel $K_{\rm Ai}(s,s)$ because the dashed line represents the GUE Tracy-Widom distribution (see Sec. VI and Fig. S-4 of SM~\cite{SM}).

{\it Discussion.--}
We discuss two topics: (i) the universal behavior of $P(x,t)$ in the XXZ model and (ii) experimental possibilities of observing our theoretical prediction.

As to (i), we discuss the signature that the curve of $P(\lfloor -t - s (t/2)^{1/3} \rfloor,t)$ for large $s>0$ is independent of $\Delta$ as pointed out in Fig.~\ref{fig2}. We here discuss its origin analytically in the two limiting cases, namely $\Delta = 0$ and large $\Delta$. First we consider the case with $\Delta =0$. As derived exactly in Sec. VI of SM~\cite{SM}, we have $\lim_{t \rightarrow \infty}F(2 + \lfloor -t - s (t/2)^{1/3} \rfloor,t) = F_{\rm Ai}(s,1/2)$ with $F_{\rm Ai}(s,a) \coloneqq {\rm Det} \left[ 1 - aK_{\rm Ai}(x,y) \right]_{\mathbb{L}^2(s,\infty)}$. Then, the distribution function approximately is approximately to be $F_{\rm Ai}(s,1/2) \simeq 1 -  \int^{\infty}_s dy K_{\rm Ai}(y,y)/2$ for large $s$ because $K_{\rm Ai}(x,y)$ is small. Hence, the rescaled probability of finding the left-most up-spin is described by $d F_{\rm Ai}(s,1/2)/ds \simeq K_{\rm Ai}(s,s)/2$ (see Fig. S-4 of SM~\cite{SM}). Second, we consider the folded XXZ model corresponding to large $\Delta$. Using Eq.~\eqref{Esol8}, we similarly approximate the probability distribution function as $F_2(s) \simeq 1 -  \int^{\infty}_s dy K_{\rm Ai}(y,y)$ for large $s$, obtaining that the rescaled probability is characterized by $K_{\rm Ai}(s,s)$. Combing our analytical discussions and the numerical findings of Fig.~\ref{fig2}, we find the signature that the probability of finding the left-most up-spin for large $s$ is universally characterized by the diagonal Airy kernel $K_{\rm Ai}(s,s)$ regardless of $\Delta$. Finally, we mention that the conjecture by Saenz, Tracy, and Widom for the XXZ model (see Conjecture 1 of Ref.~\cite{Saenz2022}) may be useful for proving the diagonal Airy kernel in the dynamics of the XXZ model (see Sec. VII of SM~\cite{SM}).

As to (ii) we discuss experimental possibilities of observing the GUE Tracy-Widom distribution function in quantum spin transport on one-dimensional systems.  
As shown in Fig.~\ref{fig2}, the probability $P(x,t)$ in the XXZ model with $\Delta \gg 1$ shows the signature of the GUE Tracy-Widom behavior in the {\it finite time regions}, and its time scale is about $5$ times spin flipping. On the experimental side, previous literature in cold atom experiments~\cite{Wei2022,Jepsen2020,Jepsen2021,Jepsen2022} realized the XXZ model as an effective description for a two-component Bose-Hubbard model under the hard-core limit~\cite{Altman2003,Duan2003,Kuklov2003}, and also an experiment using superconducting qubits~\cite{Rosenberg2024} simulates the XXZ model by periodical applications of 2-qubit unitary gates. These experiments accessed the spin transport where the spins flip more than 5 times. Taking these experimental achievements into account, we expect that signatures of the GUE Tracy-Widom distribution may be observed in state-of-the-art experiments. 

{\it Conclusions and future prospects.--}
We theoretically studied the one-dimensional quantum spin transport described by the folded XXZ model with the alternating domain-wall initial state by focusing on the probability $P(x,t)$ of finding the left-most up-spin. Employing the exact method based on the Bethe ansatz, we exactly demonstrated that the rescaled probability converges to the probability density function for the GUE Tracy-Widom distribution function in the long-time limit. Beyond the folded XXZ model, we numerically studied the XXZ model via the TEBD method, discussing the universal behavior in terms of the diagonal Airy Kernel. 

As a future prospect, it is interesting to explore the Tracy-Widom distribution in the quantum transport of non-integrable models. So far, the GUE Tracy-Widom distribution in quantum transport has been investigated in integrable models, and thus it is intriguing to uncover the role of the integrability and understand how universal the Tracy-Widom distribution is in quantum transport. Also, studying the Tracy-Widom distribution in other integral models is interesting. One of such examples is a phase model. This model describes strongly interacting bosons on a one-dimensional lattice~\cite{Bogoliubov1993,Bogoliubov1998,Pozsgay_2016} and have scattering amplitudes being similar to the folded XXZ model. As another prospect, it is fundamentally important to understand the universal behavior of the XXZ model analytically by using the generalized hydrodynamics~\cite{Olalla2016,Bertini2016,Doyon2017,Doyon2017_2,Bulchandani2017,Bulchandani18,Doyon2018,Collura2018,Jacopo2018,Sarang2019,Schemmer2019,Doyon2020_rev,Alba2021,Malvania2021,Bouchoule2022,Essler2023} and ballistic macroscopic fluctuation theory~\cite{BMFT1,BMFT2}.

\begin{acknowledgments}
KF and ST are grateful to Ryusuke Hamazaki and Yuki Kawaguchi for the valuable comments on the manuscript, and Ippei Danshita for the helpful comments on experiments of ultracold atoms.   
The work of KF has been supported by JSPS KAKENHI Grant No. JP23K13029. 
The work of TS has been supported by JSPS KAKENHI Grants No. JP21H04432, No. JP22H01143.
\end{acknowledgments}

\bibliography{reference}

\begin{thebibliography}{89}%
\makeatletter
\providecommand \@ifxundefined [1]{%
 \@ifx{#1\undefined}
}%
\providecommand \@ifnum [1]{%
 \ifnum #1\expandafter \@firstoftwo
 \else \expandafter \@secondoftwo
 \fi
}%
\providecommand \@ifx [1]{%
 \ifx #1\expandafter \@firstoftwo
 \else \expandafter \@secondoftwo
 \fi
}%
\providecommand \natexlab [1]{#1}%
\providecommand \enquote  [1]{``#1''}%
\providecommand \bibnamefont  [1]{#1}%
\providecommand \bibfnamefont [1]{#1}%
\providecommand \citenamefont [1]{#1}%
\providecommand \href@noop [0]{\@secondoftwo}%
\providecommand \href [0]{\begingroup \@sanitize@url \@href}%
\providecommand \@href[1]{\@@startlink{#1}\@@href}%
\providecommand \@@href[1]{\endgroup#1\@@endlink}%
\providecommand \@sanitize@url [0]{\catcode `\\12\catcode `\$12\catcode
  `\&12\catcode `\#12\catcode `\^12\catcode `\_12\catcode `\%12\relax}%
\providecommand \@@startlink[1]{}%
\providecommand \@@endlink[0]{}%
\providecommand \url  [0]{\begingroup\@sanitize@url \@url }%
\providecommand \@url [1]{\endgroup\@href {#1}{\urlprefix }}%
\providecommand \urlprefix  [0]{URL }%
\providecommand \Eprint [0]{\href }%
\providecommand \doibase [0]{https://doi.org/}%
\providecommand \selectlanguage [0]{\@gobble}%
\providecommand \bibinfo  [0]{\@secondoftwo}%
\providecommand \bibfield  [0]{\@secondoftwo}%
\providecommand \translation [1]{[#1]}%
\providecommand \BibitemOpen [0]{}%
\providecommand \bibitemStop [0]{}%
\providecommand \bibitemNoStop [0]{.\EOS\space}%
\providecommand \EOS [0]{\spacefactor3000\relax}%
\providecommand \BibitemShut  [1]{\csname bibitem#1\endcsname}%
\let\auto@bib@innerbib\@empty
\bibitem [{\citenamefont {Dhar}(2008)}]{Dhar2008}%
  \BibitemOpen
  \bibfield  {author} {\bibinfo {author} {\bibfnamefont {A.}~\bibnamefont
  {Dhar}},\ }\bibfield  {title} {\bibinfo {title} {Heat transport in
  low-dimensional systems},\ }\href {https://doi.org/10.1080/00018730802538522}
  {\bibfield  {journal} {\bibinfo  {journal} {Advances in Physics}\ }\textbf
  {\bibinfo {volume} {57}},\ \bibinfo {pages} {457} (\bibinfo {year}
  {2008})}\BibitemShut {NoStop}%
\bibitem [{\citenamefont {Nagaosa}\ \emph {et~al.}(2010)\citenamefont
  {Nagaosa}, \citenamefont {Sinova}, \citenamefont {Onoda}, \citenamefont
  {MacDonald},\ and\ \citenamefont {Ong}}]{Nagaosa2010}%
  \BibitemOpen
  \bibfield  {author} {\bibinfo {author} {\bibfnamefont {N.}~\bibnamefont
  {Nagaosa}}, \bibinfo {author} {\bibfnamefont {J.}~\bibnamefont {Sinova}},
  \bibinfo {author} {\bibfnamefont {S.}~\bibnamefont {Onoda}}, \bibinfo
  {author} {\bibfnamefont {A.~H.}\ \bibnamefont {MacDonald}},\ and\ \bibinfo
  {author} {\bibfnamefont {N.~P.}\ \bibnamefont {Ong}},\ }\bibfield  {title}
  {\bibinfo {title} {Anomalous hall effect},\ }\href
  {https://doi.org/10.1103/RevModPhys.82.1539} {\bibfield  {journal} {\bibinfo
  {journal} {Rev. Mod. Phys.}\ }\textbf {\bibinfo {volume} {82}},\ \bibinfo
  {pages} {1539} (\bibinfo {year} {2010})}\BibitemShut {NoStop}%
\bibitem [{\citenamefont {Marchetti}\ \emph {et~al.}(2013)\citenamefont
  {Marchetti}, \citenamefont {Joanny}, \citenamefont {Ramaswamy}, \citenamefont
  {Liverpool}, \citenamefont {Prost}, \citenamefont {Rao},\ and\ \citenamefont
  {Simha}}]{Marchetti2013}%
  \BibitemOpen
  \bibfield  {author} {\bibinfo {author} {\bibfnamefont {M.~C.}\ \bibnamefont
  {Marchetti}}, \bibinfo {author} {\bibfnamefont {J.~F.}\ \bibnamefont
  {Joanny}}, \bibinfo {author} {\bibfnamefont {S.}~\bibnamefont {Ramaswamy}},
  \bibinfo {author} {\bibfnamefont {T.~B.}\ \bibnamefont {Liverpool}}, \bibinfo
  {author} {\bibfnamefont {J.}~\bibnamefont {Prost}}, \bibinfo {author}
  {\bibfnamefont {M.}~\bibnamefont {Rao}},\ and\ \bibinfo {author}
  {\bibfnamefont {R.~A.}\ \bibnamefont {Simha}},\ }\bibfield  {title} {\bibinfo
  {title} {Hydrodynamics of soft active matter},\ }\href
  {https://doi.org/10.1103/RevModPhys.85.1143} {\bibfield  {journal} {\bibinfo
  {journal} {Rev. Mod. Phys.}\ }\textbf {\bibinfo {volume} {85}},\ \bibinfo
  {pages} {1143} (\bibinfo {year} {2013})}\BibitemShut {NoStop}%
\bibitem [{\citenamefont {Bertini}\ \emph {et~al.}(2021)\citenamefont
  {Bertini}, \citenamefont {Heidrich-Meisner}, \citenamefont {Karrasch},
  \citenamefont {Prosen}, \citenamefont {Steinigeweg},\ and\ \citenamefont
  {\ifmmode \check{Z}\else \v{Z}\fi{}nidari\ifmmode~\check{c}\else
  \v{c}\fi{}}}]{Bertini1}%
  \BibitemOpen
  \bibfield  {author} {\bibinfo {author} {\bibfnamefont {B.}~\bibnamefont
  {Bertini}}, \bibinfo {author} {\bibfnamefont {F.}~\bibnamefont
  {Heidrich-Meisner}}, \bibinfo {author} {\bibfnamefont {C.}~\bibnamefont
  {Karrasch}}, \bibinfo {author} {\bibfnamefont {T.}~\bibnamefont {Prosen}},
  \bibinfo {author} {\bibfnamefont {R.}~\bibnamefont {Steinigeweg}},\ and\
  \bibinfo {author} {\bibfnamefont {M.}~\bibnamefont {\ifmmode \check{Z}\else
  \v{Z}\fi{}nidari\ifmmode~\check{c}\else \v{c}\fi{}}},\ }\bibfield  {title}
  {\bibinfo {title} {Finite-temperature transport in one-dimensional quantum
  lattice models},\ }\href {https://doi.org/10.1103/RevModPhys.93.025003}
  {\bibfield  {journal} {\bibinfo  {journal} {Rev. Mod. Phys.}\ }\textbf
  {\bibinfo {volume} {93}},\ \bibinfo {pages} {025003} (\bibinfo {year}
  {2021})}\BibitemShut {NoStop}%
\bibitem [{\citenamefont {Sasamoto}(2007)}]{Sasamoto2007}%
  \BibitemOpen
  \bibfield  {author} {\bibinfo {author} {\bibfnamefont {T.}~\bibnamefont
  {Sasamoto}},\ }\bibfield  {title} {\bibinfo {title} {Fluctuations of the
  one-dimensional asymmetric exclusion process using random matrix
  techniques},\ }\href {https://doi.org/10.1088/1742-5468/2007/07/P07007}
  {\bibfield  {journal} {\bibinfo  {journal} {Journal of Statistical Mechanics:
  Theory and Experiment}\ }\textbf {\bibinfo {volume} {2007}},\ \bibinfo
  {pages} {P07007} (\bibinfo {year} {2007})}\BibitemShut {NoStop}%
\bibitem [{\citenamefont {Kriecherbauer}\ and\ \citenamefont
  {Krug}(2010)}]{Krug2010}%
  \BibitemOpen
  \bibfield  {author} {\bibinfo {author} {\bibfnamefont {T.}~\bibnamefont
  {Kriecherbauer}}\ and\ \bibinfo {author} {\bibfnamefont {J.}~\bibnamefont
  {Krug}},\ }\bibfield  {title} {\bibinfo {title} {A pedestrian's view on
  interacting particle systems, {{KPZ}} universality and random matrices},\
  }\href {https://doi.org/10.1088/1751-8113/43/40/403001} {\bibfield  {journal}
  {\bibinfo  {journal} {Journal of Physics A: Mathematical and Theoretical}\
  }\textbf {\bibinfo {volume} {43}},\ \bibinfo {pages} {403001} (\bibinfo
  {year} {2010})}\BibitemShut {NoStop}%
\bibitem [{\citenamefont {Corwin}(2012)}]{CORWIN2012}%
  \BibitemOpen
  \bibfield  {author} {\bibinfo {author} {\bibfnamefont {I.}~\bibnamefont
  {Corwin}},\ }\bibfield  {title} {\bibinfo {title} {The
  {{K}}ardar-{{P}}arisi-{{Z}}hang equation and universality class},\ }\href
  {https://doi.org/10.1142/S2010326311300014} {\bibfield  {journal} {\bibinfo
  {journal} {Random Matrices: Theory and Applications}\ }\textbf {\bibinfo
  {volume} {01}},\ \bibinfo {pages} {1130001} (\bibinfo {year}
  {2012})}\BibitemShut {NoStop}%
\bibitem [{\citenamefont {Quastel}\ and\ \citenamefont
  {Spohn}(2015)}]{Quastel2015}%
  \BibitemOpen
  \bibfield  {author} {\bibinfo {author} {\bibfnamefont {J.}~\bibnamefont
  {Quastel}}\ and\ \bibinfo {author} {\bibfnamefont {H.}~\bibnamefont
  {Spohn}},\ }\bibfield  {title} {\bibinfo {title} {The one-dimensional {{KPZ}}
  equation and its universality class},\ }\href
  {https://doi.org/10.1007/s10955-015-1250-9} {\bibfield  {journal} {\bibinfo
  {journal} {Journal of Statistical Physics}\ }\textbf {\bibinfo {volume}
  {160}},\ \bibinfo {pages} {965} (\bibinfo {year} {2015})}\BibitemShut
  {NoStop}%
\bibitem [{\citenamefont {Takeuchi}(2018)}]{takeuchi2018}%
  \BibitemOpen
  \bibfield  {author} {\bibinfo {author} {\bibfnamefont {K.~A.}\ \bibnamefont
  {Takeuchi}},\ }\bibfield  {title} {\bibinfo {title} {An appetizer to modern
  developments on the {{K}}ardar-{{P}}arisi-{{Z}}hang universality class},\
  }\href {https://doi.org/https://doi.org/10.1016/j.physa.2018.03.009}
  {\bibfield  {journal} {\bibinfo  {journal} {Physica A: Statistical Mechanics
  and its Applications}\ }\textbf {\bibinfo {volume} {504}},\ \bibinfo {pages}
  {77} (\bibinfo {year} {2018})},\ \bibinfo {note} {lecture Notes of the 14th
  International Summer School on Fundamental Problems in Statistical
  Physics}\BibitemShut {NoStop}%
\bibitem [{\citenamefont {Barab{\'a}si}\ and\ \citenamefont
  {Stanley}(1995)}]{barabasi1995}%
  \BibitemOpen
  \bibfield  {author} {\bibinfo {author} {\bibfnamefont {A.-L.}\ \bibnamefont
  {Barab{\'a}si}}\ and\ \bibinfo {author} {\bibfnamefont {H.~E.}\ \bibnamefont
  {Stanley}},\ }\href@noop {} {\emph {\bibinfo {title} {Fractal concepts in
  surface growth}}}\ (\bibinfo  {publisher} {Cambridge university press},\
  \bibinfo {year} {1995})\BibitemShut {NoStop}%
\bibitem [{\citenamefont {Schmittmann}\ and\ \citenamefont
  {Zia}(1995)}]{Schmittmann1995}%
  \BibitemOpen
  \bibfield  {author} {\bibinfo {author} {\bibfnamefont {B.}~\bibnamefont
  {Schmittmann}}\ and\ \bibinfo {author} {\bibfnamefont {R.}~\bibnamefont
  {Zia}},\ }\bibfield  {title} {\bibinfo {title} {Statistical mechanics of
  driven diffusive systems},\ }in\ \href
  {https://doi.org/https://doi.org/10.1016/S1062-7901(06)80014-5} {\emph
  {\bibinfo {booktitle} {Statistical Mechanics of Driven Diffusive System}}},\
  \bibinfo {series} {Phase Transitions and Critical Phenomena}, Vol.~\bibinfo
  {volume} {17},\ \bibinfo {editor} {edited by\ \bibinfo {editor}
  {\bibfnamefont {B.}~\bibnamefont {Schmittmann}}\ and\ \bibinfo {editor}
  {\bibfnamefont {R.}~\bibnamefont {Zia}}}\ (\bibinfo  {publisher} {Academic
  Press},\ \bibinfo {year} {1995})\ pp.\ \bibinfo {pages} {3--214}\BibitemShut
  {NoStop}%
\bibitem [{\citenamefont {Liggett}(2013)}]{liggett2013}%
  \BibitemOpen
  \bibfield  {author} {\bibinfo {author} {\bibfnamefont {T.~M.}\ \bibnamefont
  {Liggett}},\ }\href@noop {} {\emph {\bibinfo {title} {Stochastic interacting
  systems: contact, voter and exclusion processes}}},\ Vol.\ \bibinfo {volume}
  {324}\ (\bibinfo  {publisher} {springer science \& Business Media},\ \bibinfo
  {year} {2013})\BibitemShut {NoStop}%
\bibitem [{\citenamefont {Tracy}\ and\ \citenamefont
  {Widom}(1993)}]{Tracy1993}%
  \BibitemOpen
  \bibfield  {author} {\bibinfo {author} {\bibfnamefont {C.~A.}\ \bibnamefont
  {Tracy}}\ and\ \bibinfo {author} {\bibfnamefont {H.}~\bibnamefont {Widom}},\
  }\bibfield  {title} {\bibinfo {title} {Level-spacing distributions and the
  {{A}}iry kernel},\ }\href
  {https://doi.org/https://doi.org/10.1016/0370-2693(93)91114-3} {\bibfield
  {journal} {\bibinfo  {journal} {Physics Letters B}\ }\textbf {\bibinfo
  {volume} {305}},\ \bibinfo {pages} {115} (\bibinfo {year}
  {1993})}\BibitemShut {NoStop}%
\bibitem [{\citenamefont {Tracy}\ and\ \citenamefont
  {Widom}(1996)}]{Tracy1996}%
  \BibitemOpen
  \bibfield  {author} {\bibinfo {author} {\bibfnamefont {C.~A.}\ \bibnamefont
  {Tracy}}\ and\ \bibinfo {author} {\bibfnamefont {H.}~\bibnamefont {Widom}},\
  }\bibfield  {title} {\bibinfo {title} {On orthogonal and symplectic matrix
  ensembles},\ }\href {https://doi.org/10.1007/BF02099545} {\bibfield
  {journal} {\bibinfo  {journal} {Communications in Mathematical Physics}\
  }\textbf {\bibinfo {volume} {177}},\ \bibinfo {pages} {727} (\bibinfo {year}
  {1996})}\BibitemShut {NoStop}%
\bibitem [{\citenamefont {Forrester}(1993)}]{Forrester1993}%
  \BibitemOpen
  \bibfield  {author} {\bibinfo {author} {\bibfnamefont {P.}~\bibnamefont
  {Forrester}},\ }\bibfield  {title} {\bibinfo {title} {The spectrum edge of
  random matrix ensembles},\ }\href
  {https://doi.org/https://doi.org/10.1016/0550-3213(93)90126-A} {\bibfield
  {journal} {\bibinfo  {journal} {Nuclear Physics B}\ }\textbf {\bibinfo
  {volume} {402}},\ \bibinfo {pages} {709} (\bibinfo {year}
  {1993})}\BibitemShut {NoStop}%
\bibitem [{\citenamefont {Forrester}(2010)}]{forrester2010}%
  \BibitemOpen
  \bibfield  {author} {\bibinfo {author} {\bibfnamefont {P.~J.}\ \bibnamefont
  {Forrester}},\ }\href@noop {} {\emph {\bibinfo {title} {Log-gases and random
  matrices {{(LMS-34)}}}}}\ (\bibinfo  {publisher} {Princeton university
  press},\ \bibinfo {year} {2010})\BibitemShut {NoStop}%
\bibitem [{\citenamefont {Nahum}\ \emph {et~al.}(2017)\citenamefont {Nahum},
  \citenamefont {Ruhman}, \citenamefont {Vijay},\ and\ \citenamefont
  {Haah}}]{Nahum2017}%
  \BibitemOpen
  \bibfield  {author} {\bibinfo {author} {\bibfnamefont {A.}~\bibnamefont
  {Nahum}}, \bibinfo {author} {\bibfnamefont {J.}~\bibnamefont {Ruhman}},
  \bibinfo {author} {\bibfnamefont {S.}~\bibnamefont {Vijay}},\ and\ \bibinfo
  {author} {\bibfnamefont {J.}~\bibnamefont {Haah}},\ }\bibfield  {title}
  {\bibinfo {title} {Quantum entanglement growth under random unitary
  dynamics},\ }\href {https://doi.org/10.1103/PhysRevX.7.031016} {\bibfield
  {journal} {\bibinfo  {journal} {Phys. Rev. X}\ }\textbf {\bibinfo {volume}
  {7}},\ \bibinfo {pages} {031016} (\bibinfo {year} {2017})}\BibitemShut
  {NoStop}%
\bibitem [{\citenamefont {Ljubotina}\ \emph {et~al.}(2019)\citenamefont
  {Ljubotina}, \citenamefont {\ifmmode \check{Z}\else
  \v{Z}\fi{}nidari\ifmmode~\check{c}\else \v{c}\fi{}},\ and\ \citenamefont
  {Prosen}}]{Ljubotina2019}%
  \BibitemOpen
  \bibfield  {author} {\bibinfo {author} {\bibfnamefont {M.}~\bibnamefont
  {Ljubotina}}, \bibinfo {author} {\bibfnamefont {M.}~\bibnamefont {\ifmmode
  \check{Z}\else \v{Z}\fi{}nidari\ifmmode~\check{c}\else \v{c}\fi{}}},\ and\
  \bibinfo {author} {\bibfnamefont {T.}~\bibnamefont {Prosen}},\ }\bibfield
  {title} {\bibinfo {title} {{{K}}ardar-{{P}}arisi-{{Z}}hang physics in the
  quantum {{H}}eisenberg magnet},\ }\href
  {https://doi.org/10.1103/PhysRevLett.122.210602} {\bibfield  {journal}
  {\bibinfo  {journal} {Phys. Rev. Lett.}\ }\textbf {\bibinfo {volume} {122}},\
  \bibinfo {pages} {210602} (\bibinfo {year} {2019})}\BibitemShut {NoStop}%
\bibitem [{\citenamefont {Dupont}\ and\ \citenamefont
  {Moore}(2020)}]{Dupont2020}%
  \BibitemOpen
  \bibfield  {author} {\bibinfo {author} {\bibfnamefont {M.}~\bibnamefont
  {Dupont}}\ and\ \bibinfo {author} {\bibfnamefont {J.~E.}\ \bibnamefont
  {Moore}},\ }\bibfield  {title} {\bibinfo {title} {Universal spin dynamics in
  infinite-temperature one-dimensional quantum magnets},\ }\href
  {https://doi.org/10.1103/PhysRevB.101.121106} {\bibfield  {journal} {\bibinfo
   {journal} {Phys. Rev. B}\ }\textbf {\bibinfo {volume} {101}},\ \bibinfo
  {pages} {121106} (\bibinfo {year} {2020})}\BibitemShut {NoStop}%
\bibitem [{\citenamefont {De~Nardis}\ \emph {et~al.}(2020)\citenamefont
  {De~Nardis}, \citenamefont {Gopalakrishnan}, \citenamefont {Ilievski},\ and\
  \citenamefont {Vasseur}}]{Nardis2020}%
  \BibitemOpen
  \bibfield  {author} {\bibinfo {author} {\bibfnamefont {J.}~\bibnamefont
  {De~Nardis}}, \bibinfo {author} {\bibfnamefont {S.}~\bibnamefont
  {Gopalakrishnan}}, \bibinfo {author} {\bibfnamefont {E.}~\bibnamefont
  {Ilievski}},\ and\ \bibinfo {author} {\bibfnamefont {R.}~\bibnamefont
  {Vasseur}},\ }\bibfield  {title} {\bibinfo {title} {Superdiffusion from
  emergent classical solitons in quantum spin chains},\ }\href
  {https://doi.org/10.1103/PhysRevLett.125.070601} {\bibfield  {journal}
  {\bibinfo  {journal} {Phys. Rev. Lett.}\ }\textbf {\bibinfo {volume} {125}},\
  \bibinfo {pages} {070601} (\bibinfo {year} {2020})}\BibitemShut {NoStop}%
\bibitem [{\citenamefont {Fujimoto}\ \emph {et~al.}(2020)\citenamefont
  {Fujimoto}, \citenamefont {Hamazaki},\ and\ \citenamefont
  {Kawaguchi}}]{Fujimoto2020}%
  \BibitemOpen
  \bibfield  {author} {\bibinfo {author} {\bibfnamefont {K.}~\bibnamefont
  {Fujimoto}}, \bibinfo {author} {\bibfnamefont {R.}~\bibnamefont {Hamazaki}},\
  and\ \bibinfo {author} {\bibfnamefont {Y.}~\bibnamefont {Kawaguchi}},\
  }\bibfield  {title} {\bibinfo {title} {Family-vicsek scaling of roughness
  growth in a strongly interacting bose gas},\ }\href
  {https://doi.org/10.1103/PhysRevLett.124.210604} {\bibfield  {journal}
  {\bibinfo  {journal} {Phys. Rev. Lett.}\ }\textbf {\bibinfo {volume} {124}},\
  \bibinfo {pages} {210604} (\bibinfo {year} {2020})}\BibitemShut {NoStop}%
\bibitem [{\citenamefont {Ye}\ \emph {et~al.}(2022)\citenamefont {Ye},
  \citenamefont {Machado}, \citenamefont {Kemp}, \citenamefont {Hutson},\ and\
  \citenamefont {Yao}}]{Bingtian2022}%
  \BibitemOpen
  \bibfield  {author} {\bibinfo {author} {\bibfnamefont {B.}~\bibnamefont
  {Ye}}, \bibinfo {author} {\bibfnamefont {F.}~\bibnamefont {Machado}},
  \bibinfo {author} {\bibfnamefont {J.}~\bibnamefont {Kemp}}, \bibinfo {author}
  {\bibfnamefont {R.~B.}\ \bibnamefont {Hutson}},\ and\ \bibinfo {author}
  {\bibfnamefont {N.~Y.}\ \bibnamefont {Yao}},\ }\bibfield  {title} {\bibinfo
  {title} {Universal {{K}}ardar-{{P}}arisi-{{Z}}hang dynamics in integrable
  quantum systems},\ }\href {https://doi.org/10.1103/PhysRevLett.129.230602}
  {\bibfield  {journal} {\bibinfo  {journal} {Phys. Rev. Lett.}\ }\textbf
  {\bibinfo {volume} {129}},\ \bibinfo {pages} {230602} (\bibinfo {year}
  {2022})}\BibitemShut {NoStop}%
\bibitem [{\citenamefont {Moca}\ \emph {et~al.}(2023)\citenamefont {Moca},
  \citenamefont {Werner}, \citenamefont {Valli}, \citenamefont {Prosen},\ and\
  \citenamefont {Zar\'and}}]{Moca2023}%
  \BibitemOpen
  \bibfield  {author} {\bibinfo {author} {\bibfnamefont {C.~P.}\ \bibnamefont
  {Moca}}, \bibinfo {author} {\bibfnamefont {M.~A.}\ \bibnamefont {Werner}},
  \bibinfo {author} {\bibfnamefont {A.}~\bibnamefont {Valli}}, \bibinfo
  {author} {\bibfnamefont {T.}~\bibnamefont {Prosen}},\ and\ \bibinfo {author}
  {\bibfnamefont {G.}~\bibnamefont {Zar\'and}},\ }\bibfield  {title} {\bibinfo
  {title} {{{K}}ardar-{{P}}arisi-{{Z}}hang scaling in the hubbard model},\
  }\href {https://doi.org/10.1103/PhysRevB.108.235139} {\bibfield  {journal}
  {\bibinfo  {journal} {Phys. Rev. B}\ }\textbf {\bibinfo {volume} {108}},\
  \bibinfo {pages} {235139} (\bibinfo {year} {2023})}\BibitemShut {NoStop}%
\bibitem [{\citenamefont {De~Nardis}\ \emph {et~al.}(2023)\citenamefont
  {De~Nardis}, \citenamefont {Gopalakrishnan},\ and\ \citenamefont
  {Vasseur}}]{Nardis2023}%
  \BibitemOpen
  \bibfield  {author} {\bibinfo {author} {\bibfnamefont {J.}~\bibnamefont
  {De~Nardis}}, \bibinfo {author} {\bibfnamefont {S.}~\bibnamefont
  {Gopalakrishnan}},\ and\ \bibinfo {author} {\bibfnamefont {R.}~\bibnamefont
  {Vasseur}},\ }\bibfield  {title} {\bibinfo {title} {Nonlinear fluctuating
  hydrodynamics for {{K}}ardar-{{P}}arisi-{{Z}}hang scaling in isotropic spin
  chains},\ }\href {https://doi.org/10.1103/PhysRevLett.131.197102} {\bibfield
  {journal} {\bibinfo  {journal} {Phys. Rev. Lett.}\ }\textbf {\bibinfo
  {volume} {131}},\ \bibinfo {pages} {197102} (\bibinfo {year}
  {2023})}\BibitemShut {NoStop}%
\bibitem [{\citenamefont {Mu}\ \emph {et~al.}(2024)\citenamefont {Mu},
  \citenamefont {Gong},\ and\ \citenamefont {Lemari\'e}}]{Sen2024}%
  \BibitemOpen
  \bibfield  {author} {\bibinfo {author} {\bibfnamefont {S.}~\bibnamefont
  {Mu}}, \bibinfo {author} {\bibfnamefont {J.}~\bibnamefont {Gong}},\ and\
  \bibinfo {author} {\bibfnamefont {G.}~\bibnamefont {Lemari\'e}},\ }\bibfield
  {title} {\bibinfo {title} {{{K}}ardar-{{P}}arisi-{{Z}}hang physics in the
  density fluctuations of localized two-dimensional wave packets},\ }\href
  {https://doi.org/10.1103/PhysRevLett.132.046301} {\bibfield  {journal}
  {\bibinfo  {journal} {Phys. Rev. Lett.}\ }\textbf {\bibinfo {volume} {132}},\
  \bibinfo {pages} {046301} (\bibinfo {year} {2024})}\BibitemShut {NoStop}%
\bibitem [{\citenamefont {Cecile}\ \emph {et~al.}(2024)\citenamefont {Cecile},
  \citenamefont {De~Nardis},\ and\ \citenamefont {Ilievski}}]{Cecile2024}%
  \BibitemOpen
  \bibfield  {author} {\bibinfo {author} {\bibfnamefont {G.}~\bibnamefont
  {Cecile}}, \bibinfo {author} {\bibfnamefont {J.}~\bibnamefont {De~Nardis}},\
  and\ \bibinfo {author} {\bibfnamefont {E.}~\bibnamefont {Ilievski}},\
  }\bibfield  {title} {\bibinfo {title} {Squeezed ensembles and anomalous
  dynamic roughening in interacting integrable chains},\ }\href
  {https://doi.org/10.1103/PhysRevLett.132.130401} {\bibfield  {journal}
  {\bibinfo  {journal} {Phys. Rev. Lett.}\ }\textbf {\bibinfo {volume} {132}},\
  \bibinfo {pages} {130401} (\bibinfo {year} {2024})}\BibitemShut {NoStop}%
\bibitem [{\citenamefont {Gopalakrishnan}\ and\ \citenamefont
  {Vasseur}(2024)}]{Gopalakrishnan2024_r}%
  \BibitemOpen
  \bibfield  {author} {\bibinfo {author} {\bibfnamefont {S.}~\bibnamefont
  {Gopalakrishnan}}\ and\ \bibinfo {author} {\bibfnamefont {R.}~\bibnamefont
  {Vasseur}},\ }\bibfield  {title} {\bibinfo {title} {Superdiffusion from
  nonabelian symmetries in nearly integrable systems},\ }\href
  {https://doi.org/https://doi.org/10.1146/annurev-conmatphys-032922-110710}
  {\bibfield  {journal} {\bibinfo  {journal} {Annual Review of Condensed Matter
  Physics}\ }\textbf {\bibinfo {volume} {15}},\ \bibinfo {pages} {159}
  (\bibinfo {year} {2024})}\BibitemShut {NoStop}%
\bibitem [{\citenamefont {Aditya}\ and\ \citenamefont
  {Roy}(2024)}]{Aditya2024}%
  \BibitemOpen
  \bibfield  {author} {\bibinfo {author} {\bibfnamefont {S.}~\bibnamefont
  {Aditya}}\ and\ \bibinfo {author} {\bibfnamefont {N.}~\bibnamefont {Roy}},\
  }\bibfield  {title} {\bibinfo {title} {Family-vicsek dynamical scaling and
  {{K}}ardar-{{P}}arisi-{{Z}}hang-like superdiffusive growth of surface
  roughness in a driven one-dimensional quasiperiodic model},\ }\href
  {https://doi.org/10.1103/PhysRevB.109.035164} {\bibfield  {journal} {\bibinfo
   {journal} {Phys. Rev. B}\ }\textbf {\bibinfo {volume} {109}},\ \bibinfo
  {pages} {035164} (\bibinfo {year} {2024})}\BibitemShut {NoStop}%
\bibitem [{\citenamefont {Scheie}\ \emph {et~al.}(2021)\citenamefont {Scheie},
  \citenamefont {Sherman}, \citenamefont {Dupont}, \citenamefont {Nagler},
  \citenamefont {Stone}, \citenamefont {Granroth}, \citenamefont {Moore},\ and\
  \citenamefont {Tennant}}]{Scheie2021}%
  \BibitemOpen
  \bibfield  {author} {\bibinfo {author} {\bibfnamefont {A.}~\bibnamefont
  {Scheie}}, \bibinfo {author} {\bibfnamefont {N.~E.}\ \bibnamefont {Sherman}},
  \bibinfo {author} {\bibfnamefont {M.}~\bibnamefont {Dupont}}, \bibinfo
  {author} {\bibfnamefont {S.~E.}\ \bibnamefont {Nagler}}, \bibinfo {author}
  {\bibfnamefont {M.~B.}\ \bibnamefont {Stone}}, \bibinfo {author}
  {\bibfnamefont {G.~E.}\ \bibnamefont {Granroth}}, \bibinfo {author}
  {\bibfnamefont {J.~E.}\ \bibnamefont {Moore}},\ and\ \bibinfo {author}
  {\bibfnamefont {D.~A.}\ \bibnamefont {Tennant}},\ }\bibfield  {title}
  {\bibinfo {title} {Detection of {{K}}ardar--{{P}}arisi--{{Z}}hang
  hydrodynamics in a quantum {{H}}eisenberg spin-1/2 chain},\ }\href
  {https://doi.org/10.1038/s41567-021-01191-6} {\bibfield  {journal} {\bibinfo
  {journal} {Nature Physics}\ }\textbf {\bibinfo {volume} {17}},\ \bibinfo
  {pages} {726} (\bibinfo {year} {2021})}\BibitemShut {NoStop}%
\bibitem [{\citenamefont {Wei}\ \emph {et~al.}(2022)\citenamefont {Wei},
  \citenamefont {Rubio-Abadal}, \citenamefont {Ye}, \citenamefont {Machado},
  \citenamefont {Kemp}, \citenamefont {Srakaew}, \citenamefont {Hollerith},
  \citenamefont {Rui}, \citenamefont {Gopalakrishnan}, \citenamefont {Yao},
  \citenamefont {Bloch},\ and\ \citenamefont {Zeiher}}]{Wei2022}%
  \BibitemOpen
  \bibfield  {author} {\bibinfo {author} {\bibfnamefont {D.}~\bibnamefont
  {Wei}}, \bibinfo {author} {\bibfnamefont {A.}~\bibnamefont {Rubio-Abadal}},
  \bibinfo {author} {\bibfnamefont {B.}~\bibnamefont {Ye}}, \bibinfo {author}
  {\bibfnamefont {F.}~\bibnamefont {Machado}}, \bibinfo {author} {\bibfnamefont
  {J.}~\bibnamefont {Kemp}}, \bibinfo {author} {\bibfnamefont {K.}~\bibnamefont
  {Srakaew}}, \bibinfo {author} {\bibfnamefont {S.}~\bibnamefont {Hollerith}},
  \bibinfo {author} {\bibfnamefont {J.}~\bibnamefont {Rui}}, \bibinfo {author}
  {\bibfnamefont {S.}~\bibnamefont {Gopalakrishnan}}, \bibinfo {author}
  {\bibfnamefont {N.~Y.}\ \bibnamefont {Yao}}, \bibinfo {author} {\bibfnamefont
  {I.}~\bibnamefont {Bloch}},\ and\ \bibinfo {author} {\bibfnamefont
  {J.}~\bibnamefont {Zeiher}},\ }\bibfield  {title} {\bibinfo {title} {Quantum
  gas microscopy of {{K}}ardar-{{P}}arisi-{{Z}}hang superdiffusion},\ }\href
  {https://doi.org/10.1126/science.abk2397} {\bibfield  {journal} {\bibinfo
  {journal} {Science}\ }\textbf {\bibinfo {volume} {376}},\ \bibinfo {pages}
  {716} (\bibinfo {year} {2022})}\BibitemShut {NoStop}%
\bibitem [{\citenamefont {Rosenberg}\ \emph {et~al.}(2024)\citenamefont
  {Rosenberg}, \citenamefont {Andersen}, \citenamefont {Samajdar},
  \citenamefont {Petukhov}, \citenamefont {Hoke}, \citenamefont {Abanin},
  \citenamefont {Bengtsson}, \citenamefont {Drozdov}, \citenamefont {Erickson},
  \citenamefont {Klimov}, \citenamefont {Mi}, \citenamefont {Morvan},
  \citenamefont {Neeley}, \citenamefont {Neill}, \citenamefont {Acharya},
  \citenamefont {Allen}, \citenamefont {Anderson}, \citenamefont {Ansmann},
  \citenamefont {Arute}, \citenamefont {Arya}, \citenamefont {Asfaw},
  \citenamefont {Atalaya}, \citenamefont {Bardin}, \citenamefont {Bilmes},
  \citenamefont {Bortoli}, \citenamefont {Bourassa}, \citenamefont {Bovaird},
  \citenamefont {Brill}, \citenamefont {Broughton}, \citenamefont {Buckley},
  \citenamefont {Buell}, \citenamefont {Burger}, \citenamefont {Burkett},
  \citenamefont {Bushnell}, \citenamefont {Campero}, \citenamefont {Chang},
  \citenamefont {Chen}, \citenamefont {Chiaro}, \citenamefont {Chik},
  \citenamefont {Cogan}, \citenamefont {Collins}, \citenamefont {Conner},
  \citenamefont {Courtney}, \citenamefont {Crook}, \citenamefont {Curtin},
  \citenamefont {Debroy}, \citenamefont {Barba}, \citenamefont {Demura},
  \citenamefont {Paolo}, \citenamefont {Dunsworth}, \citenamefont {Earle},
  \citenamefont {Faoro}, \citenamefont {Farhi}, \citenamefont {Fatemi},
  \citenamefont {Ferreira}, \citenamefont {Burgos}, \citenamefont {Forati},
  \citenamefont {Fowler}, \citenamefont {Foxen}, \citenamefont {Garcia},
  \citenamefont {Genois}, \citenamefont {Giang}, \citenamefont {Gidney},
  \citenamefont {Gilboa}, \citenamefont {Giustina}, \citenamefont {Gosula},
  \citenamefont {Dau}, \citenamefont {Gross}, \citenamefont {Habegger},
  \citenamefont {Hamilton}, \citenamefont {Hansen}, \citenamefont {Harrigan},
  \citenamefont {Harrington}, \citenamefont {Heu}, \citenamefont {Hill},
  \citenamefont {Hoffmann}, \citenamefont {Hong}, \citenamefont {Huang},
  \citenamefont {Huff}, \citenamefont {Huggins}, \citenamefont {Ioffe},
  \citenamefont {Isakov}, \citenamefont {Iveland}, \citenamefont {Jeffrey},
  \citenamefont {Jiang}, \citenamefont {Jones}, \citenamefont {Juhas},
  \citenamefont {Kafri}, \citenamefont {Khattar}, \citenamefont {Khezri},
  \citenamefont {Kieferová}, \citenamefont {Kim}, \citenamefont {Kitaev},
  \citenamefont {Klots}, \citenamefont {Korotkov}, \citenamefont {Kostritsa},
  \citenamefont {Kreikebaum}, \citenamefont {Landhuis}, \citenamefont {Laptev},
  \citenamefont {Lau}, \citenamefont {Laws}, \citenamefont {Lee}, \citenamefont
  {Lee}, \citenamefont {Lensky}, \citenamefont {Lester}, \citenamefont {Lill},
  \citenamefont {Liu}, \citenamefont {Locharla}, \citenamefont {Mandrà},
  \citenamefont {Martin}, \citenamefont {Martin}, \citenamefont {McClean},
  \citenamefont {McEwen}, \citenamefont {Meeks}, \citenamefont {Miao},
  \citenamefont {Mieszala}, \citenamefont {Montazeri}, \citenamefont
  {Movassagh}, \citenamefont {Mruczkiewicz}, \citenamefont {Nersisyan},
  \citenamefont {Newman}, \citenamefont {Ng}, \citenamefont {Nguyen},
  \citenamefont {Nguyen}, \citenamefont {Niu}, \citenamefont {O’Brien},
  \citenamefont {Omonije}, \citenamefont {Opremcak}, \citenamefont {Potter},
  \citenamefont {Pryadko}, \citenamefont {Quintana}, \citenamefont {Rhodes},
  \citenamefont {Rocque}, \citenamefont {Rubin}, \citenamefont {Saei},
  \citenamefont {Sank}, \citenamefont {Sankaragomathi}, \citenamefont
  {Satzinger}, \citenamefont {Schurkus}, \citenamefont {Schuster},
  \citenamefont {Shearn}, \citenamefont {Shorter}, \citenamefont {Shutty},
  \citenamefont {Shvarts}, \citenamefont {Sivak}, \citenamefont {Skruzny},
  \citenamefont {Smith}, \citenamefont {Somma}, \citenamefont {Sterling},
  \citenamefont {Strain}, \citenamefont {Szalay}, \citenamefont {Thor},
  \citenamefont {Torres}, \citenamefont {Vidal}, \citenamefont {Villalonga},
  \citenamefont {Heidweiller}, \citenamefont {White}, \citenamefont {Woo},
  \citenamefont {Xing}, \citenamefont {Yao}, \citenamefont {Yeh}, \citenamefont
  {Yoo}, \citenamefont {Young}, \citenamefont {Zalcman}, \citenamefont {Zhang},
  \citenamefont {Zhu}, \citenamefont {Zobrist}, \citenamefont {Neven},
  \citenamefont {Babbush}, \citenamefont {Bacon}, \citenamefont {Boixo},
  \citenamefont {Hilton}, \citenamefont {Lucero}, \citenamefont {Megrant},
  \citenamefont {Kelly}, \citenamefont {Chen}, \citenamefont {Smelyanskiy},
  \citenamefont {Khemani}, \citenamefont {Gopalakrishnan}, \citenamefont
  {Prosen},\ and\ \citenamefont {Roushan}}]{Rosenberg2024}%
  \BibitemOpen
  \bibfield  {author} {\bibinfo {author} {\bibfnamefont {E.}~\bibnamefont
  {Rosenberg}}, \bibinfo {author} {\bibfnamefont {T.~I.}\ \bibnamefont
  {Andersen}}, \bibinfo {author} {\bibfnamefont {R.}~\bibnamefont {Samajdar}},
  \bibinfo {author} {\bibfnamefont {A.}~\bibnamefont {Petukhov}}, \bibinfo
  {author} {\bibfnamefont {J.~C.}\ \bibnamefont {Hoke}}, \bibinfo {author}
  {\bibfnamefont {D.}~\bibnamefont {Abanin}}, \bibinfo {author} {\bibfnamefont
  {A.}~\bibnamefont {Bengtsson}}, \bibinfo {author} {\bibfnamefont {I.~K.}\
  \bibnamefont {Drozdov}}, \bibinfo {author} {\bibfnamefont {C.}~\bibnamefont
  {Erickson}}, \bibinfo {author} {\bibfnamefont {P.~V.}\ \bibnamefont
  {Klimov}}, \bibinfo {author} {\bibfnamefont {X.}~\bibnamefont {Mi}}, \bibinfo
  {author} {\bibfnamefont {A.}~\bibnamefont {Morvan}}, \bibinfo {author}
  {\bibfnamefont {M.}~\bibnamefont {Neeley}}, \bibinfo {author} {\bibfnamefont
  {C.}~\bibnamefont {Neill}}, \bibinfo {author} {\bibfnamefont
  {R.}~\bibnamefont {Acharya}}, \bibinfo {author} {\bibfnamefont
  {R.}~\bibnamefont {Allen}}, \bibinfo {author} {\bibfnamefont
  {K.}~\bibnamefont {Anderson}}, \bibinfo {author} {\bibfnamefont
  {M.}~\bibnamefont {Ansmann}}, \bibinfo {author} {\bibfnamefont
  {F.}~\bibnamefont {Arute}}, \bibinfo {author} {\bibfnamefont
  {K.}~\bibnamefont {Arya}}, \bibinfo {author} {\bibfnamefont {A.}~\bibnamefont
  {Asfaw}}, \bibinfo {author} {\bibfnamefont {J.}~\bibnamefont {Atalaya}},
  \bibinfo {author} {\bibfnamefont {J.~C.}\ \bibnamefont {Bardin}}, \bibinfo
  {author} {\bibfnamefont {A.}~\bibnamefont {Bilmes}}, \bibinfo {author}
  {\bibfnamefont {G.}~\bibnamefont {Bortoli}}, \bibinfo {author} {\bibfnamefont
  {A.}~\bibnamefont {Bourassa}}, \bibinfo {author} {\bibfnamefont
  {J.}~\bibnamefont {Bovaird}}, \bibinfo {author} {\bibfnamefont
  {L.}~\bibnamefont {Brill}}, \bibinfo {author} {\bibfnamefont
  {M.}~\bibnamefont {Broughton}}, \bibinfo {author} {\bibfnamefont {B.~B.}\
  \bibnamefont {Buckley}}, \bibinfo {author} {\bibfnamefont {D.~A.}\
  \bibnamefont {Buell}}, \bibinfo {author} {\bibfnamefont {T.}~\bibnamefont
  {Burger}}, \bibinfo {author} {\bibfnamefont {B.}~\bibnamefont {Burkett}},
  \bibinfo {author} {\bibfnamefont {N.}~\bibnamefont {Bushnell}}, \bibinfo
  {author} {\bibfnamefont {J.}~\bibnamefont {Campero}}, \bibinfo {author}
  {\bibfnamefont {H.-S.}\ \bibnamefont {Chang}}, \bibinfo {author}
  {\bibfnamefont {Z.}~\bibnamefont {Chen}}, \bibinfo {author} {\bibfnamefont
  {B.}~\bibnamefont {Chiaro}}, \bibinfo {author} {\bibfnamefont
  {D.}~\bibnamefont {Chik}}, \bibinfo {author} {\bibfnamefont {J.}~\bibnamefont
  {Cogan}}, \bibinfo {author} {\bibfnamefont {R.}~\bibnamefont {Collins}},
  \bibinfo {author} {\bibfnamefont {P.}~\bibnamefont {Conner}}, \bibinfo
  {author} {\bibfnamefont {W.}~\bibnamefont {Courtney}}, \bibinfo {author}
  {\bibfnamefont {A.~L.}\ \bibnamefont {Crook}}, \bibinfo {author}
  {\bibfnamefont {B.}~\bibnamefont {Curtin}}, \bibinfo {author} {\bibfnamefont
  {D.~M.}\ \bibnamefont {Debroy}}, \bibinfo {author} {\bibfnamefont {A.~D.~T.}\
  \bibnamefont {Barba}}, \bibinfo {author} {\bibfnamefont {S.}~\bibnamefont
  {Demura}}, \bibinfo {author} {\bibfnamefont {A.~D.}\ \bibnamefont {Paolo}},
  \bibinfo {author} {\bibfnamefont {A.}~\bibnamefont {Dunsworth}}, \bibinfo
  {author} {\bibfnamefont {C.}~\bibnamefont {Earle}}, \bibinfo {author}
  {\bibfnamefont {L.}~\bibnamefont {Faoro}}, \bibinfo {author} {\bibfnamefont
  {E.}~\bibnamefont {Farhi}}, \bibinfo {author} {\bibfnamefont
  {R.}~\bibnamefont {Fatemi}}, \bibinfo {author} {\bibfnamefont {V.~S.}\
  \bibnamefont {Ferreira}}, \bibinfo {author} {\bibfnamefont {L.~F.}\
  \bibnamefont {Burgos}}, \bibinfo {author} {\bibfnamefont {E.}~\bibnamefont
  {Forati}}, \bibinfo {author} {\bibfnamefont {A.~G.}\ \bibnamefont {Fowler}},
  \bibinfo {author} {\bibfnamefont {B.}~\bibnamefont {Foxen}}, \bibinfo
  {author} {\bibfnamefont {G.}~\bibnamefont {Garcia}}, \bibinfo {author}
  {\bibfnamefont {É.}\ \bibnamefont {Genois}}, \bibinfo {author}
  {\bibfnamefont {W.}~\bibnamefont {Giang}}, \bibinfo {author} {\bibfnamefont
  {C.}~\bibnamefont {Gidney}}, \bibinfo {author} {\bibfnamefont
  {D.}~\bibnamefont {Gilboa}}, \bibinfo {author} {\bibfnamefont
  {M.}~\bibnamefont {Giustina}}, \bibinfo {author} {\bibfnamefont
  {R.}~\bibnamefont {Gosula}}, \bibinfo {author} {\bibfnamefont {A.~G.}\
  \bibnamefont {Dau}}, \bibinfo {author} {\bibfnamefont {J.~A.}\ \bibnamefont
  {Gross}}, \bibinfo {author} {\bibfnamefont {S.}~\bibnamefont {Habegger}},
  \bibinfo {author} {\bibfnamefont {M.~C.}\ \bibnamefont {Hamilton}}, \bibinfo
  {author} {\bibfnamefont {M.}~\bibnamefont {Hansen}}, \bibinfo {author}
  {\bibfnamefont {M.~P.}\ \bibnamefont {Harrigan}}, \bibinfo {author}
  {\bibfnamefont {S.~D.}\ \bibnamefont {Harrington}}, \bibinfo {author}
  {\bibfnamefont {P.}~\bibnamefont {Heu}}, \bibinfo {author} {\bibfnamefont
  {G.}~\bibnamefont {Hill}}, \bibinfo {author} {\bibfnamefont {M.~R.}\
  \bibnamefont {Hoffmann}}, \bibinfo {author} {\bibfnamefont {S.}~\bibnamefont
  {Hong}}, \bibinfo {author} {\bibfnamefont {T.}~\bibnamefont {Huang}},
  \bibinfo {author} {\bibfnamefont {A.}~\bibnamefont {Huff}}, \bibinfo {author}
  {\bibfnamefont {W.~J.}\ \bibnamefont {Huggins}}, \bibinfo {author}
  {\bibfnamefont {L.~B.}\ \bibnamefont {Ioffe}}, \bibinfo {author}
  {\bibfnamefont {S.~V.}\ \bibnamefont {Isakov}}, \bibinfo {author}
  {\bibfnamefont {J.}~\bibnamefont {Iveland}}, \bibinfo {author} {\bibfnamefont
  {E.}~\bibnamefont {Jeffrey}}, \bibinfo {author} {\bibfnamefont
  {Z.}~\bibnamefont {Jiang}}, \bibinfo {author} {\bibfnamefont
  {C.}~\bibnamefont {Jones}}, \bibinfo {author} {\bibfnamefont
  {P.}~\bibnamefont {Juhas}}, \bibinfo {author} {\bibfnamefont
  {D.}~\bibnamefont {Kafri}}, \bibinfo {author} {\bibfnamefont
  {T.}~\bibnamefont {Khattar}}, \bibinfo {author} {\bibfnamefont
  {M.}~\bibnamefont {Khezri}}, \bibinfo {author} {\bibfnamefont
  {M.}~\bibnamefont {Kieferová}}, \bibinfo {author} {\bibfnamefont
  {S.}~\bibnamefont {Kim}}, \bibinfo {author} {\bibfnamefont {A.}~\bibnamefont
  {Kitaev}}, \bibinfo {author} {\bibfnamefont {A.~R.}\ \bibnamefont {Klots}},
  \bibinfo {author} {\bibfnamefont {A.~N.}\ \bibnamefont {Korotkov}}, \bibinfo
  {author} {\bibfnamefont {F.}~\bibnamefont {Kostritsa}}, \bibinfo {author}
  {\bibfnamefont {J.~M.}\ \bibnamefont {Kreikebaum}}, \bibinfo {author}
  {\bibfnamefont {D.}~\bibnamefont {Landhuis}}, \bibinfo {author}
  {\bibfnamefont {P.}~\bibnamefont {Laptev}}, \bibinfo {author} {\bibfnamefont
  {K.-M.}\ \bibnamefont {Lau}}, \bibinfo {author} {\bibfnamefont
  {L.}~\bibnamefont {Laws}}, \bibinfo {author} {\bibfnamefont {J.}~\bibnamefont
  {Lee}}, \bibinfo {author} {\bibfnamefont {K.~W.}\ \bibnamefont {Lee}},
  \bibinfo {author} {\bibfnamefont {Y.~D.}\ \bibnamefont {Lensky}}, \bibinfo
  {author} {\bibfnamefont {B.~J.}\ \bibnamefont {Lester}}, \bibinfo {author}
  {\bibfnamefont {A.~T.}\ \bibnamefont {Lill}}, \bibinfo {author}
  {\bibfnamefont {W.}~\bibnamefont {Liu}}, \bibinfo {author} {\bibfnamefont
  {A.}~\bibnamefont {Locharla}}, \bibinfo {author} {\bibfnamefont
  {S.}~\bibnamefont {Mandrà}}, \bibinfo {author} {\bibfnamefont
  {O.}~\bibnamefont {Martin}}, \bibinfo {author} {\bibfnamefont
  {S.}~\bibnamefont {Martin}}, \bibinfo {author} {\bibfnamefont {J.~R.}\
  \bibnamefont {McClean}}, \bibinfo {author} {\bibfnamefont {M.}~\bibnamefont
  {McEwen}}, \bibinfo {author} {\bibfnamefont {S.}~\bibnamefont {Meeks}},
  \bibinfo {author} {\bibfnamefont {K.~C.}\ \bibnamefont {Miao}}, \bibinfo
  {author} {\bibfnamefont {A.}~\bibnamefont {Mieszala}}, \bibinfo {author}
  {\bibfnamefont {S.}~\bibnamefont {Montazeri}}, \bibinfo {author}
  {\bibfnamefont {R.}~\bibnamefont {Movassagh}}, \bibinfo {author}
  {\bibfnamefont {W.}~\bibnamefont {Mruczkiewicz}}, \bibinfo {author}
  {\bibfnamefont {A.}~\bibnamefont {Nersisyan}}, \bibinfo {author}
  {\bibfnamefont {M.}~\bibnamefont {Newman}}, \bibinfo {author} {\bibfnamefont
  {J.~H.}\ \bibnamefont {Ng}}, \bibinfo {author} {\bibfnamefont
  {A.}~\bibnamefont {Nguyen}}, \bibinfo {author} {\bibfnamefont
  {M.}~\bibnamefont {Nguyen}}, \bibinfo {author} {\bibfnamefont {M.~Y.}\
  \bibnamefont {Niu}}, \bibinfo {author} {\bibfnamefont {T.~E.}\ \bibnamefont
  {O’Brien}}, \bibinfo {author} {\bibfnamefont {S.}~\bibnamefont {Omonije}},
  \bibinfo {author} {\bibfnamefont {A.}~\bibnamefont {Opremcak}}, \bibinfo
  {author} {\bibfnamefont {R.}~\bibnamefont {Potter}}, \bibinfo {author}
  {\bibfnamefont {L.~P.}\ \bibnamefont {Pryadko}}, \bibinfo {author}
  {\bibfnamefont {C.}~\bibnamefont {Quintana}}, \bibinfo {author}
  {\bibfnamefont {D.~M.}\ \bibnamefont {Rhodes}}, \bibinfo {author}
  {\bibfnamefont {C.}~\bibnamefont {Rocque}}, \bibinfo {author} {\bibfnamefont
  {N.~C.}\ \bibnamefont {Rubin}}, \bibinfo {author} {\bibfnamefont
  {N.}~\bibnamefont {Saei}}, \bibinfo {author} {\bibfnamefont {D.}~\bibnamefont
  {Sank}}, \bibinfo {author} {\bibfnamefont {K.}~\bibnamefont
  {Sankaragomathi}}, \bibinfo {author} {\bibfnamefont {K.~J.}\ \bibnamefont
  {Satzinger}}, \bibinfo {author} {\bibfnamefont {H.~F.}\ \bibnamefont
  {Schurkus}}, \bibinfo {author} {\bibfnamefont {C.}~\bibnamefont {Schuster}},
  \bibinfo {author} {\bibfnamefont {M.~J.}\ \bibnamefont {Shearn}}, \bibinfo
  {author} {\bibfnamefont {A.}~\bibnamefont {Shorter}}, \bibinfo {author}
  {\bibfnamefont {N.}~\bibnamefont {Shutty}}, \bibinfo {author} {\bibfnamefont
  {V.}~\bibnamefont {Shvarts}}, \bibinfo {author} {\bibfnamefont
  {V.}~\bibnamefont {Sivak}}, \bibinfo {author} {\bibfnamefont
  {J.}~\bibnamefont {Skruzny}}, \bibinfo {author} {\bibfnamefont {W.~C.}\
  \bibnamefont {Smith}}, \bibinfo {author} {\bibfnamefont {R.~D.}\ \bibnamefont
  {Somma}}, \bibinfo {author} {\bibfnamefont {G.}~\bibnamefont {Sterling}},
  \bibinfo {author} {\bibfnamefont {D.}~\bibnamefont {Strain}}, \bibinfo
  {author} {\bibfnamefont {M.}~\bibnamefont {Szalay}}, \bibinfo {author}
  {\bibfnamefont {D.}~\bibnamefont {Thor}}, \bibinfo {author} {\bibfnamefont
  {A.}~\bibnamefont {Torres}}, \bibinfo {author} {\bibfnamefont
  {G.}~\bibnamefont {Vidal}}, \bibinfo {author} {\bibfnamefont
  {B.}~\bibnamefont {Villalonga}}, \bibinfo {author} {\bibfnamefont {C.~V.}\
  \bibnamefont {Heidweiller}}, \bibinfo {author} {\bibfnamefont
  {T.}~\bibnamefont {White}}, \bibinfo {author} {\bibfnamefont {B.~W.~K.}\
  \bibnamefont {Woo}}, \bibinfo {author} {\bibfnamefont {C.}~\bibnamefont
  {Xing}}, \bibinfo {author} {\bibfnamefont {Z.~J.}\ \bibnamefont {Yao}},
  \bibinfo {author} {\bibfnamefont {P.}~\bibnamefont {Yeh}}, \bibinfo {author}
  {\bibfnamefont {J.}~\bibnamefont {Yoo}}, \bibinfo {author} {\bibfnamefont
  {G.}~\bibnamefont {Young}}, \bibinfo {author} {\bibfnamefont
  {A.}~\bibnamefont {Zalcman}}, \bibinfo {author} {\bibfnamefont
  {Y.}~\bibnamefont {Zhang}}, \bibinfo {author} {\bibfnamefont
  {N.}~\bibnamefont {Zhu}}, \bibinfo {author} {\bibfnamefont {N.}~\bibnamefont
  {Zobrist}}, \bibinfo {author} {\bibfnamefont {H.}~\bibnamefont {Neven}},
  \bibinfo {author} {\bibfnamefont {R.}~\bibnamefont {Babbush}}, \bibinfo
  {author} {\bibfnamefont {D.}~\bibnamefont {Bacon}}, \bibinfo {author}
  {\bibfnamefont {S.}~\bibnamefont {Boixo}}, \bibinfo {author} {\bibfnamefont
  {J.}~\bibnamefont {Hilton}}, \bibinfo {author} {\bibfnamefont
  {E.}~\bibnamefont {Lucero}}, \bibinfo {author} {\bibfnamefont
  {A.}~\bibnamefont {Megrant}}, \bibinfo {author} {\bibfnamefont
  {J.}~\bibnamefont {Kelly}}, \bibinfo {author} {\bibfnamefont
  {Y.}~\bibnamefont {Chen}}, \bibinfo {author} {\bibfnamefont {V.}~\bibnamefont
  {Smelyanskiy}}, \bibinfo {author} {\bibfnamefont {V.}~\bibnamefont
  {Khemani}}, \bibinfo {author} {\bibfnamefont {S.}~\bibnamefont
  {Gopalakrishnan}}, \bibinfo {author} {\bibfnamefont {T.}~\bibnamefont
  {Prosen}},\ and\ \bibinfo {author} {\bibfnamefont {P.}~\bibnamefont
  {Roushan}},\ }\bibfield  {title} {\bibinfo {title} {Dynamics of magnetization
  at infinite temperature in a {{H}}eisenberg spin chain},\ }\href
  {https://doi.org/10.1126/science.adi7877} {\bibfield  {journal} {\bibinfo
  {journal} {Science}\ }\textbf {\bibinfo {volume} {384}},\ \bibinfo {pages}
  {48} (\bibinfo {year} {2024})}\BibitemShut {NoStop}%
\bibitem [{\citenamefont {Eisler}\ and\ \citenamefont
  {R\'acz}(2013)}]{Eisler2013}%
  \BibitemOpen
  \bibfield  {author} {\bibinfo {author} {\bibfnamefont {V.}~\bibnamefont
  {Eisler}}\ and\ \bibinfo {author} {\bibfnamefont {Z.}~\bibnamefont
  {R\'acz}},\ }\bibfield  {title} {\bibinfo {title} {Full counting statistics
  in a propagating quantum front and random matrix spectra},\ }\href
  {https://doi.org/10.1103/PhysRevLett.110.060602} {\bibfield  {journal}
  {\bibinfo  {journal} {Phys. Rev. Lett.}\ }\textbf {\bibinfo {volume} {110}},\
  \bibinfo {pages} {060602} (\bibinfo {year} {2013})}\BibitemShut {NoStop}%
\bibitem [{\citenamefont {Saenz}\ \emph {et~al.}(2022)\citenamefont {Saenz},
  \citenamefont {Tracy},\ and\ \citenamefont {Widom}}]{Saenz2022}%
  \BibitemOpen
  \bibfield  {author} {\bibinfo {author} {\bibfnamefont {A.}~\bibnamefont
  {Saenz}}, \bibinfo {author} {\bibfnamefont {C.~A.}\ \bibnamefont {Tracy}},\
  and\ \bibinfo {author} {\bibfnamefont {H.}~\bibnamefont {Widom}},\ }\bibinfo
  {title} {Domain {{W}}alls in the {{H}}eisenberg-{{I}}sing {{S}}pin-1/2
  {{C}}hain},\ in\ \href {https://doi.org/10.1007/978-3-031-13851-5_2} {\emph
  {\bibinfo {booktitle} {Toeplitz Operators and Random Matrices: In Memory of
  Harold Widom}}},\ \bibinfo {editor} {edited by\ \bibinfo {editor}
  {\bibfnamefont {E.}~\bibnamefont {Basor}}, \bibinfo {editor} {\bibfnamefont
  {A.}~\bibnamefont {B{\"o}ttcher}}, \bibinfo {editor} {\bibfnamefont
  {T.}~\bibnamefont {Ehrhardt}},\ and\ \bibinfo {editor} {\bibfnamefont
  {C.~A.}\ \bibnamefont {Tracy}}}\ (\bibinfo  {publisher} {Springer
  International Publishing},\ \bibinfo {address} {Cham},\ \bibinfo {year}
  {2022})\ pp.\ \bibinfo {pages} {9--47}\BibitemShut {NoStop}%
\bibitem [{\citenamefont {Collura}\ \emph {et~al.}(2018)\citenamefont
  {Collura}, \citenamefont {De~Luca},\ and\ \citenamefont
  {Viti}}]{Collura2018}%
  \BibitemOpen
  \bibfield  {author} {\bibinfo {author} {\bibfnamefont {M.}~\bibnamefont
  {Collura}}, \bibinfo {author} {\bibfnamefont {A.}~\bibnamefont {De~Luca}},\
  and\ \bibinfo {author} {\bibfnamefont {J.}~\bibnamefont {Viti}},\ }\bibfield
  {title} {\bibinfo {title} {Analytic solution of the domain-wall
  nonequilibrium stationary state},\ }\href
  {https://doi.org/10.1103/PhysRevB.97.081111} {\bibfield  {journal} {\bibinfo
  {journal} {Phys. Rev. B}\ }\textbf {\bibinfo {volume} {97}},\ \bibinfo
  {pages} {081111} (\bibinfo {year} {2018})}\BibitemShut {NoStop}%
\bibitem [{\citenamefont {Stéphan}(2019)}]{Stephan2019}%
  \BibitemOpen
  \bibfield  {author} {\bibinfo {author} {\bibfnamefont {J.-M.}\ \bibnamefont
  {Stéphan}},\ }\bibfield  {title} {\bibinfo {title} {{Free fermions at the
  edge of interacting systems}},\ }\href
  {https://doi.org/10.21468/SciPostPhys.6.5.057} {\bibfield  {journal}
  {\bibinfo  {journal} {SciPost Phys.}\ }\textbf {\bibinfo {volume} {6}},\
  \bibinfo {pages} {057} (\bibinfo {year} {2019})}\BibitemShut {NoStop}%
\bibitem [{\citenamefont {Bulchandani}\ and\ \citenamefont
  {Karrasch}(2019)}]{Bulchandani2019}%
  \BibitemOpen
  \bibfield  {author} {\bibinfo {author} {\bibfnamefont {V.~B.}\ \bibnamefont
  {Bulchandani}}\ and\ \bibinfo {author} {\bibfnamefont {C.}~\bibnamefont
  {Karrasch}},\ }\bibfield  {title} {\bibinfo {title} {Subdiffusive front
  scaling in interacting integrable models},\ }\href
  {https://doi.org/10.1103/PhysRevB.99.121410} {\bibfield  {journal} {\bibinfo
  {journal} {Phys. Rev. B}\ }\textbf {\bibinfo {volume} {99}},\ \bibinfo
  {pages} {121410} (\bibinfo {year} {2019})}\BibitemShut {NoStop}%
\bibitem [{\citenamefont {Takahashi}(1999)}]{Takahashi1999}%
  \BibitemOpen
  \bibfield  {author} {\bibinfo {author} {\bibfnamefont {M.}~\bibnamefont
  {Takahashi}},\ }\bibfield  {title} {\bibinfo {title} {Thermodynamics of
  one-dimensional solvable models},\ }\href@noop {} {\  (\bibinfo {year}
  {1999})}\BibitemShut {NoStop}%
\bibitem [{\citenamefont {Franchini}\ \emph {et~al.}(2017)\citenamefont
  {Franchini} \emph {et~al.}}]{Franchini2017}%
  \BibitemOpen
  \bibfield  {author} {\bibinfo {author} {\bibfnamefont {F.}~\bibnamefont
  {Franchini}} \emph {et~al.},\ }\href@noop {} {\emph {\bibinfo {title} {An
  introduction to integrable techniques for one-dimensional quantum
  systems}}},\ Vol.\ \bibinfo {volume} {940}\ (\bibinfo  {publisher}
  {Springer},\ \bibinfo {year} {2017})\BibitemShut {NoStop}%
\bibitem [{\citenamefont {Yang}\ \emph {et~al.}(2020)\citenamefont {Yang},
  \citenamefont {Liu}, \citenamefont {Gorshkov},\ and\ \citenamefont
  {Iadecola}}]{Yang2020}%
  \BibitemOpen
  \bibfield  {author} {\bibinfo {author} {\bibfnamefont {Z.-C.}\ \bibnamefont
  {Yang}}, \bibinfo {author} {\bibfnamefont {F.}~\bibnamefont {Liu}}, \bibinfo
  {author} {\bibfnamefont {A.~V.}\ \bibnamefont {Gorshkov}},\ and\ \bibinfo
  {author} {\bibfnamefont {T.}~\bibnamefont {Iadecola}},\ }\bibfield  {title}
  {\bibinfo {title} {Hilbert-space fragmentation from strict confinement},\
  }\href {https://doi.org/10.1103/PhysRevLett.124.207602} {\bibfield  {journal}
  {\bibinfo  {journal} {Phys. Rev. Lett.}\ }\textbf {\bibinfo {volume} {124}},\
  \bibinfo {pages} {207602} (\bibinfo {year} {2020})}\BibitemShut {NoStop}%
\bibitem [{\citenamefont {Zadnik}\ and\ \citenamefont
  {Fagotti}(2021)}]{Zadnik2021_1}%
  \BibitemOpen
  \bibfield  {author} {\bibinfo {author} {\bibfnamefont {L.}~\bibnamefont
  {Zadnik}}\ and\ \bibinfo {author} {\bibfnamefont {M.}~\bibnamefont
  {Fagotti}},\ }\bibfield  {title} {\bibinfo {title} {{The Folded Spin-1/2
  {{XXZ}} Model: I. {{D}}iagonalisation, Jamming, and Ground State
  Properties}},\ }\href {https://doi.org/10.21468/SciPostPhysCore.4.2.010}
  {\bibfield  {journal} {\bibinfo  {journal} {SciPost Phys. Core}\ }\textbf
  {\bibinfo {volume} {4}},\ \bibinfo {pages} {010} (\bibinfo {year}
  {2021})}\BibitemShut {NoStop}%
\bibitem [{\citenamefont {Zadnik}\ \emph {et~al.}(2021)\citenamefont {Zadnik},
  \citenamefont {Bidzhiev},\ and\ \citenamefont {Fagotti}}]{Zadnik2021_2}%
  \BibitemOpen
  \bibfield  {author} {\bibinfo {author} {\bibfnamefont {L.}~\bibnamefont
  {Zadnik}}, \bibinfo {author} {\bibfnamefont {K.}~\bibnamefont {Bidzhiev}},\
  and\ \bibinfo {author} {\bibfnamefont {M.}~\bibnamefont {Fagotti}},\
  }\bibfield  {title} {\bibinfo {title} {{The folded spin-1/2 {{XXZ}} model:
  II. {{T}}hermodynamics and hydrodynamics with a minimal set of charges}},\
  }\href {https://doi.org/10.21468/SciPostPhys.10.5.099} {\bibfield  {journal}
  {\bibinfo  {journal} {SciPost Phys.}\ }\textbf {\bibinfo {volume} {10}},\
  \bibinfo {pages} {099} (\bibinfo {year} {2021})}\BibitemShut {NoStop}%
\bibitem [{\citenamefont {Pozsgay}\ \emph
  {et~al.}(2021{\natexlab{a}})\citenamefont {Pozsgay}, \citenamefont {Gombor},
  \citenamefont {Hutsalyuk}, \citenamefont {Jiang}, \citenamefont
  {Pristy\'ak},\ and\ \citenamefont {Vernier}}]{Pozsgay2021}%
  \BibitemOpen
  \bibfield  {author} {\bibinfo {author} {\bibfnamefont {B.}~\bibnamefont
  {Pozsgay}}, \bibinfo {author} {\bibfnamefont {T.}~\bibnamefont {Gombor}},
  \bibinfo {author} {\bibfnamefont {A.}~\bibnamefont {Hutsalyuk}}, \bibinfo
  {author} {\bibfnamefont {Y.}~\bibnamefont {Jiang}}, \bibinfo {author}
  {\bibfnamefont {L.}~\bibnamefont {Pristy\'ak}},\ and\ \bibinfo {author}
  {\bibfnamefont {E.}~\bibnamefont {Vernier}},\ }\bibfield  {title} {\bibinfo
  {title} {Integrable spin chain with {{H}}ilbert space fragmentation and
  solvable real-time dynamics},\ }\href
  {https://doi.org/10.1103/PhysRevE.104.044106} {\bibfield  {journal} {\bibinfo
   {journal} {Phys. Rev. E}\ }\textbf {\bibinfo {volume} {104}},\ \bibinfo
  {pages} {044106} (\bibinfo {year} {2021}{\natexlab{a}})}\BibitemShut
  {NoStop}%
\bibitem [{\citenamefont {Pozsgay}\ \emph
  {et~al.}(2021{\natexlab{b}})\citenamefont {Pozsgay}, \citenamefont {Gombor},\
  and\ \citenamefont {Hutsalyuk}}]{Pozsgay2021_2}%
  \BibitemOpen
  \bibfield  {author} {\bibinfo {author} {\bibfnamefont {B.}~\bibnamefont
  {Pozsgay}}, \bibinfo {author} {\bibfnamefont {T.}~\bibnamefont {Gombor}},\
  and\ \bibinfo {author} {\bibfnamefont {A.}~\bibnamefont {Hutsalyuk}},\
  }\bibfield  {title} {\bibinfo {title} {Integrable hard-rod deformation of the
  {{H}}eisenberg spin chains},\ }\href
  {https://doi.org/10.1103/PhysRevE.104.064124} {\bibfield  {journal} {\bibinfo
   {journal} {Phys. Rev. E}\ }\textbf {\bibinfo {volume} {104}},\ \bibinfo
  {pages} {064124} (\bibinfo {year} {2021}{\natexlab{b}})}\BibitemShut
  {NoStop}%
\bibitem [{\citenamefont {Borsi}\ \emph {et~al.}(2023)\citenamefont {Borsi},
  \citenamefont {Pristy\'ak},\ and\ \citenamefont {Pozsgay}}]{Borsi2023}%
  \BibitemOpen
  \bibfield  {author} {\bibinfo {author} {\bibfnamefont {M.}~\bibnamefont
  {Borsi}}, \bibinfo {author} {\bibfnamefont {L.}~\bibnamefont {Pristy\'ak}},\
  and\ \bibinfo {author} {\bibfnamefont {B.}~\bibnamefont {Pozsgay}},\
  }\bibfield  {title} {\bibinfo {title} {Matrix product symmetries and
  breakdown of thermalization from hard rod deformations},\ }\href
  {https://doi.org/10.1103/PhysRevLett.131.037101} {\bibfield  {journal}
  {\bibinfo  {journal} {Phys. Rev. Lett.}\ }\textbf {\bibinfo {volume} {131}},\
  \bibinfo {pages} {037101} (\bibinfo {year} {2023})}\BibitemShut {NoStop}%
\bibitem [{\citenamefont {Sachdev}(1999)}]{Sachdev1999}%
  \BibitemOpen
  \bibfield  {author} {\bibinfo {author} {\bibfnamefont {S.}~\bibnamefont
  {Sachdev}},\ }\bibfield  {title} {\bibinfo {title} {Quantum phase
  transitions},\ }\href@noop {} {\bibfield  {journal} {\bibinfo  {journal}
  {Physics world}\ }\textbf {\bibinfo {volume} {12}},\ \bibinfo {pages} {33}
  (\bibinfo {year} {1999})}\BibitemShut {NoStop}%
\bibitem [{\citenamefont {Lewenstein}\ \emph {et~al.}(2012)\citenamefont
  {Lewenstein}, \citenamefont {Sanpera},\ and\ \citenamefont
  {Ahufinger}}]{Lewenstein2012}%
  \BibitemOpen
  \bibfield  {author} {\bibinfo {author} {\bibfnamefont {M.}~\bibnamefont
  {Lewenstein}}, \bibinfo {author} {\bibfnamefont {A.}~\bibnamefont
  {Sanpera}},\ and\ \bibinfo {author} {\bibfnamefont {V.}~\bibnamefont
  {Ahufinger}},\ }\href
  {https://doi.org/10.1093/acprof:oso/9780199573127.001.0001} {\emph {\bibinfo
  {title} {{Ultracold Atoms in Optical Lattices: Simulating quantum many-body
  systems}}}}\ (\bibinfo  {publisher} {Oxford University Press},\ \bibinfo
  {year} {2012})\BibitemShut {NoStop}%
\bibitem [{\citenamefont {Vidal}(2003)}]{TEBD1}%
  \BibitemOpen
  \bibfield  {author} {\bibinfo {author} {\bibfnamefont {G.}~\bibnamefont
  {Vidal}},\ }\bibfield  {title} {\bibinfo {title} {Efficient classical
  simulation of slightly entangled quantum computations},\ }\href
  {https://doi.org/10.1103/PhysRevLett.91.147902} {\bibfield  {journal}
  {\bibinfo  {journal} {Phys. Rev. Lett.}\ }\textbf {\bibinfo {volume} {91}},\
  \bibinfo {pages} {147902} (\bibinfo {year} {2003})}\BibitemShut {NoStop}%
\bibitem [{\citenamefont {Vidal}(2004)}]{TEBD2}%
  \BibitemOpen
  \bibfield  {author} {\bibinfo {author} {\bibfnamefont {G.}~\bibnamefont
  {Vidal}},\ }\bibfield  {title} {\bibinfo {title} {Efficient simulation of
  one-dimensional quantum many-body systems},\ }\href
  {https://doi.org/10.1103/PhysRevLett.93.040502} {\bibfield  {journal}
  {\bibinfo  {journal} {Phys. Rev. Lett.}\ }\textbf {\bibinfo {volume} {93}},\
  \bibinfo {pages} {040502} (\bibinfo {year} {2004})}\BibitemShut {NoStop}%
\bibitem [{\citenamefont {Schollwöck}(2011)}]{TEBD3}%
  \BibitemOpen
  \bibfield  {author} {\bibinfo {author} {\bibfnamefont {U.}~\bibnamefont
  {Schollwöck}},\ }\bibfield  {title} {\bibinfo {title} {The density-matrix
  renormalization group in the age of matrix product states},\ }\href
  {https://doi.org/https://doi.org/10.1016/j.aop.2010.09.012} {\bibfield
  {journal} {\bibinfo  {journal} {Annals of Physics}\ }\textbf {\bibinfo
  {volume} {326}},\ \bibinfo {pages} {96} (\bibinfo {year} {2011})}\BibitemShut
  {NoStop}%
\bibitem [{\citenamefont {Paeckel}\ \emph {et~al.}(2019)\citenamefont
  {Paeckel}, \citenamefont {Köhler}, \citenamefont {Swoboda}, \citenamefont
  {Manmana}, \citenamefont {Schollwöck},\ and\ \citenamefont {Hubig}}]{TEBD4}%
  \BibitemOpen
  \bibfield  {author} {\bibinfo {author} {\bibfnamefont {S.}~\bibnamefont
  {Paeckel}}, \bibinfo {author} {\bibfnamefont {T.}~\bibnamefont {Köhler}},
  \bibinfo {author} {\bibfnamefont {A.}~\bibnamefont {Swoboda}}, \bibinfo
  {author} {\bibfnamefont {S.~R.}\ \bibnamefont {Manmana}}, \bibinfo {author}
  {\bibfnamefont {U.}~\bibnamefont {Schollwöck}},\ and\ \bibinfo {author}
  {\bibfnamefont {C.}~\bibnamefont {Hubig}},\ }\bibfield  {title} {\bibinfo
  {title} {Time-evolution methods for matrix-product states},\ }\href
  {https://doi.org/https://doi.org/10.1016/j.aop.2019.167998} {\bibfield
  {journal} {\bibinfo  {journal} {Annals of Physics}\ }\textbf {\bibinfo
  {volume} {411}},\ \bibinfo {pages} {167998} (\bibinfo {year}
  {2019})}\BibitemShut {NoStop}%
\bibitem [{\citenamefont {Sch{\"u}tz}(1997)}]{Schutz1997}%
  \BibitemOpen
  \bibfield  {author} {\bibinfo {author} {\bibfnamefont {G.~M.}\ \bibnamefont
  {Sch{\"u}tz}},\ }\bibfield  {title} {\bibinfo {title} {Exact solution of the
  master equation for the asymmetric exclusion process},\ }\href
  {https://doi.org/10.1007/BF02508478} {\bibfield  {journal} {\bibinfo
  {journal} {Journal of Statistical Physics}\ }\textbf {\bibinfo {volume}
  {88}},\ \bibinfo {pages} {427} (\bibinfo {year} {1997})}\BibitemShut
  {NoStop}%
\bibitem [{\citenamefont {Tracy}\ and\ \citenamefont
  {Widom}(2008)}]{Tracy2008}%
  \BibitemOpen
  \bibfield  {author} {\bibinfo {author} {\bibfnamefont {C.~A.}\ \bibnamefont
  {Tracy}}\ and\ \bibinfo {author} {\bibfnamefont {H.}~\bibnamefont {Widom}},\
  }\bibfield  {title} {\bibinfo {title} {Integral formulas for the asymmetric
  simple exclusion process},\ }\href
  {https://doi.org/10.1007/s00220-008-0443-3} {\bibfield  {journal} {\bibinfo
  {journal} {Communications in Mathematical Physics}\ }\textbf {\bibinfo
  {volume} {279}},\ \bibinfo {pages} {815} (\bibinfo {year}
  {2008})}\BibitemShut {NoStop}%
\bibitem [{SM()}]{SM}%
  \BibitemOpen
  \href@noop {} {\bibinfo  {journal} {See Supplemental Material for (I) Proof
  of Eq. (5) in the main text, (II) Determinantal formula for Eq. (5) in the
  main text, (III) Proof of Eq. (13) in the main text, (IV) Asymptotic analysis
  to the GUE Tracy-Widom ditribution with fast convergence, (V) Numerical
  truncation in the TEBD method, (VI) Limiting probability distribution
  function for $\Delta = 0$, and (VII) Conjecture for the XXZ model by Saenz,
  Tracy, and Widom}\ }\BibitemShut {NoStop}%
\bibitem [{\citenamefont {Cantini}\ \emph {et~al.}(2020)\citenamefont
  {Cantini}, \citenamefont {Colomo},\ and\ \citenamefont
  {Pronko}}]{Cantini2020}%
  \BibitemOpen
\bibfield  {journal} {  }\bibfield  {author} {\bibinfo {author} {\bibfnamefont
  {L.}~\bibnamefont {Cantini}}, \bibinfo {author} {\bibfnamefont
  {F.}~\bibnamefont {Colomo}},\ and\ \bibinfo {author} {\bibfnamefont {A.~G.}\
  \bibnamefont {Pronko}},\ }\bibfield  {title} {\bibinfo {title} {Integral
  formulas and antisymmetrization relations for the six-vertex model},\ }\href
  {https://doi.org/10.1007/s00023-019-00856-6} {\bibfield  {journal} {\bibinfo
  {journal} {Annales Henri Poincar{\'e}}\ }\textbf {\bibinfo {volume} {21}},\
  \bibinfo {pages} {865} (\bibinfo {year} {2020})}\BibitemShut {NoStop}%
\bibitem [{\citenamefont {Petrov}(2021)}]{Petrov2021}%
  \BibitemOpen
  \bibfield  {author} {\bibinfo {author} {\bibfnamefont {L.}~\bibnamefont
  {Petrov}},\ }\bibfield  {title} {\bibinfo {title} {Refined {{C}}auchy
  identity for spin {{H}}all–{{L}}ittlewood symmetric rational functions},\
  }\href {https://doi.org/https://doi.org/10.1016/j.jcta.2021.105519}
  {\bibfield  {journal} {\bibinfo  {journal} {Journal of Combinatorial Theory,
  Series A}\ }\textbf {\bibinfo {volume} {184}},\ \bibinfo {pages} {105519}
  (\bibinfo {year} {2021})}\BibitemShut {NoStop}%
\bibitem [{\citenamefont {Korepin}(1982)}]{Korepin1982}%
  \BibitemOpen
  \bibfield  {author} {\bibinfo {author} {\bibfnamefont {V.~E.}\ \bibnamefont
  {Korepin}},\ }\bibfield  {title} {\bibinfo {title} {{Calculation of norms of
  {{B}}ethe wave functions}},\ }\href
  {https://projecteuclid.org/journals/communications-in-mathematical-physics/volume-86/issue-3/Calculation-of-norms-of-Bethe-wave-functions/cmp/1103921777.full}
  {\bibfield  {journal} {\bibinfo  {journal} {Communications in Mathematical
  Physics}\ }\textbf {\bibinfo {volume} {86}},\ \bibinfo {pages} {391 }
  (\bibinfo {year} {1982})}\BibitemShut {NoStop}%
\bibitem [{\citenamefont {{Izergin}}(1987)}]{Izergin1987}%
  \BibitemOpen
  \bibfield  {author} {\bibinfo {author} {\bibfnamefont {A.~G.}\ \bibnamefont
  {{Izergin}}},\ }\bibfield  {title} {\bibinfo {title} {{Partition function of
  the six-vertex model in a finite volume}},\ }\href@noop {} {\bibfield
  {journal} {\bibinfo  {journal} {Soviet Physics Doklady}\ }\textbf {\bibinfo
  {volume} {32}},\ \bibinfo {pages} {878} (\bibinfo {year} {1987})}\BibitemShut
  {NoStop}%
\bibitem [{\citenamefont {Izergin}\ \emph {et~al.}(1992)\citenamefont
  {Izergin}, \citenamefont {Coker},\ and\ \citenamefont
  {Korepin}}]{Izergin1992}%
  \BibitemOpen
  \bibfield  {author} {\bibinfo {author} {\bibfnamefont {A.~G.}\ \bibnamefont
  {Izergin}}, \bibinfo {author} {\bibfnamefont {D.~A.}\ \bibnamefont {Coker}},\
  and\ \bibinfo {author} {\bibfnamefont {V.~E.}\ \bibnamefont {Korepin}},\
  }\bibfield  {title} {\bibinfo {title} {Determinant formula for the six-vertex
  model},\ }\href {https://doi.org/10.1088/0305-4470/25/16/010} {\bibfield
  {journal} {\bibinfo  {journal} {Journal of Physics A: Mathematical and
  General}\ }\textbf {\bibinfo {volume} {25}},\ \bibinfo {pages} {4315}
  (\bibinfo {year} {1992})}\BibitemShut {NoStop}%
\bibitem [{\citenamefont {Korepin}\ \emph {et~al.}(1993)\citenamefont
  {Korepin}, \citenamefont {Bogoliubov},\ and\ \citenamefont
  {Izergin}}]{Korepin_Bogoliubov_Izergin_1993}%
  \BibitemOpen
  \bibfield  {author} {\bibinfo {author} {\bibfnamefont {V.~E.}\ \bibnamefont
  {Korepin}}, \bibinfo {author} {\bibfnamefont {N.~M.}\ \bibnamefont
  {Bogoliubov}},\ and\ \bibinfo {author} {\bibfnamefont {A.~G.}\ \bibnamefont
  {Izergin}},\ }\href@noop {} {\emph {\bibinfo {title} {Quantum Inverse
  Scattering Method and Correlation Functions}}},\ Cambridge Monographs on
  Mathematical Physics\ (\bibinfo  {publisher} {Cambridge University Press},\
  \bibinfo {year} {1993})\BibitemShut {NoStop}%
\bibitem [{\citenamefont {Mehta}(2004)}]{mehta1}%
  \BibitemOpen
  \bibfield  {author} {\bibinfo {author} {\bibfnamefont {M.~L.}\ \bibnamefont
  {Mehta}},\ }\href@noop {} {\emph {\bibinfo {title} {Random matrices}}}\
  (\bibinfo  {publisher} {Elsevier},\ \bibinfo {year} {2004})\BibitemShut
  {NoStop}%
\bibitem [{\citenamefont {Ferrari}\ and\ \citenamefont
  {Frings}(2011)}]{Ferrari2011}%
  \BibitemOpen
  \bibfield  {author} {\bibinfo {author} {\bibfnamefont {P.~L.}\ \bibnamefont
  {Ferrari}}\ and\ \bibinfo {author} {\bibfnamefont {R.}~\bibnamefont
  {Frings}},\ }\bibfield  {title} {\bibinfo {title} {Finite time corrections in
  {{KPZ}} growth models},\ }\href {https://doi.org/10.1007/s10955-011-0318-4}
  {\bibfield  {journal} {\bibinfo  {journal} {Journal of Statistical Physics}\
  }\textbf {\bibinfo {volume} {144}},\ \bibinfo {pages} {1123} (\bibinfo {year}
  {2011})}\BibitemShut {NoStop}%
\bibitem [{\citenamefont {B{\"o}ttcher}\ and\ \citenamefont
  {Silbermann}(2012)}]{bottcher2012}%
  \BibitemOpen
  \bibfield  {author} {\bibinfo {author} {\bibfnamefont {A.}~\bibnamefont
  {B{\"o}ttcher}}\ and\ \bibinfo {author} {\bibfnamefont {B.}~\bibnamefont
  {Silbermann}},\ }\href@noop {} {\emph {\bibinfo {title} {Introduction to
  large truncated Toeplitz matrices}}}\ (\bibinfo  {publisher} {Springer
  Science \& Business Media},\ \bibinfo {year} {2012})\BibitemShut {NoStop}%
\bibitem [{\citenamefont {Jepsen}\ \emph {et~al.}(2020)\citenamefont {Jepsen},
  \citenamefont {Amato-Grill}, \citenamefont {Dimitrova}, \citenamefont {Ho},
  \citenamefont {Demler},\ and\ \citenamefont {Ketterle}}]{Jepsen2020}%
  \BibitemOpen
  \bibfield  {author} {\bibinfo {author} {\bibfnamefont {P.~N.}\ \bibnamefont
  {Jepsen}}, \bibinfo {author} {\bibfnamefont {J.}~\bibnamefont {Amato-Grill}},
  \bibinfo {author} {\bibfnamefont {I.}~\bibnamefont {Dimitrova}}, \bibinfo
  {author} {\bibfnamefont {W.~W.}\ \bibnamefont {Ho}}, \bibinfo {author}
  {\bibfnamefont {E.}~\bibnamefont {Demler}},\ and\ \bibinfo {author}
  {\bibfnamefont {W.}~\bibnamefont {Ketterle}},\ }\bibfield  {title} {\bibinfo
  {title} {Spin transport in a tunable heisenberg model realized with ultracold
  atoms},\ }\href {https://doi.org/10.1038/s41586-020-3033-y} {\bibfield
  {journal} {\bibinfo  {journal} {Nature}\ }\textbf {\bibinfo {volume} {588}},\
  \bibinfo {pages} {403} (\bibinfo {year} {2020})}\BibitemShut {NoStop}%
\bibitem [{\citenamefont {Jepsen}\ \emph {et~al.}(2021)\citenamefont {Jepsen},
  \citenamefont {Ho}, \citenamefont {Amato-Grill}, \citenamefont {Dimitrova},
  \citenamefont {Demler},\ and\ \citenamefont {Ketterle}}]{Jepsen2021}%
  \BibitemOpen
  \bibfield  {author} {\bibinfo {author} {\bibfnamefont {P.~N.}\ \bibnamefont
  {Jepsen}}, \bibinfo {author} {\bibfnamefont {W.~W.}\ \bibnamefont {Ho}},
  \bibinfo {author} {\bibfnamefont {J.}~\bibnamefont {Amato-Grill}}, \bibinfo
  {author} {\bibfnamefont {I.}~\bibnamefont {Dimitrova}}, \bibinfo {author}
  {\bibfnamefont {E.}~\bibnamefont {Demler}},\ and\ \bibinfo {author}
  {\bibfnamefont {W.}~\bibnamefont {Ketterle}},\ }\bibfield  {title} {\bibinfo
  {title} {Transverse spin dynamics in the anisotropic heisenberg model
  realized with ultracold atoms},\ }\href
  {https://doi.org/10.1103/PhysRevX.11.041054} {\bibfield  {journal} {\bibinfo
  {journal} {Phys. Rev. X}\ }\textbf {\bibinfo {volume} {11}},\ \bibinfo
  {pages} {041054} (\bibinfo {year} {2021})}\BibitemShut {NoStop}%
\bibitem [{\citenamefont {Jepsen}\ \emph {et~al.}(2022)\citenamefont {Jepsen},
  \citenamefont {Lee}, \citenamefont {Lin}, \citenamefont {Dimitrova},
  \citenamefont {Margalit}, \citenamefont {Ho},\ and\ \citenamefont
  {Ketterle}}]{Jepsen2022}%
  \BibitemOpen
  \bibfield  {author} {\bibinfo {author} {\bibfnamefont {P.~N.}\ \bibnamefont
  {Jepsen}}, \bibinfo {author} {\bibfnamefont {Y.~K.~E.}\ \bibnamefont {Lee}},
  \bibinfo {author} {\bibfnamefont {H.}~\bibnamefont {Lin}}, \bibinfo {author}
  {\bibfnamefont {I.}~\bibnamefont {Dimitrova}}, \bibinfo {author}
  {\bibfnamefont {Y.}~\bibnamefont {Margalit}}, \bibinfo {author}
  {\bibfnamefont {W.~W.}\ \bibnamefont {Ho}},\ and\ \bibinfo {author}
  {\bibfnamefont {W.}~\bibnamefont {Ketterle}},\ }\bibfield  {title} {\bibinfo
  {title} {Long-lived phantom helix states in {{H}}eisenberg quantum magnets},\
  }\href {https://doi.org/10.1038/s41567-022-01651-7} {\bibfield  {journal}
  {\bibinfo  {journal} {Nature Physics}\ }\textbf {\bibinfo {volume} {18}},\
  \bibinfo {pages} {899} (\bibinfo {year} {2022})}\BibitemShut {NoStop}%
\bibitem [{\citenamefont {Altman}\ \emph {et~al.}(2003)\citenamefont {Altman},
  \citenamefont {Hofstetter}, \citenamefont {Demler},\ and\ \citenamefont
  {Lukin}}]{Altman2003}%
  \BibitemOpen
  \bibfield  {author} {\bibinfo {author} {\bibfnamefont {E.}~\bibnamefont
  {Altman}}, \bibinfo {author} {\bibfnamefont {W.}~\bibnamefont {Hofstetter}},
  \bibinfo {author} {\bibfnamefont {E.}~\bibnamefont {Demler}},\ and\ \bibinfo
  {author} {\bibfnamefont {M.~D.}\ \bibnamefont {Lukin}},\ }\bibfield  {title}
  {\bibinfo {title} {Phase diagram of two-component bosons on an optical
  lattice},\ }\href {https://doi.org/10.1088/1367-2630/5/1/113} {\bibfield
  {journal} {\bibinfo  {journal} {New Journal of Physics}\ }\textbf {\bibinfo
  {volume} {5}},\ \bibinfo {pages} {113} (\bibinfo {year} {2003})}\BibitemShut
  {NoStop}%
\bibitem [{\citenamefont {Duan}\ \emph {et~al.}(2003)\citenamefont {Duan},
  \citenamefont {Demler},\ and\ \citenamefont {Lukin}}]{Duan2003}%
  \BibitemOpen
  \bibfield  {author} {\bibinfo {author} {\bibfnamefont {L.-M.}\ \bibnamefont
  {Duan}}, \bibinfo {author} {\bibfnamefont {E.}~\bibnamefont {Demler}},\ and\
  \bibinfo {author} {\bibfnamefont {M.~D.}\ \bibnamefont {Lukin}},\ }\bibfield
  {title} {\bibinfo {title} {Controlling spin exchange interactions of
  ultracold atoms in optical lattices},\ }\href
  {https://doi.org/10.1103/PhysRevLett.91.090402} {\bibfield  {journal}
  {\bibinfo  {journal} {Phys. Rev. Lett.}\ }\textbf {\bibinfo {volume} {91}},\
  \bibinfo {pages} {090402} (\bibinfo {year} {2003})}\BibitemShut {NoStop}%
\bibitem [{\citenamefont {Kuklov}\ and\ \citenamefont
  {Svistunov}(2003)}]{Kuklov2003}%
  \BibitemOpen
  \bibfield  {author} {\bibinfo {author} {\bibfnamefont {A.~B.}\ \bibnamefont
  {Kuklov}}\ and\ \bibinfo {author} {\bibfnamefont {B.~V.}\ \bibnamefont
  {Svistunov}},\ }\bibfield  {title} {\bibinfo {title} {Counterflow
  superfluidity of two-species ultracold atoms in a commensurate optical
  lattice},\ }\href {https://doi.org/10.1103/PhysRevLett.90.100401} {\bibfield
  {journal} {\bibinfo  {journal} {Phys. Rev. Lett.}\ }\textbf {\bibinfo
  {volume} {90}},\ \bibinfo {pages} {100401} (\bibinfo {year}
  {2003})}\BibitemShut {NoStop}%
\bibitem [{\citenamefont {Bogoliubov}\ \emph {et~al.}(1993)\citenamefont
  {Bogoliubov}, \citenamefont {Bullough},\ and\ \citenamefont
  {Pang}}]{Bogoliubov1993}%
  \BibitemOpen
  \bibfield  {author} {\bibinfo {author} {\bibfnamefont {N.~M.}\ \bibnamefont
  {Bogoliubov}}, \bibinfo {author} {\bibfnamefont {R.~K.}\ \bibnamefont
  {Bullough}},\ and\ \bibinfo {author} {\bibfnamefont {G.~D.}\ \bibnamefont
  {Pang}},\ }\bibfield  {title} {\bibinfo {title} {Exact solution of a q-boson
  hopping model},\ }\href {https://doi.org/10.1103/PhysRevB.47.11495}
  {\bibfield  {journal} {\bibinfo  {journal} {Phys. Rev. B}\ }\textbf {\bibinfo
  {volume} {47}},\ \bibinfo {pages} {11495} (\bibinfo {year}
  {1993})}\BibitemShut {NoStop}%
\bibitem [{\citenamefont {Bogoliubov}\ \emph {et~al.}(1998)\citenamefont
  {Bogoliubov}, \citenamefont {Izergin},\ and\ \citenamefont
  {Kitanine}}]{Bogoliubov1998}%
  \BibitemOpen
  \bibfield  {author} {\bibinfo {author} {\bibfnamefont {N.}~\bibnamefont
  {Bogoliubov}}, \bibinfo {author} {\bibfnamefont {A.}~\bibnamefont
  {Izergin}},\ and\ \bibinfo {author} {\bibfnamefont {N.}~\bibnamefont
  {Kitanine}},\ }\bibfield  {title} {\bibinfo {title} {Correlation functions
  for a strongly correlated boson system},\ }\href
  {https://doi.org/https://doi.org/10.1016/S0550-3213(98)00038-8} {\bibfield
  {journal} {\bibinfo  {journal} {Nuclear Physics B}\ }\textbf {\bibinfo
  {volume} {516}},\ \bibinfo {pages} {501} (\bibinfo {year}
  {1998})}\BibitemShut {NoStop}%
\bibitem [{\citenamefont {Pozsgay}\ and\ \citenamefont
  {Eisler}(2016)}]{Pozsgay_2016}%
  \BibitemOpen
  \bibfield  {author} {\bibinfo {author} {\bibfnamefont {B.}~\bibnamefont
  {Pozsgay}}\ and\ \bibinfo {author} {\bibfnamefont {V.}~\bibnamefont
  {Eisler}},\ }\bibfield  {title} {\bibinfo {title} {Real-time dynamics in a
  strongly interacting bosonic hopping model: global quenches and mapping to
  the {{XX}} chain},\ }\href {https://doi.org/10.1088/1742-5468/2016/05/053107}
  {\bibfield  {journal} {\bibinfo  {journal} {Journal of Statistical Mechanics:
  Theory and Experiment}\ }\textbf {\bibinfo {volume} {2016}},\ \bibinfo
  {pages} {053107} (\bibinfo {year} {2016})}\BibitemShut {NoStop}%
\bibitem [{\citenamefont {Castro-Alvaredo}\ \emph {et~al.}(2016)\citenamefont
  {Castro-Alvaredo}, \citenamefont {Doyon},\ and\ \citenamefont
  {Yoshimura}}]{Olalla2016}%
  \BibitemOpen
  \bibfield  {author} {\bibinfo {author} {\bibfnamefont {O.~A.}\ \bibnamefont
  {Castro-Alvaredo}}, \bibinfo {author} {\bibfnamefont {B.}~\bibnamefont
  {Doyon}},\ and\ \bibinfo {author} {\bibfnamefont {T.}~\bibnamefont
  {Yoshimura}},\ }\bibfield  {title} {\bibinfo {title} {Emergent hydrodynamics
  in integrable quantum systems out of equilibrium},\ }\href
  {https://doi.org/10.1103/PhysRevX.6.041065} {\bibfield  {journal} {\bibinfo
  {journal} {Phys. Rev. X}\ }\textbf {\bibinfo {volume} {6}},\ \bibinfo {pages}
  {041065} (\bibinfo {year} {2016})}\BibitemShut {NoStop}%
\bibitem [{\citenamefont {Bertini}\ \emph {et~al.}(2016)\citenamefont
  {Bertini}, \citenamefont {Collura}, \citenamefont {De~Nardis},\ and\
  \citenamefont {Fagotti}}]{Bertini2016}%
  \BibitemOpen
  \bibfield  {author} {\bibinfo {author} {\bibfnamefont {B.}~\bibnamefont
  {Bertini}}, \bibinfo {author} {\bibfnamefont {M.}~\bibnamefont {Collura}},
  \bibinfo {author} {\bibfnamefont {J.}~\bibnamefont {De~Nardis}},\ and\
  \bibinfo {author} {\bibfnamefont {M.}~\bibnamefont {Fagotti}},\ }\bibfield
  {title} {\bibinfo {title} {Transport in out-of-equilibrium {{XXZ}} chains:
  Exact profiles of charges and currents},\ }\href
  {https://doi.org/10.1103/PhysRevLett.117.207201} {\bibfield  {journal}
  {\bibinfo  {journal} {Phys. Rev. Lett.}\ }\textbf {\bibinfo {volume} {117}},\
  \bibinfo {pages} {207201} (\bibinfo {year} {2016})}\BibitemShut {NoStop}%
\bibitem [{\citenamefont {Doyon}\ and\ \citenamefont
  {Yoshimura}(2017)}]{Doyon2017}%
  \BibitemOpen
  \bibfield  {author} {\bibinfo {author} {\bibfnamefont {B.}~\bibnamefont
  {Doyon}}\ and\ \bibinfo {author} {\bibfnamefont {T.}~\bibnamefont
  {Yoshimura}},\ }\bibfield  {title} {\bibinfo {title} {{A note on generalized
  hydrodynamics: inhomogeneous fields and other concepts}},\ }\href
  {https://doi.org/10.21468/SciPostPhys.2.2.014} {\bibfield  {journal}
  {\bibinfo  {journal} {SciPost Phys.}\ }\textbf {\bibinfo {volume} {2}},\
  \bibinfo {pages} {014} (\bibinfo {year} {2017})}\BibitemShut {NoStop}%
\bibitem [{\citenamefont {Doyon}\ \emph {et~al.}(2017)\citenamefont {Doyon},
  \citenamefont {Dubail}, \citenamefont {Konik},\ and\ \citenamefont
  {Yoshimura}}]{Doyon2017_2}%
  \BibitemOpen
  \bibfield  {author} {\bibinfo {author} {\bibfnamefont {B.}~\bibnamefont
  {Doyon}}, \bibinfo {author} {\bibfnamefont {J.}~\bibnamefont {Dubail}},
  \bibinfo {author} {\bibfnamefont {R.}~\bibnamefont {Konik}},\ and\ \bibinfo
  {author} {\bibfnamefont {T.}~\bibnamefont {Yoshimura}},\ }\bibfield  {title}
  {\bibinfo {title} {Large-scale description of interacting one-dimensional
  bose gases: Generalized hydrodynamics supersedes conventional
  hydrodynamics},\ }\href {https://doi.org/10.1103/PhysRevLett.119.195301}
  {\bibfield  {journal} {\bibinfo  {journal} {Phys. Rev. Lett.}\ }\textbf
  {\bibinfo {volume} {119}},\ \bibinfo {pages} {195301} (\bibinfo {year}
  {2017})}\BibitemShut {NoStop}%
\bibitem [{\citenamefont {Bulchandani}\ \emph {et~al.}(2017)\citenamefont
  {Bulchandani}, \citenamefont {Vasseur}, \citenamefont {Karrasch},\ and\
  \citenamefont {Moore}}]{Bulchandani2017}%
  \BibitemOpen
  \bibfield  {author} {\bibinfo {author} {\bibfnamefont {V.~B.}\ \bibnamefont
  {Bulchandani}}, \bibinfo {author} {\bibfnamefont {R.}~\bibnamefont
  {Vasseur}}, \bibinfo {author} {\bibfnamefont {C.}~\bibnamefont {Karrasch}},\
  and\ \bibinfo {author} {\bibfnamefont {J.~E.}\ \bibnamefont {Moore}},\
  }\bibfield  {title} {\bibinfo {title} {Solvable hydrodynamics of quantum
  integrable systems},\ }\href {https://doi.org/10.1103/PhysRevLett.119.220604}
  {\bibfield  {journal} {\bibinfo  {journal} {Phys. Rev. Lett.}\ }\textbf
  {\bibinfo {volume} {119}},\ \bibinfo {pages} {220604} (\bibinfo {year}
  {2017})}\BibitemShut {NoStop}%
\bibitem [{\citenamefont {Bulchandani}\ \emph {et~al.}(2018)\citenamefont
  {Bulchandani}, \citenamefont {Vasseur}, \citenamefont {Karrasch},\ and\
  \citenamefont {Moore}}]{Bulchandani18}%
  \BibitemOpen
  \bibfield  {author} {\bibinfo {author} {\bibfnamefont {V.~B.}\ \bibnamefont
  {Bulchandani}}, \bibinfo {author} {\bibfnamefont {R.}~\bibnamefont
  {Vasseur}}, \bibinfo {author} {\bibfnamefont {C.}~\bibnamefont {Karrasch}},\
  and\ \bibinfo {author} {\bibfnamefont {J.~E.}\ \bibnamefont {Moore}},\
  }\bibfield  {title} {\bibinfo {title} {Bethe-boltzmann hydrodynamics and spin
  transport in the {{XXZ}} chain},\ }\href
  {https://doi.org/10.1103/PhysRevB.97.045407} {\bibfield  {journal} {\bibinfo
  {journal} {Phys. Rev. B}\ }\textbf {\bibinfo {volume} {97}},\ \bibinfo
  {pages} {045407} (\bibinfo {year} {2018})}\BibitemShut {NoStop}%
\bibitem [{\citenamefont {Doyon}\ \emph {et~al.}(2018)\citenamefont {Doyon},
  \citenamefont {Yoshimura},\ and\ \citenamefont {Caux}}]{Doyon2018}%
  \BibitemOpen
  \bibfield  {author} {\bibinfo {author} {\bibfnamefont {B.}~\bibnamefont
  {Doyon}}, \bibinfo {author} {\bibfnamefont {T.}~\bibnamefont {Yoshimura}},\
  and\ \bibinfo {author} {\bibfnamefont {J.-S.}\ \bibnamefont {Caux}},\
  }\bibfield  {title} {\bibinfo {title} {Soliton gases and generalized
  hydrodynamics},\ }\href {https://doi.org/10.1103/PhysRevLett.120.045301}
  {\bibfield  {journal} {\bibinfo  {journal} {Phys. Rev. Lett.}\ }\textbf
  {\bibinfo {volume} {120}},\ \bibinfo {pages} {045301} (\bibinfo {year}
  {2018})}\BibitemShut {NoStop}%
\bibitem [{\citenamefont {De~Nardis}\ \emph {et~al.}(2018)\citenamefont
  {De~Nardis}, \citenamefont {Bernard},\ and\ \citenamefont
  {Doyon}}]{Jacopo2018}%
  \BibitemOpen
  \bibfield  {author} {\bibinfo {author} {\bibfnamefont {J.}~\bibnamefont
  {De~Nardis}}, \bibinfo {author} {\bibfnamefont {D.}~\bibnamefont {Bernard}},\
  and\ \bibinfo {author} {\bibfnamefont {B.}~\bibnamefont {Doyon}},\ }\bibfield
   {title} {\bibinfo {title} {Hydrodynamic diffusion in integrable systems},\
  }\href {https://doi.org/10.1103/PhysRevLett.121.160603} {\bibfield  {journal}
  {\bibinfo  {journal} {Phys. Rev. Lett.}\ }\textbf {\bibinfo {volume} {121}},\
  \bibinfo {pages} {160603} (\bibinfo {year} {2018})}\BibitemShut {NoStop}%
\bibitem [{\citenamefont {Gopalakrishnan}\ and\ \citenamefont
  {Vasseur}(2019)}]{Sarang2019}%
  \BibitemOpen
  \bibfield  {author} {\bibinfo {author} {\bibfnamefont {S.}~\bibnamefont
  {Gopalakrishnan}}\ and\ \bibinfo {author} {\bibfnamefont {R.}~\bibnamefont
  {Vasseur}},\ }\bibfield  {title} {\bibinfo {title} {Kinetic theory of spin
  diffusion and superdiffusion in {{XXZ}} spin chains},\ }\href
  {https://doi.org/10.1103/PhysRevLett.122.127202} {\bibfield  {journal}
  {\bibinfo  {journal} {Phys. Rev. Lett.}\ }\textbf {\bibinfo {volume} {122}},\
  \bibinfo {pages} {127202} (\bibinfo {year} {2019})}\BibitemShut {NoStop}%
\bibitem [{\citenamefont {Schemmer}\ \emph {et~al.}(2019)\citenamefont
  {Schemmer}, \citenamefont {Bouchoule}, \citenamefont {Doyon},\ and\
  \citenamefont {Dubail}}]{Schemmer2019}%
  \BibitemOpen
  \bibfield  {author} {\bibinfo {author} {\bibfnamefont {M.}~\bibnamefont
  {Schemmer}}, \bibinfo {author} {\bibfnamefont {I.}~\bibnamefont {Bouchoule}},
  \bibinfo {author} {\bibfnamefont {B.}~\bibnamefont {Doyon}},\ and\ \bibinfo
  {author} {\bibfnamefont {J.}~\bibnamefont {Dubail}},\ }\bibfield  {title}
  {\bibinfo {title} {Generalized hydrodynamics on an atom chip},\ }\href
  {https://doi.org/10.1103/PhysRevLett.122.090601} {\bibfield  {journal}
  {\bibinfo  {journal} {Phys. Rev. Lett.}\ }\textbf {\bibinfo {volume} {122}},\
  \bibinfo {pages} {090601} (\bibinfo {year} {2019})}\BibitemShut {NoStop}%
\bibitem [{\citenamefont {Doyon}(2020)}]{Doyon2020_rev}%
  \BibitemOpen
  \bibfield  {author} {\bibinfo {author} {\bibfnamefont {B.}~\bibnamefont
  {Doyon}},\ }\bibfield  {title} {\bibinfo {title} {{Lecture notes on
  Generalised Hydrodynamics}},\ }\href
  {https://doi.org/10.21468/SciPostPhysLectNotes.18} {\bibfield  {journal}
  {\bibinfo  {journal} {SciPost Phys. Lect. Notes}\ ,\ \bibinfo {pages} {18}}
  (\bibinfo {year} {2020})}\BibitemShut {NoStop}%
\bibitem [{\citenamefont {Alba}\ \emph {et~al.}(2021)\citenamefont {Alba},
  \citenamefont {Bertini}, \citenamefont {Fagotti}, \citenamefont {Piroli},\
  and\ \citenamefont {Ruggiero}}]{Alba2021}%
  \BibitemOpen
  \bibfield  {author} {\bibinfo {author} {\bibfnamefont {V.}~\bibnamefont
  {Alba}}, \bibinfo {author} {\bibfnamefont {B.}~\bibnamefont {Bertini}},
  \bibinfo {author} {\bibfnamefont {M.}~\bibnamefont {Fagotti}}, \bibinfo
  {author} {\bibfnamefont {L.}~\bibnamefont {Piroli}},\ and\ \bibinfo {author}
  {\bibfnamefont {P.}~\bibnamefont {Ruggiero}},\ }\bibfield  {title} {\bibinfo
  {title} {Generalized-hydrodynamic approach to inhomogeneous quenches:
  correlations, entanglement and quantum effects},\ }\href
  {https://doi.org/10.1088/1742-5468/ac257d} {\bibfield  {journal} {\bibinfo
  {journal} {Journal of Statistical Mechanics: Theory and Experiment}\ }\textbf
  {\bibinfo {volume} {2021}},\ \bibinfo {pages} {114004} (\bibinfo {year}
  {2021})}\BibitemShut {NoStop}%
\bibitem [{\citenamefont {Malvania}\ \emph {et~al.}(2021)\citenamefont
  {Malvania}, \citenamefont {Zhang}, \citenamefont {Le}, \citenamefont
  {Dubail}, \citenamefont {Rigol},\ and\ \citenamefont {Weiss}}]{Malvania2021}%
  \BibitemOpen
  \bibfield  {author} {\bibinfo {author} {\bibfnamefont {N.}~\bibnamefont
  {Malvania}}, \bibinfo {author} {\bibfnamefont {Y.}~\bibnamefont {Zhang}},
  \bibinfo {author} {\bibfnamefont {Y.}~\bibnamefont {Le}}, \bibinfo {author}
  {\bibfnamefont {J.}~\bibnamefont {Dubail}}, \bibinfo {author} {\bibfnamefont
  {M.}~\bibnamefont {Rigol}},\ and\ \bibinfo {author} {\bibfnamefont {D.~S.}\
  \bibnamefont {Weiss}},\ }\bibfield  {title} {\bibinfo {title} {Generalized
  hydrodynamics in strongly interacting 1d bose gases},\ }\href
  {https://doi.org/10.1126/science.abf0147} {\bibfield  {journal} {\bibinfo
  {journal} {Science}\ }\textbf {\bibinfo {volume} {373}},\ \bibinfo {pages}
  {1129} (\bibinfo {year} {2021})}\BibitemShut {NoStop}%
\bibitem [{\citenamefont {Bouchoule}\ and\ \citenamefont
  {Dubail}(2022)}]{Bouchoule2022}%
  \BibitemOpen
  \bibfield  {author} {\bibinfo {author} {\bibfnamefont {I.}~\bibnamefont
  {Bouchoule}}\ and\ \bibinfo {author} {\bibfnamefont {J.}~\bibnamefont
  {Dubail}},\ }\bibfield  {title} {\bibinfo {title} {Generalized hydrodynamics
  in the one-dimensional bose gas: theory and experiments},\ }\href
  {https://doi.org/10.1088/1742-5468/ac3659} {\bibfield  {journal} {\bibinfo
  {journal} {Journal of Statistical Mechanics: Theory and Experiment}\ }\textbf
  {\bibinfo {volume} {2022}},\ \bibinfo {pages} {014003} (\bibinfo {year}
  {2022})}\BibitemShut {NoStop}%
\bibitem [{\citenamefont {Essler}(2023)}]{Essler2023}%
  \BibitemOpen
  \bibfield  {author} {\bibinfo {author} {\bibfnamefont {F.~H.}\ \bibnamefont
  {Essler}},\ }\bibfield  {title} {\bibinfo {title} {A short introduction to
  generalized hydrodynamics},\ }\href
  {https://doi.org/https://doi.org/10.1016/j.physa.2022.127572} {\bibfield
  {journal} {\bibinfo  {journal} {Physica A: Statistical Mechanics and its
  Applications}\ }\textbf {\bibinfo {volume} {631}},\ \bibinfo {pages} {127572}
  (\bibinfo {year} {2023})},\ \bibinfo {note} {lecture Notes of the 15th
  International Summer School of Fundamental Problems in Statistical
  Physics}\BibitemShut {NoStop}%
\bibitem [{\citenamefont {Doyon}\ \emph
  {et~al.}(2023{\natexlab{a}})\citenamefont {Doyon}, \citenamefont {Perfetto},
  \citenamefont {Sasamoto},\ and\ \citenamefont {Yoshimura}}]{BMFT1}%
  \BibitemOpen
  \bibfield  {author} {\bibinfo {author} {\bibfnamefont {B.}~\bibnamefont
  {Doyon}}, \bibinfo {author} {\bibfnamefont {G.}~\bibnamefont {Perfetto}},
  \bibinfo {author} {\bibfnamefont {T.}~\bibnamefont {Sasamoto}},\ and\
  \bibinfo {author} {\bibfnamefont {T.}~\bibnamefont {Yoshimura}},\ }\bibfield
  {title} {\bibinfo {title} {Emergence of hydrodynamic spatial long-range
  correlations in nonequilibrium many-body systems},\ }\href
  {https://doi.org/10.1103/PhysRevLett.131.027101} {\bibfield  {journal}
  {\bibinfo  {journal} {Phys. Rev. Lett.}\ }\textbf {\bibinfo {volume} {131}},\
  \bibinfo {pages} {027101} (\bibinfo {year} {2023}{\natexlab{a}})}\BibitemShut
  {NoStop}%
\bibitem [{\citenamefont {Doyon}\ \emph
  {et~al.}(2023{\natexlab{b}})\citenamefont {Doyon}, \citenamefont {Perfetto},
  \citenamefont {Sasamoto},\ and\ \citenamefont {Yoshimura}}]{BMFT2}%
  \BibitemOpen
  \bibfield  {author} {\bibinfo {author} {\bibfnamefont {B.}~\bibnamefont
  {Doyon}}, \bibinfo {author} {\bibfnamefont {G.}~\bibnamefont {Perfetto}},
  \bibinfo {author} {\bibfnamefont {T.}~\bibnamefont {Sasamoto}},\ and\
  \bibinfo {author} {\bibfnamefont {T.}~\bibnamefont {Yoshimura}},\ }\bibfield
  {title} {\bibinfo {title} {{Ballistic macroscopic fluctuation theory}},\
  }\href {https://doi.org/10.21468/SciPostPhys.15.4.136} {\bibfield  {journal}
  {\bibinfo  {journal} {SciPost Phys.}\ }\textbf {\bibinfo {volume} {15}},\
  \bibinfo {pages} {136} (\bibinfo {year} {2023}{\natexlab{b}})}\BibitemShut
  {NoStop}%
\bibitem [{\citenamefont {Bornemann}(2010)}]{Bornemann}%
  \BibitemOpen
  \bibfield  {author} {\bibinfo {author} {\bibfnamefont {F.}~\bibnamefont
  {Bornemann}},\ }\bibfield  {title} {\bibinfo {title} {On the numerical
  evaluation of distributions in random matrix theory: A review},\ }\href@noop
  {} {\bibfield  {journal} {\bibinfo  {journal} {Markov Proc. Relat. Fields}\
  }\textbf {\bibinfo {volume} {16}},\ \bibinfo {pages} {803} (\bibinfo {year}
  {2010})}\BibitemShut {NoStop}%
\end{thebibliography}%

\clearpage
\widetext

\setcounter{equation}{0}
\setcounter{figure}{0}
\setcounter{section}{0}
\setcounter{table}{0}
\renewcommand{\theequation}{S-\arabic{equation}}
\renewcommand{\thefigure}{S-\arabic{figure}}
\renewcommand{\thetable}{S-\arabic{table}}

\section*{Supplemental Material for ``Quantum Transport in Interacting Spin Chains: Exact Derivation of the GUE Tracy-Widom Distribution''}

\centerline{Kazuya Fujimoto and Tomohiro Sasamoto}
\vspace{3mm}

\centerline{Department of Physics, Institute of Science Tokyo, 2-12-1 Ookayama, Meguro-ku, Tokyo 152-8551, Japan}
\vspace{5mm}

\par\vskip3mm \hrule\vskip.5mm\hrule \vskip.30cm
This Supplemental Material describes the following:
\begin{itemize}
\item[  ]{ (I) Proof of Eq.~(\ref{Esol1}) in the main text, } 
\item[  ]{ (II) Determinantal formula for Eq.~(\ref{Esol1}) in the main text, } 
\item[  ]{ (III) Proof of Eq.~\eqref{MBW4} in the main text, }
\item[  ]{ (IV) Asymptotic analysis to the GUE Tracy-Widom ditribution with fast convergence, } 
\item[  ]{ (V) Numerical truncation in the TEBD method, } 
\item[  ]{ (VI) Limiting probability distribution function for $\Delta = 0$, } 
\item[  ]{ (VII) Conjecture for the XXZ model by Saenz, Tracy, and Widom. }
\end{itemize}
\par\vskip1mm \hrule\vskip.5mm\hrule

\vspace{5mm}

\section{Proof of Eq.~(\ref{Esol1}) in the main text }
We prove that Eq.~\eqref{Esol1} in the main text satisfies the Schrödinger equation with $\hat{H}_{\rm fXXZ}$ and the initial state $\ket{\phi(0)} = \prod_{j=1}^N \hat{R}_{y_j} \ket{0}$ with the constraint $y_j + 2 \leq y_{j+1}$. Here, $y_j$ indicates the initial up-spin site label.
 Note that the many-body wavefunction $\Phi(x_1, ..., x_N,t)$ under this setting becomes zero if the condition $x_j + 2 \leq x_{j+1}$ does not hold, because all the matrix elements of $\hat{H}_{\rm fXXZ}$ for breaking this condition is zero. Hence, in the following, we consider $\Phi(x_1, ..., x_N,t)$ under the restricted configuration $x_j + 2 \leq x_{j+1}$. 

We first derive equations that $\Phi(x_1, ..., x_N,t)$ satisfies. From the Schrödinger equation for the folded XXZ model, we obtain the equation of motion as 
\begin{align} 
{\rm i}\dfrac{\partial}{\partial t} \Phi(x_1,...,x_N,t) = \dfrac{1}{2} \sum_{j=1}^{N} \Bigl( \Phi(x_1, ... , \underbrace{x_j+1}_{j}, ... ,x_N,t) + \Phi(x_1, ... , \underbrace{x_j-1}_{j}, ... ,x_N,t) \Bigl) 
\label{S_EoM1}
\end{align}
with the boundary condition defined by 
\begin{align} 
\Phi(x_1, ... , \underbrace{x_j+1}_{j}, \underbrace{x_j+2}_{j+1}, ... ,x_N,t) + \Phi(x_1, ... , \underbrace{x_j}_{j}, \underbrace{x_j+1}_{j+1},  ... ,x_N,t) = 0.
\label{S_EoM2}
\end{align}
Here, the boundary condition comes from the fact that breaking the condition $x_j + 2 \leq x_{j+1}$ leads to $\Phi(x_1, ..., x_N,t)=0$. 
The initial state considered here becomes
\begin{align} 
\Phi(x_1,...,x_N,0) = \prod_{j=1}^{N} \delta_{x_j, y_j}, 
\label{S_Initial1}
\end{align}
Under this setup, the exact many-body wavefunction is given by 
\begin{align} 
\Phi(x_1,...,x_N,t) &= \int_{C_r} d \bm{\xi} \sum_{\sigma \in \mathbb{S}_N }  A_{\sigma}( {\bm \xi} )  \prod_{j=1}^N\xi_{\sigma_j}^{x_{j}-y_{\sigma_j} -1} e^{- {\rm i}E_{\xi_j} t}.
\label{S_Esol3}
\end{align}
In the following subsections, we prove that Eq.~\eqref{S_Esol3} solves Eqs.~\eqref{S_EoM1}, \eqref{S_EoM2}, and \eqref{S_Initial1}.

\subsection{Proof for Eq.~(\ref{S_Esol3}) satisfying the equation~(\ref{S_EoM1}) of motion}
We substitute Eq.~\eqref{S_Esol3} into Eq.~\eqref{S_EoM1}, obtaining
\begin{align} 
{\rm LHS~of~Eq.~\eqref{S_EoM1}} &= \int_{C_r} d \bm{\xi} \sum_{\sigma \in \mathbb{S}_N }  A_{\sigma}( {\bm \xi} )  \left( \sum_{k=1}^N E_{\xi_k} \right) \left( \prod_{j=1}^N\xi_{\sigma_j}^{x_{j}-y_{\sigma_j} -1} e^{- {\rm i}E_{\xi_j} t} \right), 
\label{S_R} \\
{\rm RHS~of~Eq.~\eqref{S_EoM1}} &= \frac{1}{2} \sum_{k=1}^{N}  \int_{C_r} d \bm{\xi} \sum_{\sigma \in \mathbb{S}_N }  A_{\sigma}( {\bm \xi} )  \left( \xi_{\sigma_k} + \xi_{\sigma_k}^{-1} \right)\left( \prod_{j=1}^N\xi_{\sigma_j}^{x_{j}-y_{\sigma_j} -1} e^{- {\rm i}E_{\xi_j} t} \right) \label{S_L1}\\
&=  \int_{C_r} d \bm{\xi} \sum_{\sigma \in \mathbb{S}_N }  A_{\sigma}( {\bm \xi} )  \left\{ \frac{1}{2} \sum_{k=1}^{N} \left( \xi_{\sigma_k} + \xi_{\sigma_k}^{-1} \right) \right\} \left( \prod_{j=1}^N\xi_{\sigma_j}^{x_{j}-y_{\sigma_j} -1} e^{- {\rm i}E_{\xi_j} t} \right) \label{S_L2} \\
&=  \int_{C_r} d \bm{\xi} \sum_{\sigma \in \mathbb{S}_N }  A_{\sigma}( {\bm \xi} )  \left\{ \frac{1}{2} \sum_{k=1}^{N} \left( \xi_{k} + \xi_{k}^{-1} \right) \right\} \left( \prod_{j=1}^N\xi_{\sigma_j}^{x_{j}-y_{\sigma_j} -1} e^{- {\rm i}E_{\xi_j} t} \right) \label{S_L3}\\
&=  \int_{C_r} d \bm{\xi} \sum_{\sigma \in \mathbb{S}_N }  A_{\sigma}( {\bm \xi} )  \left( \sum_{k=1}^{N} E_{\xi_k} \right) \left( \prod_{j=1}^N\xi_{\sigma_j}^{x_{j}-y_{\sigma_j} -1} e^{- {\rm i}E_{\xi_j} t} \right) \label{S_L4}.
\end{align}
Here, we use the formula $\sum_{k=1}^N f(\xi_{\sigma_k}) = \sum_{k=1}^N f(\xi_{k}) $ with a function $f(\bullet)$ to get Eq.~\eqref{S_L3}. 
Comparing Eq.~\eqref{S_R} with Eq.~\eqref{S_L4}, we complete the proof.

\subsection{Proof for Eq.~(\ref{S_Esol3}) satisfying the boundary condition of Eq.~(\ref{S_EoM2})}
We substitute Eq.~\eqref{S_Esol3} into Eq.~\eqref{S_EoM2}, obtaining
\begin{align} 
&{\rm LHS~of~Eq.~\eqref{S_EoM2}} \label{S_LC1} \\
&=  \int_{C_r} d \bm{\xi} \sum_{\sigma \in \mathbb{S}_N }  A_{\sigma}( {\bm \xi} )   \left( \prod_{l \in \{1,.., N\}: l \neq j, j+1} \xi_{\sigma_l}^{x_{l}-y_{\sigma_l} -1} e^{- {\rm i}E_{\xi_l} t} \right) e^{- {\rm i}(E_{\xi_j} + E_{\xi_{j+1}}) t}
\left(  \xi_{\sigma_j}^{x_{j}-y_{\sigma_j} } \xi_{\sigma_{j+1}}^{x_{j}-y_{\sigma_{j+1}} + 1} +  \xi_{\sigma_j}^{x_{j}-y_{\sigma_j} -1} \xi_{\sigma_{j+1}}^{x_{j}-y_{\sigma_{j+1}} }  \right) \label{S_LC2}\\
&= \int_{C_r} d \bm{\xi} \sum_{\sigma \in \mathbb{S}_N }  A_{\sigma}( {\bm \xi} )   \left( \prod_{l \in \{1,.., N\}: l \neq j, j+1} \xi_{\sigma_l}^{x_{l}-y_{\sigma_l} -1} e^{- {\rm i}E_{\xi_l} t} \right)   e^{- {\rm i}(E_{\xi_j} + E_{\xi_{j+1}}) t}
\xi_{\sigma_j}^{x_{j}-y_{\sigma_j} } \xi_{\sigma_{j+1}}^{x_{j}-y_{\sigma_{j+1}}} \left(   \xi_{\sigma_{j+1}} +  \xi_{\sigma_j}^{ -1} \right) \label{S_LC3}\\
&= \int_{C_r} d \bm{\xi} \sum_{\sigma \in \mathbb{S}_N } G_{\sigma}(\bm{\xi},j). \label{S_LC4}
\end{align} 
Here, we define the function $G_{\sigma}(\bm{\xi},j)$ as
\begin{align} 
G_{\sigma}(\bm{\xi},j) \coloneqq A_{\sigma}( {\bm \xi} )   \left( \prod_{l \in \{1,.., N\}: l \neq j, j+1} \xi_{\sigma_l}^{x_{l}-y_{\sigma_l} -1} e^{- {\rm i}E_{\xi_l} t} \right)   e^{- {\rm i}(E_{\xi_j} + E_{\xi_{j+1}}) t}
\xi_{\sigma_j}^{x_{j}-y_{\sigma_j} } \xi_{\sigma_{j+1}}^{x_{j}-y_{\sigma_{j+1}}} \left(   \xi_{\sigma_{j+1}} +  \xi_{\sigma_j}^{ -1} \right). \label{S_LC5}
\end{align} 
Next, for a fixed permutation $\sigma$, we consider a new permutation $\mu$ defined by
\begin{align} 
\mu_l \coloneqq
\left\{
\begin{array}{ll}
\sigma_l & (l \neq j, j+1) \\
\sigma_j & (l = j + 1) \\
\sigma_{j+1} & (l = j)
\end{array}
\right.
\label{S_LC6}
\end{align} 
Then, by definition, we can show 
\begin{align} 
 A_{\mu} (\bm \xi) =  A_{\sigma} (\bm \xi) S(\xi_{\sigma_{j+1}}, \xi_{\sigma_{j}}). \label{S_LC7}
\end{align} 
Using Eqs.~\eqref{S_LC5} and \eqref{S_LC7}, we can prove 
\begin{align}
 &G_{\sigma}(\bm{\xi},j)  + G_{\mu}(\bm{\xi},j) \\
 &= A_{\sigma}( {\bm \xi} )   \left( \prod_{l \in \{1,.., N\}: l \neq j, j+1} \xi_{\sigma_l}^{x_{l}-y_{\sigma_l} -1} e^{- {\rm i}E_{\xi_l} t} \right)   e^{- {\rm i}(E_{\xi_j} + E_{\xi_{j+1}}) t}
\xi_{\sigma_j}^{x_{j}-y_{\sigma_j} } \xi_{\sigma_{j+1}}^{x_{j}-y_{\sigma_{j+1}}} 
 \left\{   \xi_{\sigma_{j+1}} +  \xi_{\sigma_j}^{ -1} -  \dfrac{\xi_{\sigma_{j+1}} }{ \xi_{\sigma_j} } \left( \xi_{\sigma_{j+1}}^{-1} + \xi_{\sigma_j} \right)  \right\} \\
 &=0.
\label{S_LC8}
\end{align} 
This completes the proof. 

\subsection{Proof for Eq.~(\ref{S_Esol3}) satisfying the initial state of Eq.~(\ref{S_Initial1}) }
We shall prove that Eq.~\eqref{S_Esol3} with $t=0$ leads to Eq.~\eqref{S_Initial1}. First, we derive the explicit expression of $A_{\sigma}(\bm{\xi})$ in Eq.~\eqref{S_Esol3} by introducing three sets $U \coloneqq \{1,..., N \}$, $P \coloneqq \{ (\alpha, \beta) | \alpha < \beta \wedge \alpha, \beta \in U \}$, $P_{\sigma}^{<} \coloneqq \{ (\alpha, \beta) | \sigma_{\alpha} < \sigma_{\beta} \wedge (\alpha, \beta) \in P \}$, and $P_{\sigma}^{>} \coloneqq \{ (\alpha, \beta) | \sigma_{\alpha} > \sigma_{\beta} \wedge (\alpha, \beta) \in P \}$. Then, we can derive
\begin{align} 
A_{\sigma}(\bm{\xi}) 
&= {\rm sgn}(\sigma) \prod_{(\alpha, \beta) \in P_{\sigma}^> }  \xi_{\sigma_{\alpha} }  \xi_{\sigma_\beta}^{-1}  \\
&= {\rm sgn}(\sigma) \left( \prod_{(\alpha, \beta) \in P_{\sigma}^> }  \xi_{\sigma_{\alpha} }  \xi_{\sigma_\beta}^{-1} \right) \left( \prod_{(\alpha, \beta) \in P_{\sigma}^< }  \xi_{\sigma_{\alpha} }  \xi_{\sigma_\alpha}^{-1} \right) \\
&= {\rm sgn}(\sigma) \left( \prod_{(\alpha, \beta) \in P }  \xi_{\sigma_{\alpha} } \right) \left( \prod_{(\alpha, \beta) \in P }  \xi_{\alpha }^{-1} \right) \\
&= {\rm sgn}(\sigma) \prod_{j=1}^N \xi_{\sigma_j}^{ \sigma_j  - j}.
\label{S_Exp_A}
\end{align}
We substitute Eq.~\eqref{S_Exp_A} into \eqref{S_Esol3}, getting
\begin{align} 
\Phi(x_1,...,x_N,0) 
&= \sum_{\sigma \in \mathbb{S}_N} {\rm sgn}(\sigma) \int d\bm{\xi} \prod_{j=1}^N \xi_{\sigma_j}^{x_j-j-y_{\sigma_j} + \sigma_j - 1} \\
&= \sum_{\sigma \in \mathbb{S}_N} {\rm sgn}(\sigma) \prod_{j=1}^N \delta_{x_j, y_{\sigma_j} + j - \sigma_j} \label{S_Initial2_sup} \\
&= \prod_{j=1}^N \delta_{x_j, y_{j}}.  \label{S_Initial2}
\end{align}
Here, the equality of Eq.~\eqref{S_Initial2} can be proved by noting that the inequalities $x_j + 2 \leq x_{j+1}$ and $y_j + 2 \leq y_{j+1}$ are violated when there are inversions in $\sigma$. 
In order to show this fact, let us consider integers $m$ and $n$ such that two conditions $m < n$ and $\sigma_m > \sigma_n$ hold. 
In our setup, we have inequalities $x_{n} - x_{m} \geq 2(n-m)$ and $y_{\sigma_m} - y_{\sigma_n} \geq  2(\sigma_{m} - \sigma_{n})$, which can be derived by the above inequalities $x_j + 2 \leq x_{j+1}$ and $y_j + 2 \leq y_{j+1}$. Using a relation $x_j = y_{\sigma_j} + j - \sigma_j$ coming from Eq.~\eqref{S_Initial2_sup}, we can derive
\begin{align} 
x_{n} - x_{m} - 2(n-m) &= y_{\sigma_n} - y_{\sigma_m} - (n - m) + \sigma_{m} - \sigma_{n} \\
&\leq   \underbrace{(m - n)}_{<0} +  \underbrace{(\sigma_{n} - \sigma_{m})}_{<0}. 
\end{align}
Thus, we obtain $x_{n} - x_{m} - 2(n-m) < 0$, proving Eq.~\eqref{S_Initial2} because this inequality is contradicted with $x_j + 2 \leq x_{j+1}$.

\section{Determinantal formula for Eq.~(\ref{Esol1}) in the main text }
We shall prove that Eq.~\eqref{Esol1} in the main text can be expressed as a determinantal form.
Using Eqs.~\eqref{S_Esol3} and \eqref{S_Exp_A}, we can obtain
\begin{align} 
\Phi(x_1, ... ,x_N,t) 
&=  \sum_{\sigma \in \mathbb{S}_N} \int_{C_r} d\bm{\xi} \prod_{j=1}^N \xi_{\sigma_j}^{x_j-j-y_{\sigma_j} + \sigma_j - 1} e^{- {\rm i} E_{\xi_j} t} \\
&= \int_{C_r} d\bm{\xi} ~\det \left(  \xi_{k}^{x_j-j-y_{k} + k - 1} e^{- {\rm i} E_{\xi_j} t}  \right)_{j,k \in \{1,...,N\}} \\
&= \int_{C_r} d\bm{\xi} ~\det \left(  \xi_{k}^{ \tilde{x}_j - \tilde{y}_{k} - 1} e^{- {\rm i} E_{\xi_j} t}  \right)_{j,k \in \{1,...,N\} } \label{det_form1}
\end{align}
with the variables $ \tilde{x}_j \coloneqq x_j - j $ and $ \tilde{y}_k \coloneqq y_k - k $.
Here, we use the definition of a determinant to derive the second line. When the initial state is the alternating domain-wall state ($y_j = 2j$), the exact many-body wavefunction becomes 
\begin{align} 
\Phi(x_1, ... ,x_N,t) = \int d\bm{\xi} ~\det \left(  \xi_{k}^{ x_j - j  - k - 1} e^{- {\rm i} E_{\xi_j} t}  \right)_{j,k \in \{1,...,N\} }.
\end{align}

We can explicitly show that the many-body wavefunction $\Phi(x_1, ... ,x_N,t) $ of Eq.~\eqref{det_form1} vanishes when there exists a site label $j$ such that the relation $x_j + 1 = x_{j+1}$ holds.
From Eq.~\eqref{det_form1}, we obtain 
\begin{align} 
\Phi(x_1, ..., \underbrace{x_j}_{j}, \underbrace{ x_j + 1 }_{j+1}, ... ,x_N,t) &= \int_{C_r} d\bm{\xi}~ \left( \prod_{k=1}^N e^{- {\rm i} E_{\xi_k} t}\right) ~\det \left(  \xi_{l}^{ \tilde{x}_k - \tilde{y}_{l} - 1}  \right)_{k,l \in \{1,...,N\} } \\
&= \int_{C_r} d\bm{\xi} ~\left( \prod_{k=1}^N e^{- {\rm i} E_{\xi_k} t}\right) ~\det 
\begin{pmatrix}
\xi_{1}^{ \tilde{x}_1 - \tilde{y}_{1} - 1}  & \cdots & \xi_{N}^{ \tilde{x}_1 - \tilde{y}_{N} - 1}   \\
\\
\vdots &   & \vdots \\ 
\\
\xi_{1}^{ \tilde{x}_j - \tilde{y}_{1} - 1} & \cdots  & \xi_{N}^{ \tilde{x}_j - \tilde{y}_{N} - 1}   \\ 
\\
\xi_{1}^{ \tilde{x}_{j} - \tilde{y}_{1} - 1}  & \cdots  & \xi_{N}^{ \tilde{x}_{j} - \tilde{y}_{N} - 1}   \\
\\
\vdots &   & \vdots \\ 
\\
\xi_{1}^{ \tilde{x}_N - \tilde{y}_{1} - 1}  & \cdots & \xi_{N}^{ \tilde{x}_N - \tilde{y}_{N} - 1}
\end{pmatrix} 
\label{S_mat}
\\
&= 0
\end{align}
To get the last line, we use the fact that the $j$ and $(j+1)$th rows of the matrix of Eq.~\eqref{S_mat} are identical.


\section{Proof of Eq.~(\ref{MBW4}) in the main text }
We derive an exact many-body wavefunction $\Phi_{\rm XX}(x_1,...,x_N,t)$ in the XX model, which is given in Eq.~(\ref{MBW4}) in the main text.
The Hamiltonian for the XX model is given as
\begin{align} 
\hat{H}_{\rm XX} = \sum_{x \in \mathbb{Z}} \left( \hat{X}_{x} \hat{X}_{x+1} + \hat{Y}_{x} \hat{Y}_{x+1}  \right).
\label{S_H_XX}
\end{align}
The initial state is 
\begin{align} 
\ket{\phi_{\rm XX}(0)} = \prod_{x=1}^{N} \hat{R}_{y_j} \ket{0}
\label{S_initialXX1}
\end{align}
with $y_j + 1 \leq y_{j+1} $. Under this setup, the many-body wavefunction $\Phi_{\rm XX}(x_1,..., x_N,t)$ obeys
\begin{align} 
{\rm i}\dfrac{\partial}{\partial t} \Phi_{\rm XX}(x_1,...,x_N,t) = \dfrac{1}{2} \sum_{j=1}^{N} \Bigl( \Phi_{\rm XX}(x_1, ... , \underbrace{x_j+1}_{j}, ... ,x_N,t) + \Phi_{\rm XX}(x_1, ... , \underbrace{x_j-1}_{j}, ... ,x_N,t) \Bigl) 
\label{S_EoMXX1}
\end{align}
with the boundary condition defined by 
\begin{align} 
\Phi_{\rm XX}(x_1, ... , \underbrace{x_j+1}_{j}, \underbrace{x_j+1}_{j+1}, ... ,x_N,t) + \Phi_{\rm XX}(x_1, ... , \underbrace{x_j}_{j}, \underbrace{x_j}_{j+1},  ... ,x_N,t) = 0.
\label{S_EoMXX2}
\end{align}
The initial state of Eq.~\eqref{S_initialXX1} reads 
\begin{align} 
\Phi_{\rm XX}(x_1,...,x_N,0) = \prod_{j=1}^{N} \delta_{x_j, y_j}. 
\label{S_InitialXX2}
\end{align}
Under this setup, the exact solutions for the many-body wavefunction becomes 
\begin{align} 
\Phi_{\rm XX}(x_1,...,x_N,t) &= \int_{C_r} d \bm{\xi} \sum_{\sigma \in \mathbb{S}_N }  {\rm sgn}(\sigma) \prod_{j=1}^N\xi_{\sigma_j}^{x_{j}-y_{\sigma_j} -1} e^{- {\rm i}E_{\xi_j} t}.
\label{S_XX_Esol1}
\end{align}
In the following subsections, we prove that Eq.~\eqref{S_XX_Esol1} solves Eqs.~\eqref{S_EoMXX1}, \eqref{S_EoMXX2}, and \eqref{S_InitialXX2}.

\subsection{Proof for Eq.~(\ref{S_XX_Esol1}) satisfying the equation~(\ref{S_EoMXX1}) of motion}
We substitute Eq.~\eqref{S_XX_Esol1} into Eq.~\eqref{S_EoMXX1}, obtaining
\begin{align} 
{\rm LHS~of~Eq.~\eqref{S_EoMXX1} } &= \int_{C_r} d \bm{\xi} \sum_{\sigma \in \mathbb{S}_N }  {\rm sgn}(\sigma)  \left( \sum_{k=1}^N E_{\xi_k}\right) \left( \prod_{j=1}^N\xi_{\sigma_j}^{x_{j}-y_{\sigma_j} -1} e^{- {\rm i}E_{\xi_j} t} \right), \\
{\rm RHS~of~Eq.~\eqref{S_EoMXX1} } &= \dfrac{1}{2} \sum_{k=1}^N \int_{C_r} d \bm{\xi} \sum_{\sigma \in \mathbb{S}_N }  {\rm sgn}(\sigma)   \left( \xi_{\sigma_k} + \xi_{\sigma_k}^{-1} \right) \left( \prod_{j=1}^N\xi_{\sigma_j}^{x_{j}-y_{\sigma_j} -1} e^{- {\rm i}E_{\xi_j} t} \right) \\
&= \int_{C_r} d \bm{\xi} \sum_{\sigma \in \mathbb{S}_N }  {\rm sgn}(\sigma)  \left\{ \dfrac{1}{2} \sum_{k=1}^N \left( \xi_{\sigma_k} + \xi_{\sigma_k}^{-1} \right)  \right\} \left( \prod_{j=1}^N\xi_{\sigma_j}^{x_{j}-y_{\sigma_j} -1} e^{- {\rm i}E_{\xi_j} t} \right)\\
&= \int_{C_r} d \bm{\xi} \sum_{\sigma \in \mathbb{S}_N }  {\rm sgn}(\sigma)  \left\{ \dfrac{1}{2} \sum_{k=1}^N \left( \xi_{k} + \xi_{k}^{-1} \right)  \right\} \left( \prod_{j=1}^N\xi_{\sigma_j}^{x_{j}-y_{\sigma_j} -1} e^{- {\rm i}E_{\xi_j} t} \right) \\
&= \int_{C_r} d \bm{\xi} \sum_{\sigma \in \mathbb{S}_N }  {\rm sgn}(\sigma)  \left( \sum_{k=1}^N E_{\xi_k}  \right) \left( \prod_{j=1}^N\xi_{\sigma_j}^{x_{j}-y_{\sigma_j} -1} e^{- {\rm i}E_{\xi_j} t} \right).
\label{S_EoMXX3}
\end{align}
Thus, we can prove that Eq.~\eqref{S_XX_Esol1} satisfies the equation~\eqref{S_EoMXX1} of motion. 

\subsection{Proof for Eq.~(\ref{S_XX_Esol1}) satisfying the boundary condition of Eq.~(\ref{S_EoMXX2}) }
We substitute Eq.~\eqref{S_XX_Esol1} into Eq.~\eqref{S_EoMXX2}, obtaining
\begin{align} 
&{\rm LHS~of~Eq.~\eqref{S_EoMXX2} } \\
&=  \int_{C_r} d \bm{\xi} \sum_{\sigma \in \mathbb{S}_N }  {\rm sgn}(\sigma)   \left( \prod_{l \in \{1,.., N\}: l \neq j, j+1} \xi_{\sigma_l}^{x_{l}-y_{\sigma_l} -1} e^{- {\rm i}E_{\xi_l} t} \right) e^{- {\rm i}(E_{\xi_j} + E_{\xi_{j+1}}) t}
\left(  \xi_{\sigma_j}^{x_{j}-y_{\sigma_j} } \xi_{\sigma_{j+1}}^{x_{j}-y_{\sigma_{j+1}} } +  \xi_{\sigma_j}^{x_{j}-y_{\sigma_j} -1} \xi_{\sigma_{j+1}}^{x_{j}-y_{\sigma_{j+1}} - 1 }  \right) \\
&=  \int_{C_r} d \bm{\xi} \sum_{\sigma \in \mathbb{S}_N }  {\rm sgn}(\sigma)   \left( \prod_{l \in \{1,.., N\}: l \neq j, j+1} \xi_{\sigma_l}^{x_{l}-y_{\sigma_l} -1} e^{- {\rm i}E_{\xi_l} t} \right) e^{- {\rm i}(E_{\xi_j} + E_{\xi_{j+1}}) t}
\xi_{\sigma_j}^{x_{j}-y_{\sigma_j} } \xi_{\sigma_{j+1}}^{x_{j}-y_{\sigma_{j+1}} } \left(  1 +  \xi_{\sigma_j}^{-1} \xi_{\sigma_{j+1}}^{- 1}  \right) \\
&= \int_{C_r} d \bm{\xi} \sum_{\sigma \in \mathbb{S}_N } G_{\sigma}^{\rm XX}(\bm{\xi},j).
\end{align}
Here, we define the function $G_{\sigma}^{\rm XX}(\bm{\xi},j) $ as 
\begin{align} 
G_{\sigma}^{\rm XX}(\bm{\xi},j) \coloneqq {\rm sgn}(\sigma)   \left( \prod_{l \in \{1,.., N\}: l \neq j, j+1} \xi_{\sigma_l}^{x_{l}-y_{\sigma_l} -1} e^{- {\rm i}E_{\xi_l} t} \right) e^{- {\rm i}(E_{\xi_j} + E_{\xi_{j+1}}) t}
\xi_{\sigma_j}^{x_{j}-y_{\sigma_j} } \xi_{\sigma_{j+1}}^{x_{j}-y_{\sigma_{j+1}} } \left(  1 +  \xi_{\sigma_j}^{-1} \xi_{\sigma_{j+1}}^{- 1}  \right).
\end{align} 
Using the permutation $\mu$ defined in Eq.~\eqref{S_LC6} for a fixed permutation $\sigma$, we can prove
\begin{align}
 &G_{\sigma}^{\rm XX}(\bm{\xi},j)  + G_{\mu}^{\rm XX}(\bm{\xi},j) \\
 &= {\rm sgn}(\sigma)   \left( \prod_{l \in \{1,.., N\}: l \neq j, j+1} \xi_{\sigma_l}^{x_{l}-y_{\sigma_l} -1} e^{- {\rm i}E_{\xi_l} t} \right)   e^{- {\rm i}(E_{\xi_j} + E_{\xi_{j+1}}) t}
\xi_{\sigma_j}^{x_{j}-y_{\sigma_j} } \xi_{\sigma_{j+1}}^{x_{j}-y_{\sigma_{j+1}}} 
 \left\{  1 +  \xi_{\sigma_j}^{-1} \xi_{\sigma_{j+1}}^{- 1} - \left( 1 + \xi_{\sigma_{j+1}}^{-1} \xi_{\sigma_{j}}^{- 1} \right) \right\} \\
 &=0.
\end{align} 
This completes the proof.

\subsection{Proof for Eq.~(\ref{S_XX_Esol1}) satisfying the initial state of Eq.~(\ref{S_InitialXX2}) }
We give a detailed proof for $\Phi_{\rm XX}(x_1,...,x_N,0) = \prod_{j=1}^N \delta_{ x_j , y_j }$ using Eq.~\eqref{S_XX_Esol1}.  
Let me put $t=0$ into Eq.~\eqref{S_XX_Esol1}:
\begin{align} 
\Phi_{\rm XX}(x_1,...,x_N,0) &=  \sum_{\sigma \in \mathbb{S}_N }  {\rm sgn}(\sigma) \prod_{j=1}^N \delta_{x_{j}, y_{\sigma_j}} \\
&= \prod_{j=1}^N \delta_{x_{j}, y_j}.
\label{S_XX_Esol2}
\end{align}
The equality of Eq.~\eqref{S_XX_Esol2} is proved by considering the inequalities $x_{j} +1 \leq x_{j+1} $ and $y_{j} +1 \leq y_{j+1} $. Suppose that there is an inversion in a given $\sigma$ such that $\sigma_m > \sigma_n$ holds for $m < n$. From the inequalities $x_{j} +1 \leq x_{j+1} $ and $y_{j} +1 \leq y_{j+1} $, elementary calculation leads to $x_{n} - x_{m} \geq n-m$ and $y_{\sigma_m} - y_{\sigma_n} \geq \sigma_m - \sigma_n$. Then we can show an inequality $x_{n} - x_{m} + m - n < 0$ by 
\begin{align} 
x_{n} - x_{m} + m - n  &= y_{\sigma_{n}} - y_{\sigma_{m}} + m - n \\&
\leq  \underbrace{\sigma_n - \sigma_m}_{<0} + \underbrace{m - n}_{<0}
\label{S_XX_Esol3}
\end{align}
This proves the equality of Eq.~\eqref{S_XX_Esol2}.

\subsection{Determinantal formula for the exact many-body wavefunction}
We derive the determinantal formulation for the many-body wavefunction. Using Eq.~\eqref{S_XX_Esol1} and the definition of a determinant, we obtain
\begin{align} 
\Phi_{\rm XX} (x_1,...,x_N,t) &= \int_{C_r} d \bm{\xi} \det \left(  \xi_{k}^{x_{j}-y_{k} -1} e^{- {\rm i}E_{\xi_j} t} \right)_{j,k \in \{1,...,N\} }.
\label{S_XX_Esol4}
\end{align}
When the initial state is the domain-wall state ($y_j=j$), we obtain
\begin{align} 
\Phi_{\rm XX} (x_1,...,x_N,t) &= \int_{C_r} d \bm{\xi} \det \left(  \xi_{k}^{x_{j}-k -1} e^{- {\rm i}E_{\xi_j} t} \right)_{j,k \in \{1,...,N\} }.
\label{S_XX_Esol5}
\end{align}
This completes the poof of Eq.~\eqref{MBW4}. 

\clearpage
\section{ Asymptotic analysis to the GUE Tracy-Widom ditribution with fast convergence }\label{sec:Asy}
In the main text, we derive the GUE Tracy-Widom distribution using the scaling variable $s$ defined through $x = 2 + \lfloor - t - (t/2)^{1/3} s \rfloor$. This definition of the scaling variable $s$ is based on the previous work~\cite{Eisler2013}. In this section, we introduce another scaling variable $u$ defined through $x = 2 + \lfloor -t - a - (t/2)^{1/3} u \rfloor$ with the fixed parameter $a$, explaining that the choice of the value $a=1/2$ leads to faster convergence of $F(x,t)$ to the GUE Tracy-Widom distribution in comparison with the previous work with $a=0$. This fast convergence was mathematically studied in the context of classical stochastic processes in Ref.~\cite{Ferrari2011}, which investigated finite-time corrections to nonequilibrium fluctuations in classical models belonging to the Kardar-Parisi-Zhang universality class. We here explain this result with language familiar with physicists. If readers prefer the mathematically rigorous treatment, we recommend them to check the proposition 3.2 of Ref.~\cite{Ferrari2011}. 

\subsection{  Explanation for the asymptotic analysis with fast convergence }
First, following the calculation of Ref.~\cite{Eisler2013}, we rewrite the kernel $K_{\rm B} (t,m,n)$ in the main text by using the formula $J_{m-1}(t) = J_m'(t) + \dfrac{m}{t} J_m(t)$:
\begin{align} 
K_{\rm B} (t,m,n)  &=  \sum_{l= -\infty}^{-1} J_{m-l}(t) J_{n-l}(t)  \\ 
&=\dfrac{t}{2(m-n)} \left(  J_{m}'(t) J_{n}(t) - J_{m}(t) J_{n}'(t) \right) - \dfrac{1}{2} J_{m}(t)  J_{n}(t). \label{S_KB1}
\end{align}
Here, $J_{m}'(t)$ is the Bessel function of the first kind differentiated with respect to $t$.

Second, we revisit the asymptotic analysis for the Bessel function of the first kind. 
We employ the asymptotic analysis for $t \gg 1$, obtaining
\begin{align} 
&J_{\lfloor t + a + (t/2)^{1/3} u \rfloor}(t) \simeq  \left( \dfrac{2}{t} \right)^{1/3}  {\rm Ai}\left( u + a \left(\dfrac{2}{t}\right)^{1/3} \right),  \label{Ai1}\\
&J'_{\lfloor t + a + (t/2)^{1/3} u \rfloor}(t) \simeq  -\left( \dfrac{2}{t} \right)^{2/3} {\rm Ai}'\left( u + a \left(\dfrac{2}{t}\right)^{1/3} \right), \label{Ai2} 
\end{align} 
where ${\rm Ai}'(x)$ denote the Airy function differentiated with respect to $x$. The derivations of these asymptotic formulae are described in the subsequent subsections.

Third, we introduce continuous variables $v$ and $w$ through $m = \lfloor t + a + (t/2)^{1/3} v \rfloor$ and $n = \lfloor t + a + (t/2)^{1/3} w \rfloor $, substituting Eqs.~\eqref{Ai1} and \eqref{Ai2} into Eq.~\eqref{S_KB1}. As a result, we obtain 
\begin{align} 
& K_{\rm B} (t, \lfloor t + a + (t/2)^{1/3} v, \lfloor t + a + (t/2)^{1/3} w \rfloor )  \nonumber \\
& \simeq \left( \dfrac{2}{t} \right)^{1/3}  \dfrac{1}{v-w} 
\left[ {\rm Ai}' \left( v + a \left(\dfrac{2}{t}\right)^{1/3} \right) {\rm Ai} \left( w + a \left(\dfrac{2}{t}\right)^{1/3} \right) - {\rm Ai} \left( v + a \left(\dfrac{2}{t}\right)^{1/3} \right) {\rm Ai}' \left( w + a \left(\dfrac{2}{t}\right)^{1/3} \right) \right] \nonumber \\
&~~~ - \dfrac{1}{2} \left( \dfrac{2}{t} \right)^{2/3}  {\rm Ai} \left( v + a \left(\dfrac{2}{t}\right)^{1/3} \right) {\rm Ai} \left( w + a \left(\dfrac{2}{t}\right)^{1/3} \right).  \label{S_KB2}
\end{align}

Finally, we expand the Airy function in Eq.~\eqref{S_KB2} with respect to the small quantity $a (2/t)^{1/3}$. 
Using the differential equation for the Airy function, namely $d^2{\rm Ai}(x)/ dx^2 = x  {\rm Ai}(x) $, we get
\begin{align} 
K_{\rm B} (t,m,n) \simeq & \left( \dfrac{2}{t} \right)^{1/3}  K_{\rm Ai}(v,w)  + \left( a - \dfrac{1}{2}  \right) \left( \dfrac{2}{t} \right)^{2/3}  {\rm Ai} \left( v \right) {\rm Ai} \left( w \right).  \label{S_KB3}
\end{align}
Here, we define the Airy kernel $K_{\rm Ai}(v,w)$ by 
\begin{align} 
K_{\rm Ai}(v,w) \coloneqq  \dfrac{ {\rm Ai} \left( v  \right) {\rm Ai}' \left( w  \right) - {\rm Ai}' \left( v \right) {\rm Ai} \left( w \right) }{v-w}.
\end{align}
The previous works~\cite{Eisler2013,Saenz2022} set $a=0$, and thus the last term of Eq.~\eqref{S_KB3}, which is proportional to $(2/t)^{2/3}$, survives in the {\it finite time regions}. 
However, when setting $a=1/2$, the term vanishes. As a result, we can expect that the kernel $K_{\rm B} (t,m,n) $ converges to the Airy kernel $K_{\rm Ai}(v,w)$ faster.

\begin{figure*}[t]
\begin{center}
\includegraphics[keepaspectratio, width=15.0cm]{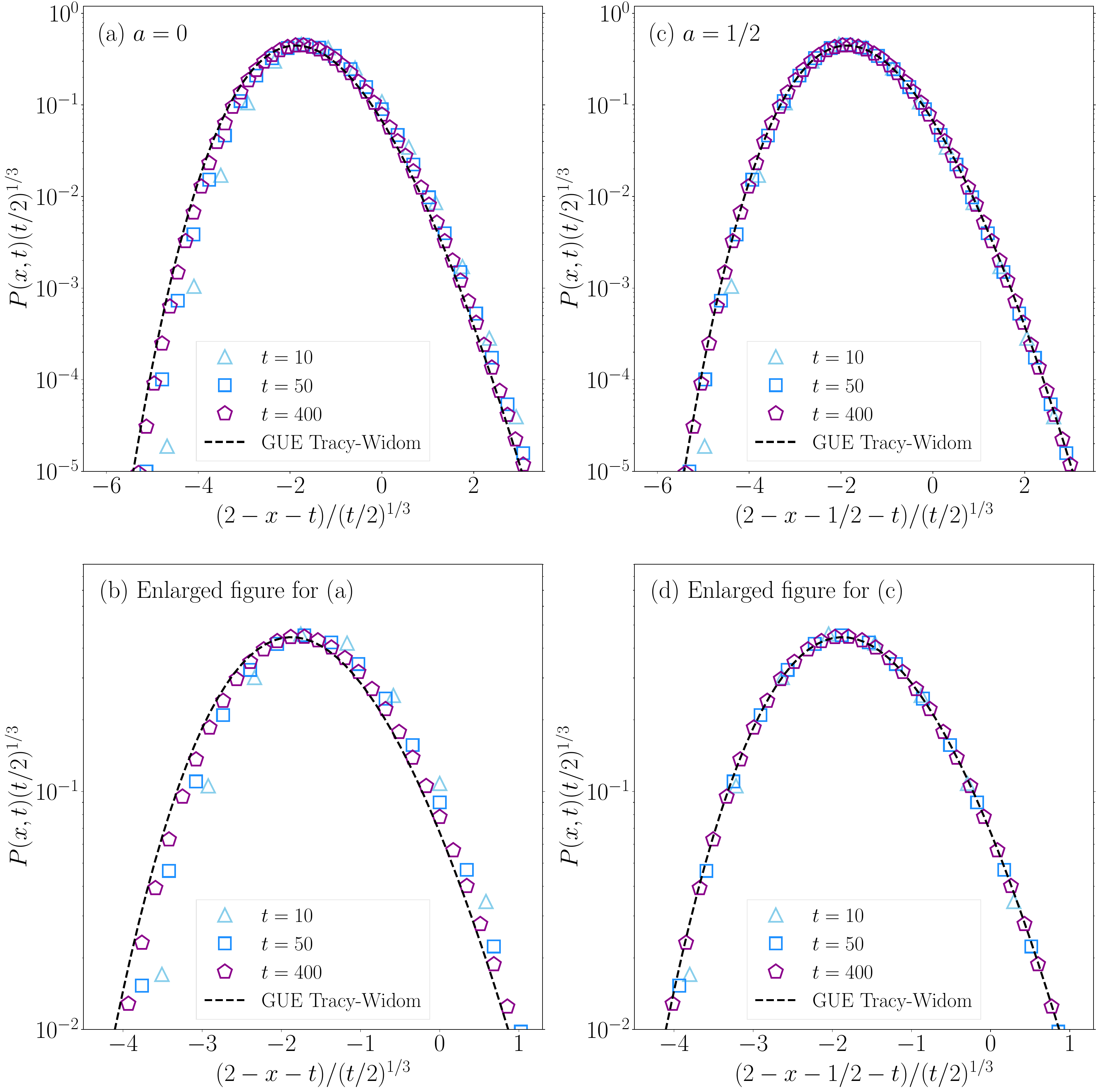}
\caption{Numerical test for the fast convergence of $P(x,t)$ to the probability density function for the GUE Tracy-Widom distribution. The probability $P(x,t)$ is numerically computed using Eq.~\eqref{S_KB1}. 
We show the data with $a=0$ [$a=1/2$] in (a) [(c)], and the corresponding enlarged figure is shown in (b) [(d)]. The dashed lines represent the probability density function $dF_2(s)/ds$ for the GUE Tracy-Widom distribution. 
} 
\label{Sfig1} 
\end{center}
\end{figure*}

\begin{figure*}[t]
\begin{center}
\includegraphics[keepaspectratio, width=18.0cm]{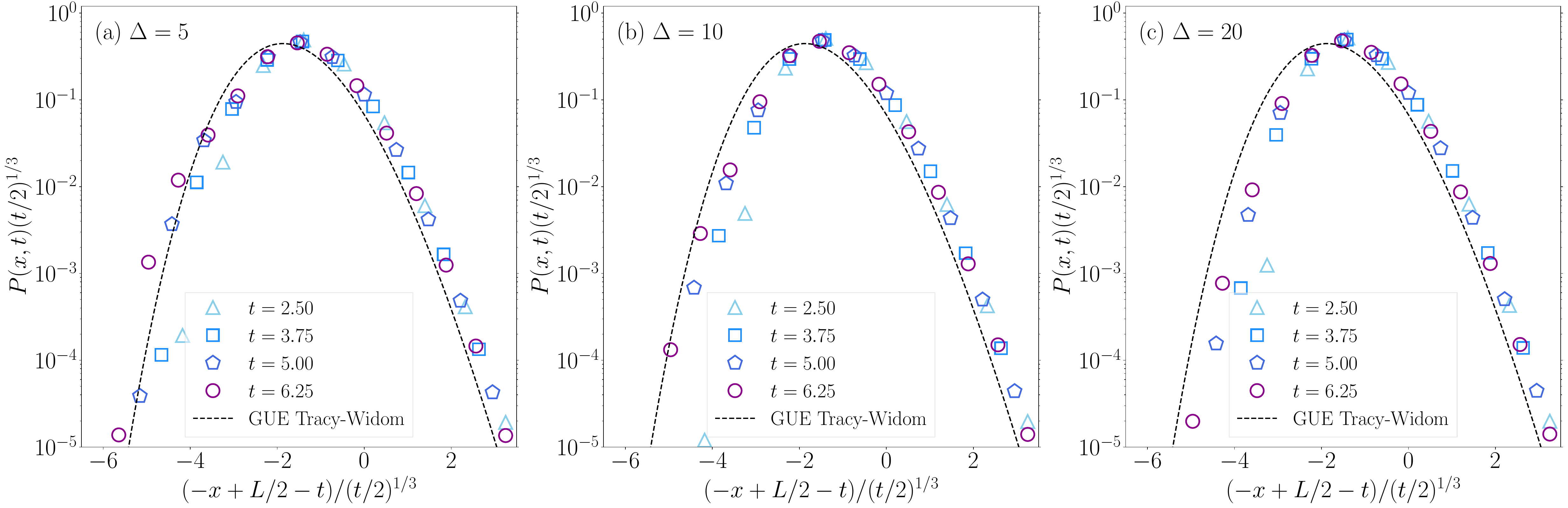}
\caption{Time evolution for the probability $P(x,t)$ in the XXZ model with $\Delta=$ (a) $5$, (b) $10$, and (c) $20$. 
The numerical data displayed in these figures are sama as those in Fig.~\ref{fig2} of the main text. A point being different from Fig.~\ref{fig2} of the main text is the abscissa. We here use the scale with $a=0$ while do the one with $a=1/2$ in Fig.~\ref{fig2} of the main text. The dashed lines represent the probability density function $dF_2(s)/ds$ for the GUE Tracy-Widom distribution. 
} 
\label{Sfig2} 
\end{center}
\end{figure*}

\subsection{ Numerical test for the fast convergence }
We numerically test the fact that the choice of $a=1/2$ leads to the fast convergence of $F(x,t)$ to the GUE Tracy-Widom distribution. 
The formula used here is Eq.~\eqref{Esol6} of the main text, which is given by 
\begin{align} 
F(x,t) = {\rm det} \left( 1 - K_{\rm B} (t,m,n) \right)_{l^2 ( \{ 2-x,3-x,.. \} ) } \label{S_KB1}
\end{align}
with 
\begin{align} 
K_{\rm B} (t,m,n) &= \dfrac{t}{2} \dfrac{ J_{m}(t) J_{n+1}(t) - J_{m+1}(t) J_{n}(t) }{m-n} \\
&= \sum_{l= -\infty}^{-1} J_{m-l}(t) J_{n-l}(t). \label{S_KB2}
\end{align}
Note that Eq.~\eqref{S_KB2} is used when calculating the diagonal element $K_{\rm B} (t,m,m) $. 
Using them, we can numerically obtain the probability $P(x,t) = F(x,t) - F(x+1,t) $ of finding the left-most up-spin at site $x$ and time $t$. 
In the case of Eq.~\eqref{S_KB1}, we use the scaling variable $u$ defined through $2 - x = \lfloor  t + a + (t/2)^{1/3} u \rfloor$, expecting that the choice of $a=1/2$ will exhibit faster convergence of $F(x,t)$ to the GUE Tray-Widom distribution in comparison with the case with $a=0$. 

Figure~\ref{Sfig1} showcases the numerical results for the time evolution of $P(x,t)$ computed by Eq.~\eqref{S_KB1}.
In Figs.~\ref{Sfig1}(a) and (b), we plot the numerical data with $a=0$ for the rescaled abscissa and ordinate, while the numerical data with $a=1/2$ are shown in Figs.~\ref{Sfig1}(c) and (d). 
All the results exhibit that the deviations from the GUE Tracy-Widom decrease in time, but one can clearly see speed of the convergence to the GUE Tracy-Widom in the case with $a=1/2$ is faster than that with $a=0$. 

In Fig.~\ref{fig2} of main text, we use $a=1/2$, finding the good agreement with the GUE Tracy-Widom distribution.
For comparison with Fig.~\ref{fig2} of main text, we also display the same data with $a=0$ in Fig,~\ref{Sfig2}, where the deviations from the GUE Tracy-Widom are large.

\subsection{Derivation of Eq.~(\ref{Ai1})}
We shall derive Eq.~\eqref{Ai1} using the asymptotic analysis. 
The integral representation for $J_{n}(t)$ is given by
\begin{align} 
J_n(t) = \dfrac{1}{2 \pi e^{{\rm i} \pi n/2}} \int_{0}^{2 \pi} d\theta ~{\rm exp} \left( {\rm i} n \theta + {\rm i} t \cos\theta   \right).
\end{align}
Putting $n=\lfloor t + a + (t/2)^{1/3} u \rfloor$ into the above for $t \gg 1$, we derive
\begin{align} 
J_{\lfloor t + a + (t/2)^{1/3} u \rfloor}(t) &\simeq \dfrac{1}{2 \pi e^{{\rm i} \pi (t + a + (t/2)^{1/3} u )/2}} \int_{0}^{2 \pi} d\theta 
~{\rm exp} \left\{ {\rm i} \left[  a + \left( \dfrac{t}{2} \right)^{1/3} u \right] \theta + {\rm i} t (\theta + \cos\theta)    \right\} \label{S_Asy1} \\
& \simeq \dfrac{1}{2 \pi e^{{\rm i} \pi (t + a + (t/2)^{1/3} u )/2}} \int_{0}^{2 \pi} d\theta ~{\rm exp} \left\{ {\rm i} \left[  a + \left( \dfrac{t}{2} \right)^{1/3} u \right] \theta + {\rm i} t \left[ \dfrac{\pi}{2}  + \dfrac{1}{6}\left(\theta - \dfrac{\pi}{2} \right)^3  \right]    \right\} \label{S_Asy2} \\
& = \dfrac{1}{2 \pi } \int_{-\pi/2}^{3\pi/2} d\phi ~{\rm exp} \left\{ {\rm i} \left[  a + \left( \dfrac{t}{2} \right)^{1/3} u \right] \phi + {\rm i} \dfrac{ t }{6} \phi^3     \right\} \label{S_Asy3} \\
& \simeq \dfrac{1}{2 \pi } \left( \dfrac{2}{t} \right)^{1/3} \int_{-\infty }^{ \infty } d\psi ~{\rm exp} \left\{ {\rm i} \left[  u + a \left( \dfrac{2}{t} \right)^{1/3} \right] \psi + {\rm i} \dfrac{ \psi^3 }{3}     \right\} \label{S_Asy4} \\
& = \left( \dfrac{2}{t} \right)^{1/3}  {\rm Ai}\left(  u + a\left( \dfrac{2}{t} \right)^{1/3}  \right). 
\end{align}
In Eq.~\eqref{S_Asy2}, we expand the function $\theta + \cos \theta $ around $\pi/2$ since $e^{{\rm i} t (\theta + \cos \theta )}$ rapidly oscillate as a function of $\theta$ for $t \gg 1$.
To derive the last equality, we use the integral formula given by 
\begin{align} 
{\rm Ai}(x) = \dfrac{1}{\pi} \int_0^{\infty}~d\theta~ \cos \left( x \theta + \dfrac{\theta^3}{3} \right).
\end{align} 
This completes the derivation of Eq.~\eqref{Ai1}.

\subsection{Derivation of Eq.~(\ref{Ai2}) }
We shall derive Eq.~\eqref{Ai2} using the asymptotic analysis. 
Following the calculation being similar to the above, we have
\begin{align} 
J'_{\lfloor t + a + (t/2)^{1/3} u \rfloor}(t) &\simeq \dfrac{1}{2 \pi e^{{\rm i} \pi (t + a + (t/2)^{1/3} u )/2}} \int_{0}^{2 \pi} d\theta 
~ \cos \theta ~{\rm exp} \left\{ {\rm i} \left[  a + \left( \dfrac{t}{2} \right)^{1/3} u \right] \theta + {\rm i} t (\theta + \cos\theta)    \right\} \label{S_Asy5} \\
& \simeq \dfrac{1}{2 \pi e^{{\rm i} \pi (t + a + (t/2)^{1/3} u )/2}} \int_{0}^{2 \pi} d\theta ~\cos \theta~{\rm exp} \left\{ {\rm i} \left[  a + \left( \dfrac{t}{2} \right)^{1/3} u \right] \theta + {\rm i} t \left[ \dfrac{\pi}{2}  + \dfrac{1}{6}\left(\theta - \dfrac{\pi}{2} \right)^3 \right]    \right\} \label{S_Asy6} \\
& = \dfrac{1}{2 \pi } \int_{-\pi/2}^{3\pi/2} d\phi ~\cos (\pi/2 + \phi) ~{\rm exp} \left\{ {\rm i} \left[  a + \left( \dfrac{t}{2} \right)^{1/3} u \right] \phi + {\rm i} \dfrac{ t }{6} \phi^3    \right\} \label{S_Asy7} \\
& \simeq -\dfrac{1}{2 \pi } \left( \dfrac{2}{t} \right)^{2/3} \int_{-\infty }^{\infty} d\psi ~ \psi~{\rm exp} \left\{ {\rm i} \left[  u + a \left( \dfrac{2}{t} \right)^{1/3} \right] \psi + {\rm i} \dfrac{ \psi^3 }{3}      \right\} \label{S_Asy8} \\
& = - \left( \dfrac{2}{t} \right)^{2/3}  {\rm Ai}'\left(  u + a\left( \dfrac{2}{t} \right)^{1/3}  \right).
\end{align}
Here, we use the integral formula given by 
\begin{align} 
{\rm Ai}'(x) = -\dfrac{1}{\pi} \int_0^{\infty}~d\theta~ \theta \sin \left( x \theta + \dfrac{\theta^3}{3} \right).
\end{align} 
This completes the derivation of Eq.~\eqref{Ai2}.

\clearpage
\section{ Numerical truncation in the TEBD method }
We show numerical evidences that our numerical results for the probability $P(x,t)$ well converge in the framework of the TEBD method~\cite{TEBD1,TEBD2,TEBD3,TEBD4}.
In this numerical method, we expand a quantum state $\ket{\psi}$ by the Vidal's canonical form given by
\begin{align} 
\ket{\psi} &= \sum_{\sigma_1 =1}^2 \cdots \sum_{\sigma_L =1}^2 \psi( \sigma_1, ... , \sigma_L) \ket{ \sigma_1, ... , \sigma_L},  \\
\nonumber \\
\psi( \sigma_1, ... , \sigma_L) &= \sum_{\sigma_1 =1}^2 \sum_{\alpha_1 =1}^{\chi_1} \cdots \sum_{\sigma_{N-1} =1}^2 \sum_{\alpha_{N-1} =1}^{\chi_{N-1}} \sum_{\sigma_{N} =1}^2 \Gamma^{[1],\sigma_1}_{\alpha_1} \lambda^{[1]}_{\alpha_1} \Gamma^{[2],\sigma_2}_{\alpha_1,\alpha_2} \lambda^{[2]}_{\alpha_2} \cdots \Gamma^{[N-1],\sigma_{N-1}}_{\alpha_{N-2},\alpha_{N-1}} \lambda^{[N-1]}_{\alpha_{N-1}} \Gamma^{[N],\sigma_N}_{\alpha_{N-1}}. 
\end{align} 
Here, $\lambda$ and $\Gamma$, indices of which are abbreviated, are real and complex coefficients for the many-body wavefunction $\psi( \sigma_1, ... , \sigma_L)$, respectively, and $\chi_{j}$ represents a bond dimension between site $j$ and site $j+1$.  
In general, the bond dimension $\chi_j$ depends on time $t$ during dynamics, and we have 
\begin{align} 
\sum_{\alpha=1}^{\chi_j} \left( \lambda^{[j]}_{\alpha} \right)^2 = 1
\end{align} 
if we do not truncate the bond dimension. In what follows, we assume that $\lambda^{[j]}_{\alpha}$ is in decreasing order in the label $\alpha$.
In numerical calculations based on the TEBD method, one usually truncates the bond dimension. 
To be more specific, we set a threshold parameter $e \in \mathbb{R}$ and discard small $\lambda^{[j]}_{\alpha}$ such that the following inequality holds:
\begin{align} 
\left|~ \sum_{\alpha=1}^{\chi_j^{\rm trun}} \left( \lambda^{[j]}_{\alpha} \right)^2 - 1~ \right|< e, 
\end{align} 
where $\chi_j^{\rm trun}$ indicates the truncated bond dimension. In this numerical method, we need to confirm that the numerical data converges by systematically changing a value of $e$ for the truncation of the bond dimension. 

\begin{figure*}[t]
\begin{center}
\includegraphics[keepaspectratio, width=18.0cm]{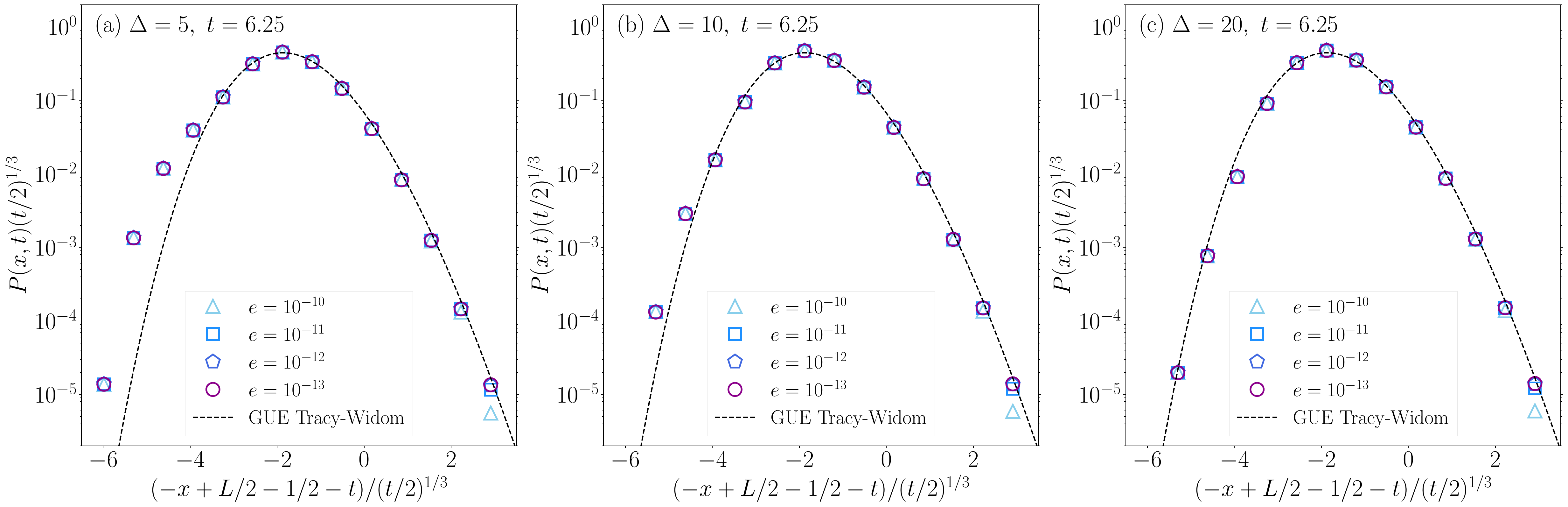}
\caption{Probability $P(x,t)$ at $t=6.25$ in the XXZ model with $\Delta=$ (a) $5$, (b) $10$, and (c) $20$. 
In each panel, the probabilities $P(x,t)$ with the threshold parameter $e=10^{-10}, 10^{-11}, 10^{-12}$ and $10^{-13}$ are shown. 
The numerical data with $e=10^{-13}$ displayed in these figures are sama as those in Fig.~\ref{fig2} of the main text. 
The dashed lines represent the probability density function $dF_2(s)/ds$ for the GUE Tracy-Widom distribution. 
} 
\label{Sfig3} 
\end{center}
\end{figure*}

Figure~\ref{Sfig3} showcases numerical data for $P(x,t)$ at $t=6.25$ obtained by changing a value of the threshold parameter $e$. 
We find that the numerical data of $P(x,t)$ with $e=10^{-13}$, which is displayed in Fig.~\ref{fig2} of the main text, are well convergent. 

\section{Limiting probability distribution function for $\Delta = 0$}
We shall derive the limiting probability distribution function for the left-most up-spin in the dynamics of the XXZ model with $\Delta=0$.
The initial state is the alternating domain-wall state $\ket{\phi(0)} = \prod_{x=1}^{\infty} \hat{R}_{2x} \ket{0}$.
Our following derivation here is based on Ref.~\cite{Eisler2013} of the main text. 

First, we analytically derive a probability distribution function $F_{\rm XX}(x,t)$ of finding the left-most up-spin at site $x$ and time $t$.  
For this purpose, we introduce a generating function $G(x,t,\lambda)$ as
\begin{align} 
G(x,t,\lambda) \coloneqq \bra{\phi(t)} e^{-\lambda \hat{N}_{x} }  \ket{\phi(t)} 
\end{align} 
with $\hat{N}_{x} \coloneqq \sum_{j=-\infty}^{x} (\hat{Z}_j + 1/2) $. 
Since the XXZ model with $\Delta = 0$ can be mapped into a Hamiltonian for noninteracting fermions via the Jordan-Wigner transformation, we can get
\begin{align} 
G(x,t,\lambda) = {\rm det} \left[ 1 + \left( e^{-\lambda}-1\right) K_{\rm XX}(t,m,n) \right]_{l^2( \{-x,-x+1,..\})}, 
\end{align} 
where the function $K_{\rm XX}(t,m,n) $ is defined by
\begin{align} 
K_{\rm XX}(t,m,n) \coloneqq \sum_{k=1}^{\infty} J_{m+2k}(t) J_{n+2k}(t) 
\end{align} 
Taking the large $\lambda$ limit, we obtain the probability $P_{0}(x,t)$ for finding no up-spins in the region $\{x,x-1,..\}$:
\begin{align} 
P_{0}(x,t) = {\rm det} \left[ 1 - K_{\rm XX}(t,m,n) \right]_{l^2( \{-x,-x+1,..\})} .
\end{align} 
This result leads to 
\begin{align} 
F_{\rm XX}(x,t) = P_{0}(x-1,t) = {\rm det} \left( 1 -  K_{\rm XX}(t,m,n) \right)_{l^2( \{-x+1,-x+2,..\})}.
\label{eq:SFXX}
\end{align}

Second, we shall take the long-time limit for $K_{\rm XX}(t,m,n)$ using scaling variables $y$ and $z$ defined through $ m = 1 - \lfloor t + (t/2)^{1/3}y \rfloor$ and $ n = 1 - \lfloor t + (t/2)^{1/3}z \rfloor$. Following the conventional asymptotic analysis described in Sec.~\ref{sec:Asy}, we obtain
\begin{align} 
K_{\rm XX} \left( t, 1 - \lfloor t + (t/2)^{1/3}y \rfloor, 1 - \lfloor t + (t/2)^{1/3}z \rfloor \right) \simeq \left( \dfrac{2}{t} \right)^{2/3} \sum_{k=1}^{\infty} {\rm Ai}\left( y + 2k \left( \dfrac{2}{t} \right)^{1/3}  \right) {\rm Ai}\left( z + 2k \left( \dfrac{2}{t} \right)^{1/3}   \right)~~~~~(t \gg1 ).
\end{align} 
Next we introduce a new variable $w_k \coloneqq 2 k  \left( \dfrac{2}{t} \right)^{1/3}$, for which the unit of this discretized variable is given by $2 \left( \dfrac{2}{t} \right)^{1/3}$.
As a result, we eventually get
\begin{align} 
K_{\rm XX} \left(t, 1 - \lfloor t + (t/2)^{1/3}y \rfloor, 1 - \lfloor t + (t/2)^{1/3}z \rfloor \right) \simeq \frac{1}{2} \left( \dfrac{2}{t} \right)^{1/3} \int_{0}^{\infty} dw ~{\rm Ai}\left( y + w \right) {\rm Ai}\left( z + w   \right)~~~~~(t \gg1 ).
\label{eq:SKXX}
\end{align} 

Third, we take the long-time limit for $F_{\rm XX}(x,t) $ itself by employing Eq.~\eqref{eq:SKXX} and a scaling variable $s$ defined through $ x = 1 - \lfloor t + (t/2)^{1/3}s \rfloor$. 
We apply the Fredholm expansion to Eq.~\eqref{eq:SFXX}, obtaining for $t \gg 1$, 
\begin{align} 
F_{\rm XX}\left( 1- \lfloor t + (t/2)^{1/3}s,t \right) &= 1 + \sum_{k=1}^{\infty} \dfrac{ (-1)^k  }{k !} \sum_{l_1=-x+1}^{\infty} \cdots \sum_{l_k=-x+1}^{\infty}~{\rm det} \left[ K_{\rm XX}(t,m,n) \right]_{m,n \in \{l_1,...,l_k \}} \\
&\simeq 1 + \sum_{k=1}^{\infty} \dfrac{ (-1)^k  }{k !} \sum_{l_1=-x+1}^{\infty} \cdots \sum_{l_k=-x+1}^{\infty} \left(\dfrac{2}{t} \right)^{k/3} {\rm det} \left[ 
\frac{1}{2}  \int_{0}^{\infty}dw~ {\rm Ai} \left( y_m + w \right) {\rm Ai}\left( y_n + w \right) \right]_{m,n \in \{1,...,k \}} \\
&\simeq 1 + \sum_{k=1}^{\infty} \dfrac{ (-1)^k  }{k !} \int_s^{\infty} dy_1 \cdots \int_s^{\infty} dy_k~  {\rm det} \left[  \frac{1}{2}  \int_{0}^{\infty} dw~ {\rm Ai} \left( y_m + w \right) {\rm Ai}\left( y_n + w \right) \right]_{m,n \in \{1,...,k \}} \\
& = {\rm Det} \left[ 1 - \dfrac{1}{2} K_{\rm Ai}(x,y)\right]_{\mathbb{L}^2(s,\infty)} \\
& =  F_{\rm Ai}(s,1/2).
\label{eq:SF1/2}
\end{align} 
Here, we use the variable $y_m$ through $ l_m = 1 - \lfloor t + (t/2)^{1/3}y_m \rfloor $ in the second line, the Airy Kernel $K_{\rm Ai}(x,y) = \left( {\rm Ai}(x) {\rm Ai}'(y) - {\rm Ai}'(x) {\rm Ai}(y) \right)/(x-y)$ in the fourth line, and $F_{\rm Ai}(s,a) \coloneqq  {\rm Det} \left[ 1 - a K_{\rm Ai}(x,y)\right]_{\mathbb{L}^2(s,\infty)}$ in the fifth line. The expression of $F_{\rm Ai}(s,1/2)$ is different from the GUE Tracy-Widom distribution function $F_2(s) = {\rm Det} \left[ 1 - K_{\rm Ai}(x,y)\right]_{\mathbb{L}^2(s,\infty)}$ by the factor $1/2$ in front of the Airy Kernel. 

\begin{figure*}[t]
\begin{center}
\includegraphics[keepaspectratio, width=14.0cm]{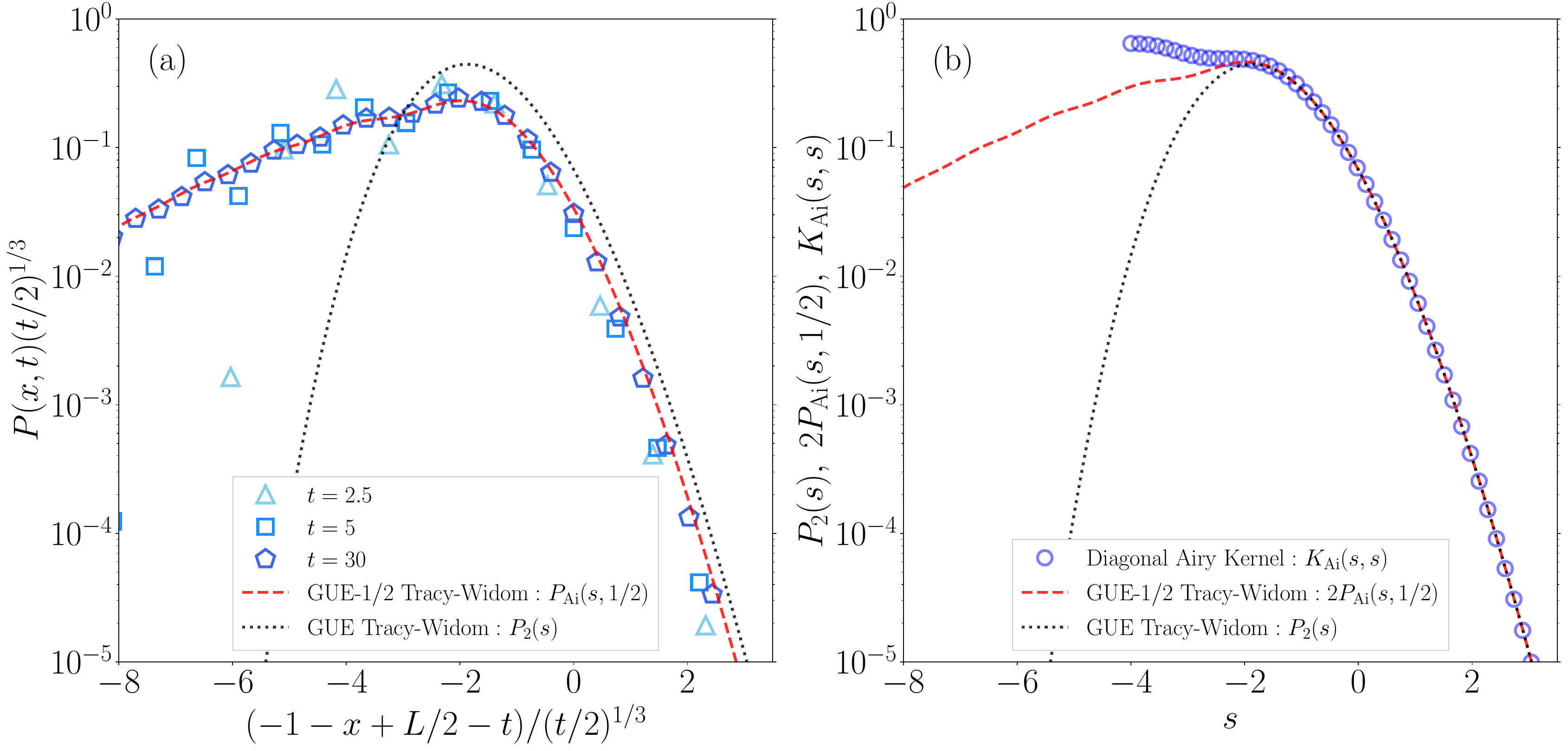}
\caption{
(a) Probability density functions $P_{2}(s)$ and $P_{\rm Ai}(s,1/2)$, and a rescaled probability $P(x,t)$ for the XXZ model with $\Delta=0$. 
The functions $P_{2}(s)$ and $P_{\rm Ai}(s,1/2)$ denote the probability density function for the GUE Tracy-Widom distribution $F_2(s)$ and $F_{\rm Ai}(s,1/2)$ of Eq.~\eqref{eq:SF1/2}. 
The numerical setup for the XXZ model with $\Delta=0$ is same as Fig.~\ref{fig2} of the main text.  
(b) Comparison for $P_2(s)$, $2 P_{\rm Ai}(s,1/2)$, and $K_{\rm Ai}(s,s)$.
} 
\label{Sfig4} 
\end{center}
\end{figure*}

Figure~\ref{Sfig4} (a) displays the probability density functions $P_{2}(s) \coloneqq d F_{2}(s)/ds $ and $P_{\rm Ai}(s,1/2) = d F_{\rm Ai}(s,1/2)/ds $, and the time dependent probability $P(x,t)$ in the XXZ model with $\Delta=0$. The probability density functions $P_{2}(s)$ and $P_{\rm Ai}(s,1/2)$ are numerically computed via the Bornemann method~\cite{Bornemann} and the probability $P(x,t)$ for the XXZ model with $\Delta=0$ is numerically calculated under the same setup of Fig.~\ref{fig2} of the main text. We find that the function $P_{\rm Ai}(s,1/2)$ shows large deviations from $P_{2}(s)$ especially in the left region and the probability $P(x,t)$ becomes closer to $P_{\rm Ai}(s,1/2)$ as time goes by. In Fig.~\ref{Sfig4} (b), we show $P_{2}(s)$, $2 P_{\rm Ai}(s,1/2)$, and $K_{\rm Ai}(s,s)$, finding that they are almost same in the right region. This can be understood by noting that the Airy function is small in the right region. Taking this fact into account, we obtain the approximated expressions in the right region, which are given by
\begin{align} 
P_{2}(s) &\simeq K_{\rm Ai}(s,s), \\
P_{\rm Ai}(s,1/2) &\simeq \dfrac{1}{2} K_{\rm Ai}(s,s).
\end{align} 
This demonstrates the behavior of Fig.~\ref{Sfig4}(b). 

Finally, we consider the dynamics of the XXZ model with $\Delta=0$ starting from the generalized domain-wall state $\ket{\phi(0)} = \prod_{x=1}^{\infty} \hat{R}_{\alpha x} \ket{0}$ characterized by a positive integer $\alpha$. Following almost the same calculations explained just above, we can derive
\begin{align} 
F_{\rm XX}\left( 1- \lfloor t + (t/2)^{1/3}s, t \right) \simeq  F_{\rm Ai}(s,1/\alpha)~~~~~(t \gg 1).
\label{eq:SF1/alpha}
\end{align} 
Then, the corresponding probability $P(x,t)$ with $x = 1- \lfloor t + (t/2)^{1/3}s \rfloor $ in the right region is approximately given by 
\begin{align} 
P_{\rm Ai}(s,1/\alpha) &\simeq \dfrac{1}{\alpha} K_{\rm Ai}(s,s).
\end{align} 
Therefore, the curve of $P(x,t)$ characterized by the diagonal Airy Kernel is universal in the sense that it appears irrespective of $\alpha$.

\section{Conjecture for the XXZ model by Saenz, Tracy, and Widom}
We explain our discussion concerning the universal behavior in the XXZ model, which is described in the main text, after introducing the conjecture given by Saenz, Tracy, and Widom in Ref.~\cite{Saenz2022}.

We first describe the conjecture. The Hamiltonian $\hat{H}$ used in Ref.~\cite{Saenz2022} is 
\begin{align} 
\hat{H} = 2 \sum_{x \in \mathbb{Z}} \left( \hat{X}_{x+1} \hat{X}_{x} + \hat{Y}_{x+1} \hat{Y}_{x} + \Delta \hat{Z}_{x+1} \hat{Z}_{x} \right). 
\end{align}
The initial state is $ \prod_{j=1}^N \hat{R}_{y_j} \ket{0}$ with the integer $y_j~(j \in \{1,2, ..., N)$. 
The quantity of the interest is a probability $\mathcal{F}_N(x,t)$ for finding the left-most up-spin smaller than $x$ at time $t$. 
Under this setup, Saenz, Tracy, and Widom proposed the following conjecture:
\begin{conj}[Ref.~\cite{Saenz2022} of the main text]
As $t \gg N \rightarrow \infty$, $\mathcal{F}_N(x,t)$, with $x = -2t - st^{-1/3}$ and $y_j + 1= v_jt^{-1/3}$, equals to the limit of
\begin{align} 
\sum_{\sigma \in \mathbb{S}_{N}} (-1)^{\sigma} \sum_{S \subset [N]}  (-1)^{|S|} t^{-|S|/3} F(\sigma, S) \prod_{k \in S} \bold{K}_{\rm Ai}( s + v_{\sigma(k)}, s + v_{k})
\end{align} 
with the set $[N] = \{1,2,...,N \}$. The functions $\bold{K}_{\rm Ai}(x,y)$ and $F(\sigma, S)$ are defined by
\begin{align} 
\bold{K}_{\rm Ai}(x,y) &\coloneqq \int_{\infty e^{-2\pi {\rm i}/3}}^{\infty e^{2\pi {\rm i}/3}} d\eta \int_{\infty e^{-\pi {\rm i}/3}}^{\infty e^{\pi {\rm i}/3}} d\xi \dfrac{ \exp \left( \xi^3/3 - \eta^3/3 - x \xi + z \eta \right) }{\xi - \eta}, \\ 
F(\sigma, S) &\coloneqq {\rm i}^{|S^{\rm c}|} \oint_{\hat{\Gamma}} \cdots \oint_{\hat{\Gamma}} B(\xi; \sigma, S) \prod_{j \in S^{\rm c}} \left( \dfrac{ \xi_{\sigma(j)} - (2 \Delta + {\rm i} )   }{ (2 {\rm i} \Delta + 1) \xi_{\sigma(j)} - {\rm i} } \right)^{\nu(\sigma, S)} \times \prod_{j \in S^{\rm c}} ({\rm i} \xi_{\sigma(j)})^{y_j - y_{\tau(j)}-1} d^{\sigma(S^{\rm c})} \xi.
\end{align} 
The definitions of $B(\xi; \sigma, S)$, $\nu(\sigma, S)$, and $\hat{\Gamma}$ are given by Eq.~(96), Eq.~(97), and Fig.~4 of Ref.~\cite{Saenz2022}, respectively. 
\end{conj}

When $s$ is large, the conjecture approximately leads to
\begin{align} 
\mathcal{F}_N(x,t) 
& \sim \sum_{\sigma \in \mathbb{S}_{N}} (-1)^{\sigma} F(\sigma, \phi) + \sum_{\sigma \in \mathbb{S}_{N}} (-1)^{\sigma} \sum_{j=1}^{N}  (-1) t^{-1/3} F(\sigma, \{ j \})  \bold{K}_{\rm Ai}( s + v_{\sigma(j)}, s + v_{j}), 
\label{STWc}
\end{align} 
where the first term is unity as proved in Ref.~\cite{Saenz2022}. Thus, the probability $P(\lfloor -t - s (t/2)^{1/3} \rfloor,t)$ of finding the left-most up-spin for large $s$ can be characterized by the sum of the Airy Kernel. This may be useful for confirming the signature based on our numerical results that this quantity seems to be described by the diagonal Airy Kernel.

\end{document}